%% file: ms.tex
\documentclass[useAMS,usenatbib,usegraphicx]{mn2e}
\usepackage{amsmath,amsfonts,amssymb,url,pslatex,color}
\usepackage{caption}
\input macros.tex
\title[Two Super-Earths in the 3:2 MMR around KOI-1599]%
{Two Super-Earths in the 3:2 MMR around KOI-1599}
\author[F.~Panichi, C.~Migaszewski, K.~Go\'zdziewski]{
F. Panichi$^{1}$\thanks{e-mail: federico.panichi@stud.usz.edu.pl}
,
C. Migaszewski$^{2}$
\thanks{e-mail: migaszewski@umk.pl}
\&
K. Go\'zdziewski$^{2}$
\thanks{e-mail: krzysztof.gozdziewski@umk.pl}
\\
$^{1}$Institute of Physics and CASA*, Faculty of Mathematics and Physics,
University of Szczecin,  Wielkopolska 15, 70-451 Szczecin, Poland\\
$^{2}$Centre for Astronomy, Faculty of Physics, Astronomy and Informatics,
Nicolaus Copernicus University, Grudziadzka 5, 87-100 Toru\'n, Poland \\
}
\begin{document}
%
\date{Accepted 2019 March 5; Received 2019 March 3; in original form 2019 January 5}
\pagerange{\pageref{firstpage}--\pageref{lastpage}} \pubyear{2019}
\maketitle
\label{firstpage}
\begin{abstract}
We validate the planetary origin of the KOI-1599  transit time variations (TTVs) with statistical and dynamical tests. We re-analysed KEPLER Q1-Q17 light-curves of the star, and we independently derived the TTVs. They appear as strongly anti-correlated, suggestive of two mutually interacting planets. We found similar radii of the candidates, $1.9 \pm 0.2 \RE$ for the inner KOI-1599.02, and $1.9 \pm 0.3 \RE$ for the outer KOI-1599.01.  The standard MCMC TTV analysis constrains the planet masses safely below the dynamical instability limit of  $\simeq 3\,\mJ$. The best-fitting MCMC model yields $(9.0\pm0.3)\,\mE$, and $(4.6\pm0.3)\,\mE$, for the inner and the outer planet, respectively. The planets are trapped in 3:2 mean motion resonance (MMR) with anti-aligned apsides ($\Delta \varpi = 180^{\circ}$) at low-eccentric ($e\simeq 0.01)$ orbits. However, we found that the TTV mass determination depends on eccentricity priors with the dispersion in the (0.01,0.05) range. They permit a second family of TTV models with smaller masses of $\simeq 7\,\mE$, and $\simeq 3.6\,\mE$, respectively, exhibiting two modes of $\Delta \varpi = 0^{\circ},180^{\circ}$ librations. The 3:2 MMR is dynamically robust and persists for both modes. In order to resolve the mass duality, we re-analysed the TTV data with a quasi-analytic model of resonant TTV signals. This model favours the smaller masses. We also reproduced this model with simulating the migration capture of the system into the 3:2 MMR. 
\end{abstract}
\begin{keywords}
celestial mechanics -- planetary systems -- stars: individual: KOI-1599
\end{keywords}

\section{Introduction} 

The \kepler{} mission \citep[][]{Borucki2011} discovered hundreds of extra-solar multiple planetary systems (\citep[][]{Akeson2013}, \url{https://exoplanetarchive.ipac.caltech.edu/}).  The distribution of the period-ratio in the \kepler{} sample \citep{Lissauer2011b,Fabrycky2012b,Delisle2014a} shows a paucity of systems near to first-order mean motion resonances (MMRs), with a significant peak close to the 3:2~MMR. Such features of the period-ratio distribution may be related to the formation history and dynamic evolution of multiple planet systems \citep[e.g.,][]{Lithwick2012,Batygin2013,Papaloizou2015}.  It is therefore critical to determine whether a multiple planetary system is dynamically resonant or only close to a strictly resonant configuration \citep[e.g.,][]{Petrovich2013,Goldreich2014}.
                
A crucial data source regarding the multiple \kepler{} planet configurations are the TTV measurements published in recent catalogues by \cite{Rowe2015} and \cite{Holczer2016} (furthermore, H16).  The data span the Q1-Q16 quarters of the \kepler{} light-curves (LCs). Through inspecting these measurements, we selected KOI-1599 with two putative planetary companions, with the inner candidate marked as a possible planet and the outer one not yet examined in the NASA Exoplanet Archive. The TTVs of KOI-1599 exhibit an anti-correlated sinusoidal trend, indicative of two gravitationally interacting objects \citep[e.g.,][]{Steffen2012a,Steffen2015}. Our primary motivation for investigating this putative 2-planet configuration is the proximity of their orbital periods to the 3:2 MMR. We did not find any studies aiming to characterise this interesting and likely resonant system. 

Since the star is dim ($V \simeq15$~mag),  it would be a difficult target for a spectroscopic follow-up, and we aim to constrain masses of the planetary companions with the TTV orbital model \citep[e.g.][]{Agol2005,Holman2010}. Recently, \cite{Baranec2016} imaged $\sim 1000$ dim \kepler{} stars unsuitable for the spectroscopic follow-up. They detected two nearby dim field stars, yet with a substantial  angular separation of $\sim3$ arcsec from KOI-1599, which may dismiss the blend effect. We aim to verify this furthermore on the dynamical grounds, re-compute the planet-to-star radius-ratio, and the density estimates for the candidate planets.

In this work, we follow \cite{Holman2010,Nesvorny2013,MacDonald2016},  as well as \cite{Panichi2018}, regarding the dynamical photometry method. In Sect.~\ref{section1}, we re-analysed the whole Q1-Q17 DR-25 \kepler{} LCs of KOI-1599, and we update the TTVs measurements. In Sect.~\ref{section2}, we validated the two transiting objects as planets. In Sect.~\ref{section3}, we derive the orbital model and masses of the 2-planet configuration.  In Sect.~\ref{section4}, we characterise the 3:2~MMR resonant architecture. In Sect.~\ref{section6} we show that the best-fitting configurations may be interpreted as the natural outcome of planetary migration. We discuss the internal compositions of the planets in Sect.~\ref{section5}. We present our conclusions in Sect.~\ref{conclusion}. Supplementary Material (SM) with source TTV data is presented on-line.

\section{The light-curve analysis and TTVs }
\label{section1} 
Aiming to validate the KOI-1599 planets, we used the \kepler{} photometric data only. We re-analysed the corrected, de-trended LC-INIT light-curves (LCs) from the DR-25 \kepler{} release, spanning the whole Q1-Q17 quarters, also in order to verify and, possibly, refine the previous TTVs measurements.  To avoid confusion, the inner planet KOI-1599.02 has index ``$1$'' and the outer planet KOI-1599.01 is labeled with ``$2$'', respectively. 
With the box-least-squares (BLS) algorithm \citep{Kovacs2002}, we searched for periodic signals in the LCs. The two transiting objects are apparently close to the 3:2~MMR, thus their mid-transit times may be significantly shifted from the linear ephemeris. 

We followed \cite{Panichi2018}, to extract the TTVs from the LC-INIT LCs. We split the LCs in fragments and we selected a narrow window ($\pm 0.5$~days) at multiples of the two periods obtained with the BLS search. Each of them should contain only one transit-like signature.  Once we re-normalised the out-of-transit parts of each fragments, we superimposed them in order to obtain a folded LC for both planets. We used the \code{exonailer} package of \cite{Espinoza2016} for preliminary estimations of the planet-to-star radius-ratio ($p$) and the orbital inclination ($I$). We interpolated the quadratic limb-darkening coefficients based on data in the NASA archive. We kept the photometric noise $\sigma_w$ fixed and we estimated it from the off-transit fragments of the LCs. After fixing these parameters, we fitted mid-transit moments ($T$) for each of the LC fragments. We applied this preliminary list of mid-transit times for re-folding each of the fragments, and we iterated the same procedure until no significant differences  in the best-fitting parameters are present. The inferred parameters and related uncertainties are listed in Tab.~\ref{table:tab1}. In this way, we controlled the derived TTVs, radii $R_i$, $(1.9\pm0.2)$ and $(1.9\pm0.3$) Earth radii, as well as orbital inclinations $I_i$. The system appears as almost co-planar, since the inner planet has $I_{\rm 1}= 88.60\pm0.06$ [deg], while for the outer planet $I_{\rm 2}= 89.78\pm0.1$ [deg]. We checked that star mass and radius from \citep{Rowe2015} are in agreement, within the $1\sigma$ uncertainties, when compared with recent estimates in \cite{Johnson2017} and \cite{SanchisOjeda2012}.
\begin{table}
\centering
\caption{
Parameters inferred from the analysis of the folded light-curve and their uncertainties. The mid-transit time $T_{\rm mid.}$ and mean period $P_{\rm mean}$ from linear ephemeris, planet-to-star radii ratio $p$, semi-major axis in stellar units $a/R_{\star}$, and we report the limb darkening coefficients $q_{\rm 1},q_{\rm 2}$. For all parameters, we estimate the uncertainties as the 16-th, and 84-th percentile of the MCMC samples.}
\label{table:tab1}
\begin{tabular}{l r r}
\hline
Planet & KOI-1599.02 & KOI-1599.01  \\
\hline
\hline
 $P_{\rm mean}\,$[d] & 13.6164 $\pm$0.0001 & 20.408 $\pm$ 0.0004  \\
$T_{\rm mid.}\,$[$\mbox{BJD}$-2454900] & 60.32 $\pm$ 0.01 & 52.73 $\pm$ 0.02 \\
$p$ &  0.0181 $\pm$ 0.0003 & 0.0180 $\pm$ 0.0004 \\
$a/R_{\star}$ & 24.89 $\pm$ 0.11 & 32.63 $\pm$ 0.42 \\
$I$\,[deg] & 88.60$\pm$0.06 & 89.78$\pm$0.16 \\
$q_1$ &  \multicolumn{2}{c}{0.2602 (fixed)}    \\
$q_2$ &  \multicolumn{2}{c}{0.4096 (fixed)} \\
\hline
\end{tabular}
\end{table}

We derived the median values of the  photometric parameters and their  uncertainties with the Markov Chain Monte Carlo (MCMC) affine sampler,  developed in the \code{emcee} package by \cite{ForemanMackey2013}. We reported the mid-transit times and TTV measurements in the SM on-line.

\section{Validation of the planetary TTV origin}
\label{section2} 

\begin{figure*}
\centering
\includegraphics[height=0.21\textheight]{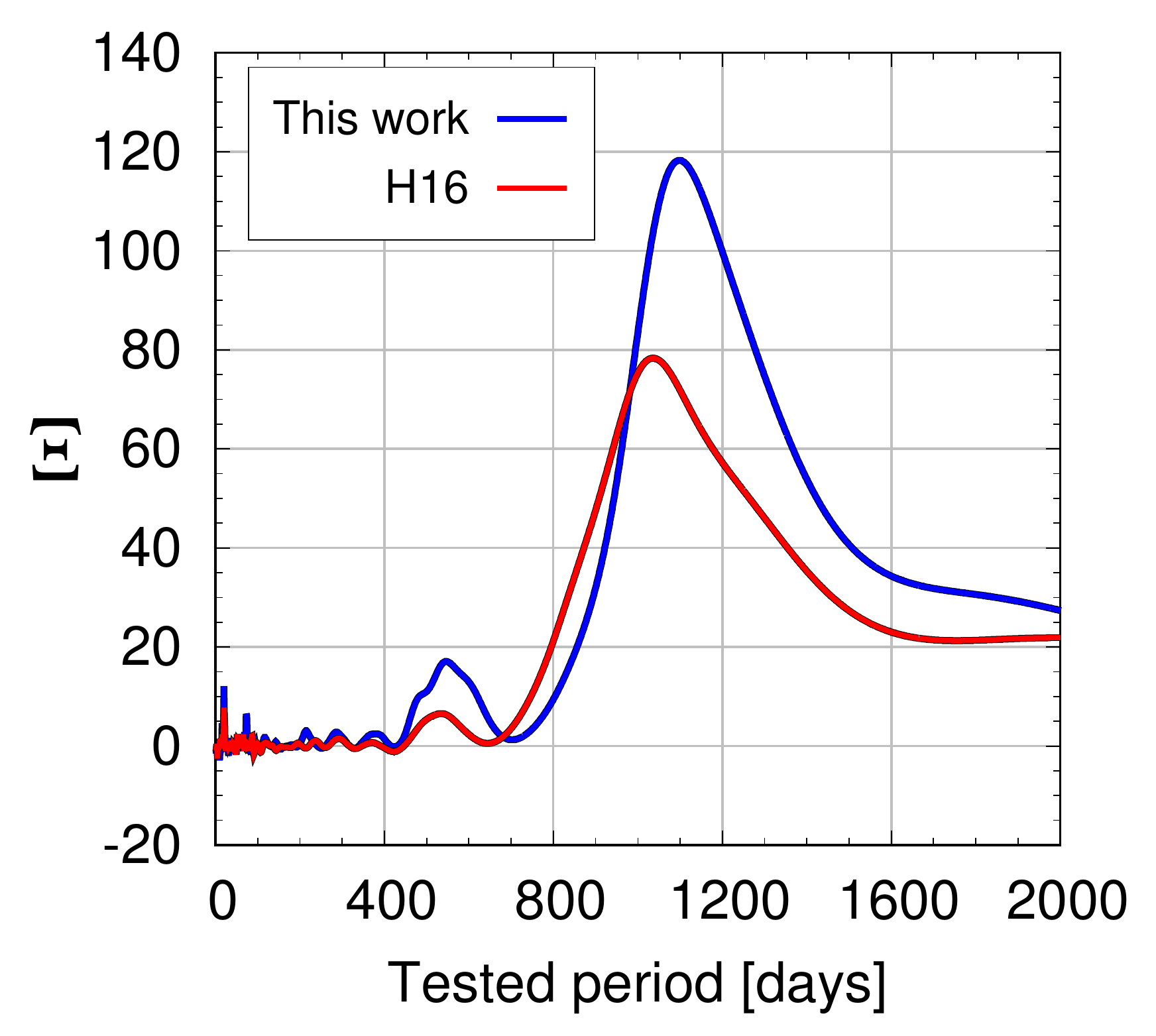}
\includegraphics[height=0.21\textheight]{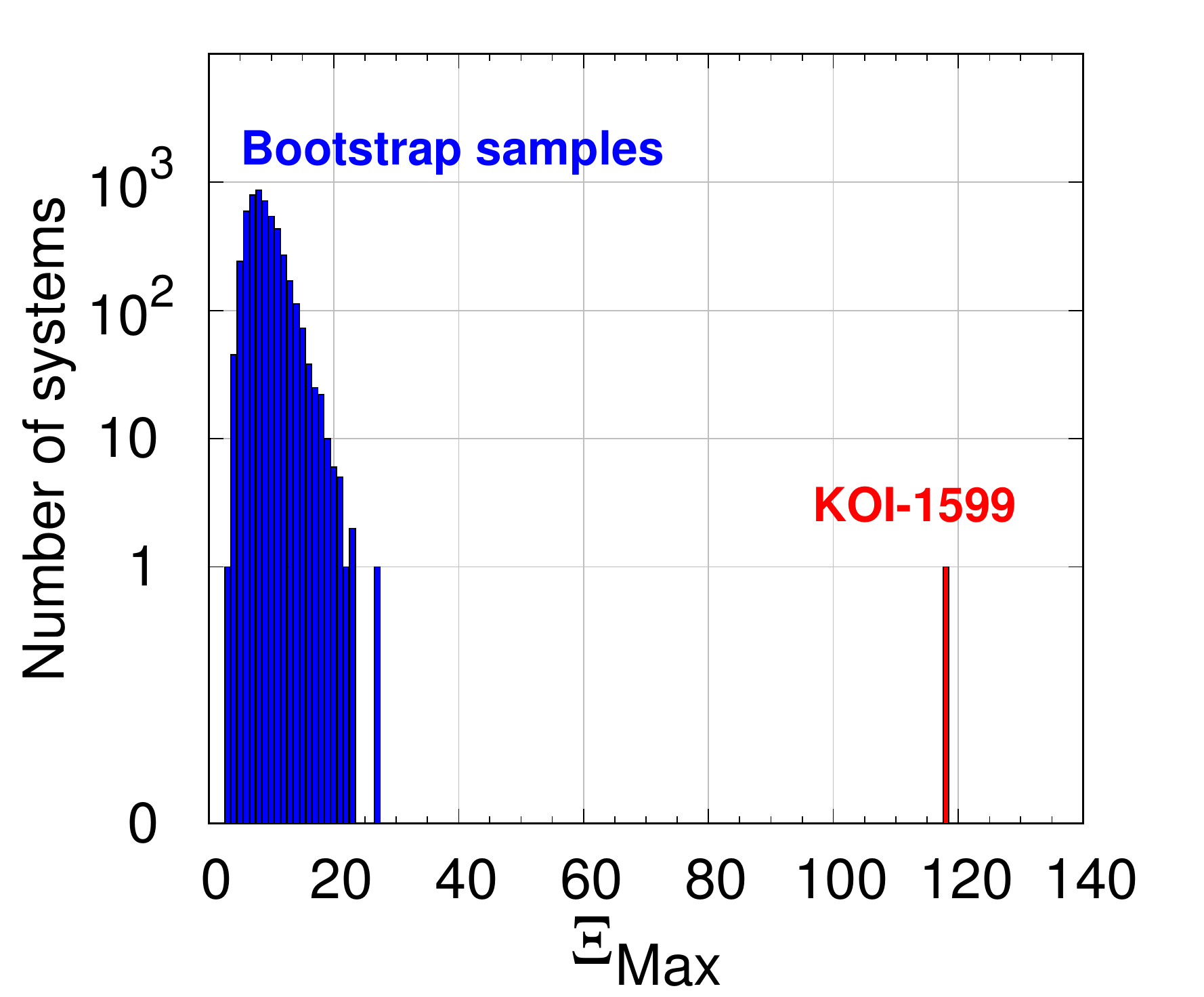}
\includegraphics[height=0.21\textheight]{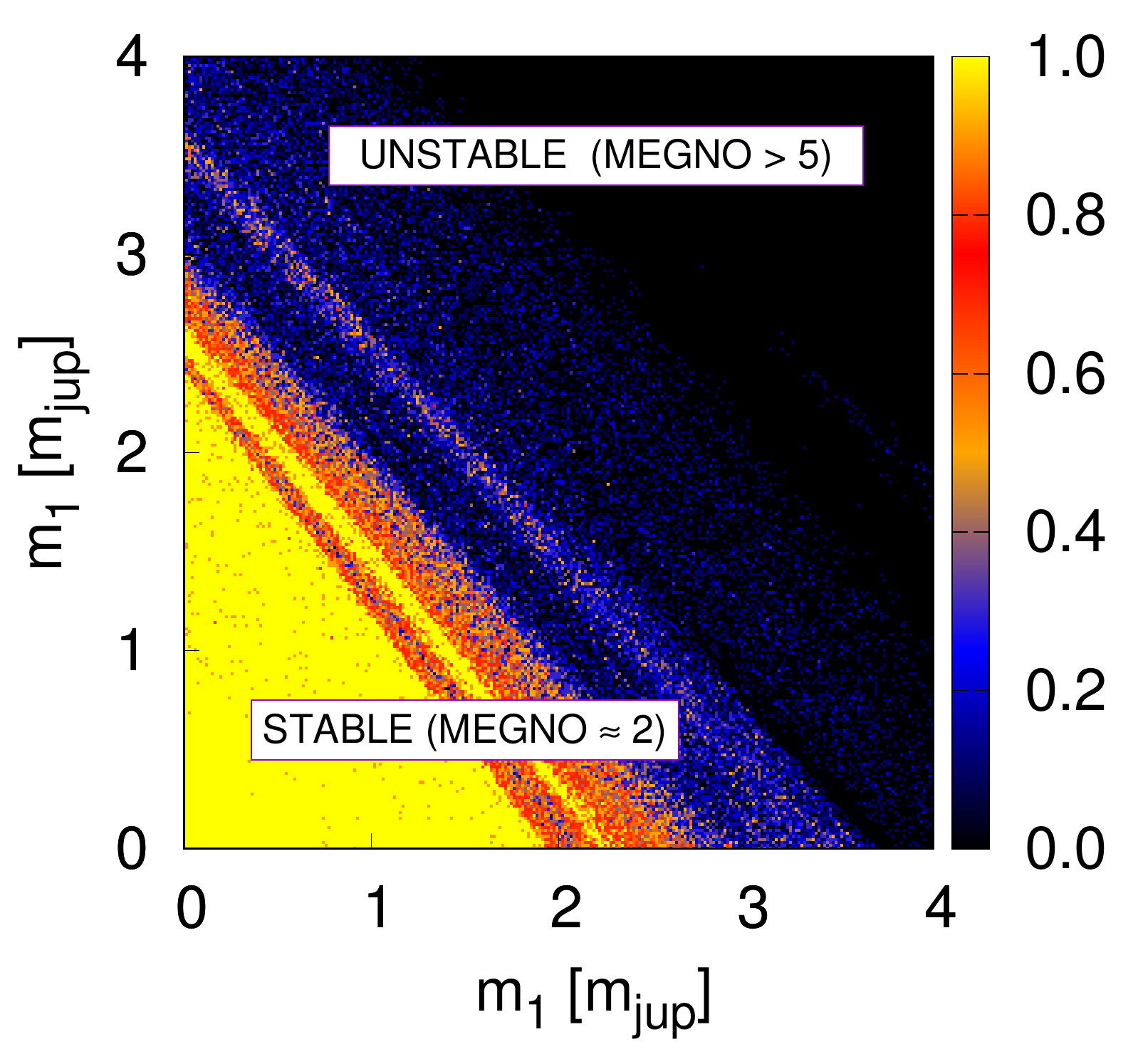}
\caption{
{\em Left}: Cross-correlation in terms of the $\Xi$ statistics, for our new TTV data (blue line) and H16 dataset (red line), respectively. The modulation period's peak of $\simeq 1100$~d exceeds other peaks.  {\em  Middle}: Monte Carlo bootstrap analysis of KOI-1599. The maximum of $\Xi$ for the observed system (red) is much larger than for any of the synthetic configurations (blue). {\em Right}: the probability of stable systems in the $(m_1,m_2)$-plane. We used the Mean Exponential Growth factor of Nearby Orbits (MEGNO) to compute the dynamical stability of each initial conditions. See Sect.~\ref{section2} for details.
}
\label{figure:fig1}
\end{figure*} 

We applied tests originally proposed by the \kepler{} team to validate the majority of candidate planets in multiple systems \citep[e.g.][]{Lissauer2011b,Steffen2012a,Rowe2014}.
 
The anti-correlation (cross-correlation) of the TTVs can be used to confirm that two objects orbit the same star \citep{Steffen2012a} and their observed anti-correlated TTV signals cannot be a random noise. This method relies on the Monte Carlo bootstrap analysis. We fitted a sinusoidal signal with a fixed modulation period to the TTV series of each planet with the Levenberg-Marquardt algorithm \citep{Press1986}.  For each of the sampled periods, the cross-correlation statistics $\Xi$ \citep{Steffen2012a} measures the fit quality --- the larger the $\Xi$, the better the quality.  Figure~\ref{figure:fig1} shows  $\Xi$ for the KOI-1599 TTV signals, as a function of the modulation period. Our new TTVs dataset exhibits a peak at a position similar to the one in the H16 data, but it seems to have even larger significance.

The next step consists of the Monte Carlo bootstrap analysis \citep[e.g.,][]{Press1986}. We tested $5\times10^4$ synthetic  TTV datasets obtained by random shuffling of the original TTVs, with their uncertainties, and the mid-transit times. As for the real TTV data, we fitted each of the randomly generated datasets with the sine function, and the maximum value of its  $\Xi$ is recorded.  We illustrate the results in Fig.~\ref{figure:fig1} (middle panel).  The False Alarm Probability (FAP)  is defined as the ratio of synthetic systems with $\Xi$ larger than  that of the observed system to the total number of samples (here, $5\times10^4$).  The histogram indicates that  unlikely  the sinusoidal TTVs are  artefacts. The FAP is smaller than  $10^{-3}$ adopted by \cite{Steffen2012a} for validating other \kepler{} Objects of Interest  (KOIs).  We conclude that the KOI-1599 data passes the TTV anti-correlation test.

As a second validation test, we used the dynamical stability constraints. We note that this experiment was not based on the TTVs measurements, rather than on canonical information regarding the orbital  periods of transiting objects inferred from the LCs.

In order to conduct the test, we changed the masses, $m_{\rm 1}$ and $m_{\rm 2}$, of the two putative transiting objects in a two--dimensional grid of $512\times512$ points. At each point, we constructed a number of synthetic configurations, by fixing their orbital periods (and semi-major axes) in accordance with the mean photometric periods. The mean anomalies and  arguments of periastrons are random in the $[0,360^{\circ}]$ range, and the eccentricities are randomly sampled from $[0,0.1]$. In order to check  the stability of these synthetic systems, we used the Lyapunov-based fast indicator MEGNO \citep{Cincotta2000,Gozdziewski2008}.

We present the results in Fig.~\ref{figure:fig1} for the probability of picking up a stable system in the ($m_{\rm 1}$,$m_{\rm 2}$)-plane, for 100 sampled configurations with a fixed pair of masses. Clearly,  stable systems are possible unless the masses are larger than $2$--$3$~m$_{\rm Jup}$. Beyond this limit, the probability of guessing a stable configuration sharply decreases. Moreover, the masses in stable systems are well below the planetary threshold of $\sim 14$~m$_{\rm Jup}$ \citep[e.g.,][]{Spiegel2011}. 

\section{The best-fitting TTV models}
\label{section3} 

We applied the same orbital model and the TTV model optimisation as in our earlier papers \citep[e.g.][]{Gozdziewski2016}. We assumed a co-planar system, and the model parameters $\vec{p} =  \{m_i,P_i,x_i\equiv e_i \cos\varpi_i, y_i\equiv e_i\sin\varpi_i,T_i\}$ for $i=1,2$, where $P_i$, $e_i$, $\varpi_i$ and $T_i$ stand for the orbital period, eccentricity, longitude of pericenter and the moment of the first transit, respectively, w.r.t. the initial epoch of $T_0=\mbox{BKJD}-139$~d. 

We performed a preliminary optimisation of the likelihood function $\cal L(\vec{p}$) with the evolutionary algorithms (GEA), but, as anticipated, constraining the eccentricities is difficult due to the mass-eccentricity degeneration  \citep[e.g.,][]{Hadden2014,Deck2015,JontofHutter2016}. 
Therefore, we restrict eccentricities of the two planets in the GEA search, in order to avoid the ``over-fitting'' of the TTVs \citep[e.g.,][]{Migaszewski2017a,Migaszewski2018,MacDonald2016}. 
The limit $e_i < 0.05$ ($i=1,2$) is typical for Earth-like multiple \kepler{} planets near to or involved in MMR \citep[e.g.,][]{Kane2012,Kipping2014,Xie2016,JontofHutter2016,Shabram2016}.

\begin{figure}
\centerline{
\includegraphics[width=0.47\textwidth]{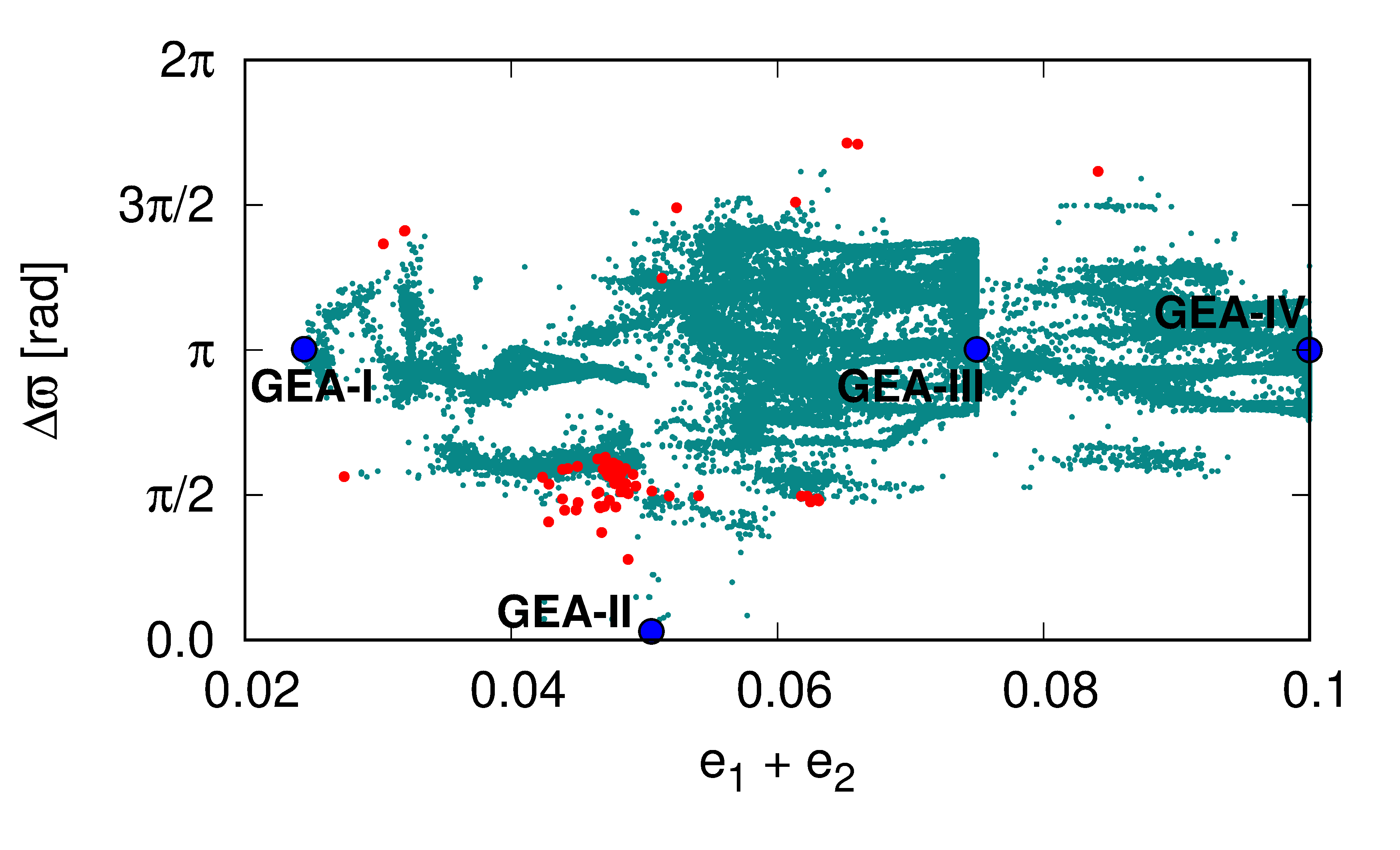}
}
\caption{
Distribution in the $(e_{\rm 1}+e_{\rm 2},\Delta\varpi)$-plane of the statistics obtained with the GEA sampling. We collected $2.5 \times 10^{5}$ best-fitting solutions with $\Chi < 1.15$. The solutions with critical angles librating are shown as gray dots, and solutions with circulating resonant angles are marked with red dots. The integration time for each solution is equal to $1000$~yrs.  Four representative solutions selected and listed in Tab.~\ref{table:tab2} are marked with blue circles. 
}
\label{figure:fig2}
\end{figure}

In Fig.~\ref{figure:fig2}, we project the best-fitting GEA solutions yielding $\chi^2<1.15$ on the $(e_{\rm 1}+e_{\rm 2},\Delta\varpi)$-plane of the osculating elements. For these solutions, we also computed the amplitude of the secular angle $\Delta\varpi=\varpi_2-\varpi_1$, as well as of the critical angles of the 3:2~MMR, 
$ \phi_{\rm 3:2,1} =  2\lambda_{\rm 1} - 3 \lambda_{\rm 2} + \varpi_{\rm 1}$, 
$\phi_{\rm 3:2,2} =  2\lambda_{\rm 1} - 3 \lambda_{\rm 2} + \varpi_{\rm 2}$,
where $\lambda_i$ is the mean longitude of the $i-th$ planet, $\varpi_i$ its longitude of periastron, and the indexes $1,2$ are for the inner and outer planet, respectively. 

The distribution of $2.5\times 10^{5}$ GEA models illustrated in Fig.~\ref{figure:fig2} clusters around $\Delta\varpi \simeq 180^{\circ}$, and is qualitatively different from that one of Kepler-29, for which most of the eccentricity-unconstrained GEA models are characterised by $\Delta\varpi \sim 0^{\circ}$ \citep{Migaszewski2017a}.  
For the set of models illustrated in Fig.~\ref{figure:fig2}, we calculated the amplitude of the critical and secular angles for a fixed integration time equal to $1000$ yrs, equivalent to $\sim2\times10^5$ dynamical (outer) periods. We selected four representative solutions marked in Fig.~\ref{figure:fig2}. We listed their Keplerian osculating orbital elements in Tab.~\ref{table:tab2}.  The GEA-I solution in Tab.~\ref{table:tab2} represents low-eccentricity anti-aligned configurations, the GEA-II model is an example of moderate-eccentricity, aligned ($\Delta \varpi =0^{\circ}$) systems. The GEA-III model is a representative configuration for anti-aligned solutions with moderate eccentricity, while the GEA-IV model is a representative of the solutions with eccentricities close to the upper limit imposed in the GEA search. The anti-aligned configurations  are the most frequent solutions inferred from the GEA search. 
We found that almost all solutions exhibit one or two critical angles librating, while only 172 models (marked in red) exhibit circulating critical angles. We distinguish between aligned models ($\Delta\varpi\simeq 0^{\circ})$ for which only one critical angle librate, while both angles librate in anti-aligned solutions ($\Delta\varpi\simeq 180^{\circ})$. Remarkably, such behaviour is independent of eccentricities and masses. Moreover, the circulating critical angles remain highly coherent. 
Furthermore, we found that such coherent circulations may be related to the proximity of the solutions to the separatrices of the 3:2 MMR, as will be demonstrated in Sect.~\ref{section4}. Therefore, the GEA experiment results indicate that the system can be considered as resonant in statistical sense. 

\begin{table*}
\centering
\caption{
Keplerian osculating orbital elements and masses of four representative solutions for KOI-1599  derived by the GEA optimisation algorithm. We also report the best-fitting period and initial mid-transit time for each solution. The reference epoch is Barycentric Kepler Julian Day (BKJD) $-$ 139~days. The mass of the KOI-1599 star is $1.02\,\msun$ \citep{Rowe2015}.
}
\label{table:tab2}
\begin{tabular}{l r r r r r r r r}
\hline\hline
Model/ 
& \multicolumn{2}{c}{{\bf GEA~I}  (low $e$, $\Delta\omega \simeq \pi$)}
& \multicolumn{2}{c}{{\bf GEA~II} (low $e$, $\Delta\omega \simeq 0$)}
& \multicolumn{2}{c}{{\bf GEA~III} (moderate $e$, $\Delta\omega \simeq \pi$)}
& \multicolumn{2}{c}{{\bf GEA~IV}  (high $e$, $\Delta\omega \simeq \pi$)}\\
Planet & KOI-1599.02 & KOI-1599.01 & KOI-1599.02 & KOI-1599.01 & 
         KOI-1599.02 & KOI-1599.01 & KOI-1599.02 & KOI-1599.01 \\
\hline
\hline
$P\,$[d] & 
13.6094 & 20.4386 & 13.6099 & 20.4376 & 13.6099  & 20.4361 & 13.6100 & 20.4355\\
$T\,$[d] & 
74.011  & 72.961 & 74.009 &  72.951 &  74.012 & 72.952 & 74.012 &  72.953 \\
$m_p\,[\mE]$ & 
 8.3 & 4.3 &  7.6 & 3.7 &  7.0 & 3.6 &  6.8 & 3.5\\
$a\,[\au]$  &
 0.11230 & 0.14727 & 0.11230 & 0.14726 & 0.11230 & 0.14725  &  0.11230 & 0.14725 \\
$e$  &
 0.0067 & 0.0177 &  0.0062 & 0.0444 & 0.0250 & 0.0500 &  0.050 & 0.050 \\
$\omega\,$[deg] & 
 40.6 & -139.9 & -177.9 & 176.7 &  -24.3 & 155.6 & -31.2 & 148.8 \\
${\cal M}\,$[deg] & 
 176.8 & 31.4 &  34.0 & 71.4 & -116.3 & 92.3 & -107.2 & 99.4 \\
\hline\hline
\end{tabular}
\end{table*}

Due to the short observational window, it is not possible to constrain the architecture of the system by inspecting the GEA models statistics. However, having the best-fitting GEA solutions as the initial guess, we used the MCMC sampling for characterising these solutions through imposing Bayesian priors on the TTV  model parameters. In particular, we set Gaussian priors ${\cal N}(\mu,\sigma)$ for the $(x_i,y_i)$-variables with  $\mu_i=0$ and the same $\sigma_{x_i,y_i}$ for both planets. In order to assess proper values of these priors, we performed the MCMC sampling for $\sigma_{x_i,y_i} \in [0.001,0.12]$, with small steps, following a similar strategy as in \cite{Migaszewski2018}. The priors were set uniform for all other parameters. We iterated 1024 {\tt emcee} walkers  around  selected GEA models for up to 256,000 samples each, aiming to keep the acceptance rate between 0.2 and 0.5.

\begin{figure*}
\centerline{
\hbox{
\quad\includegraphics[height=5cm]{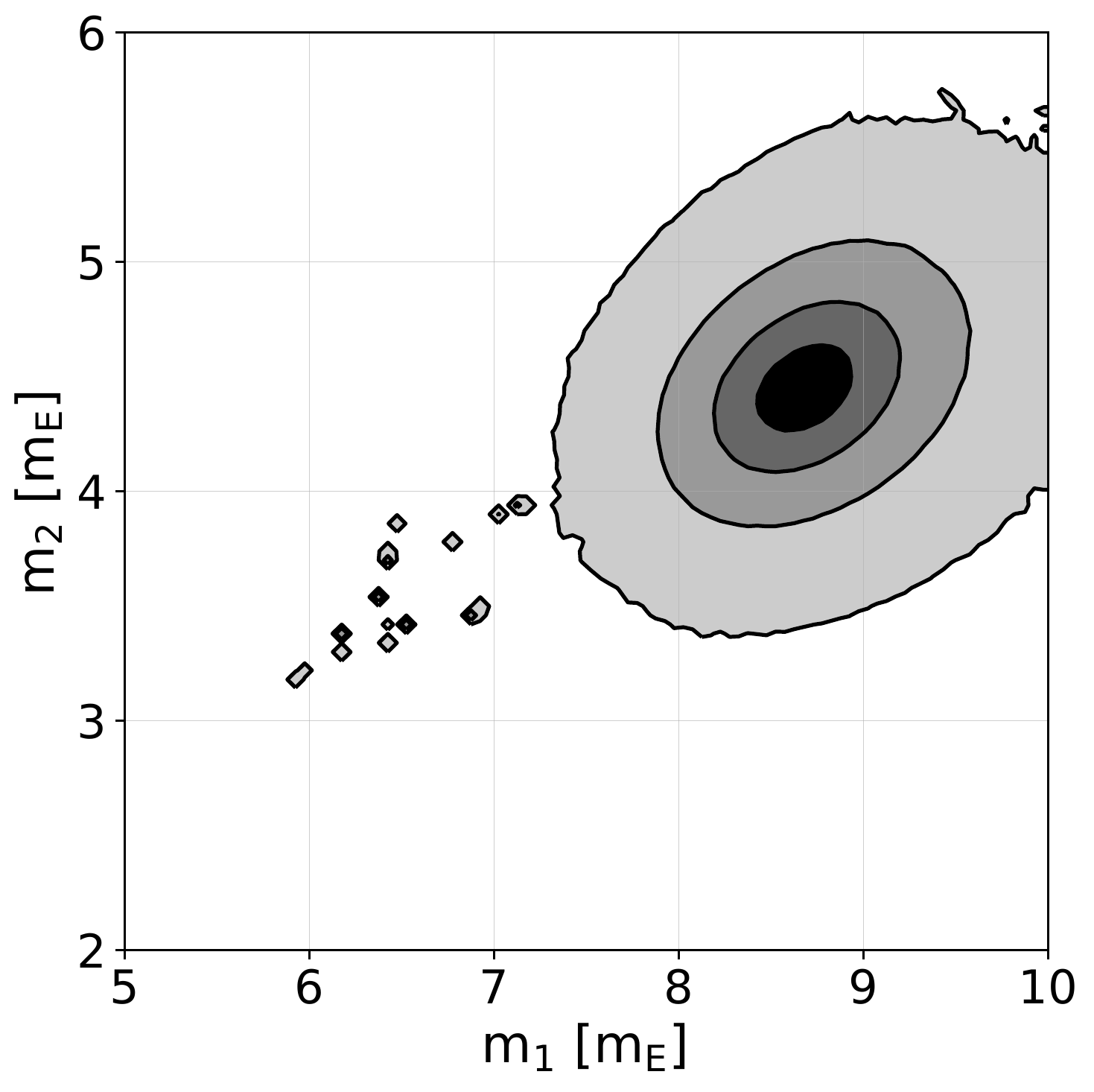}
\quad\includegraphics[height=5cm]{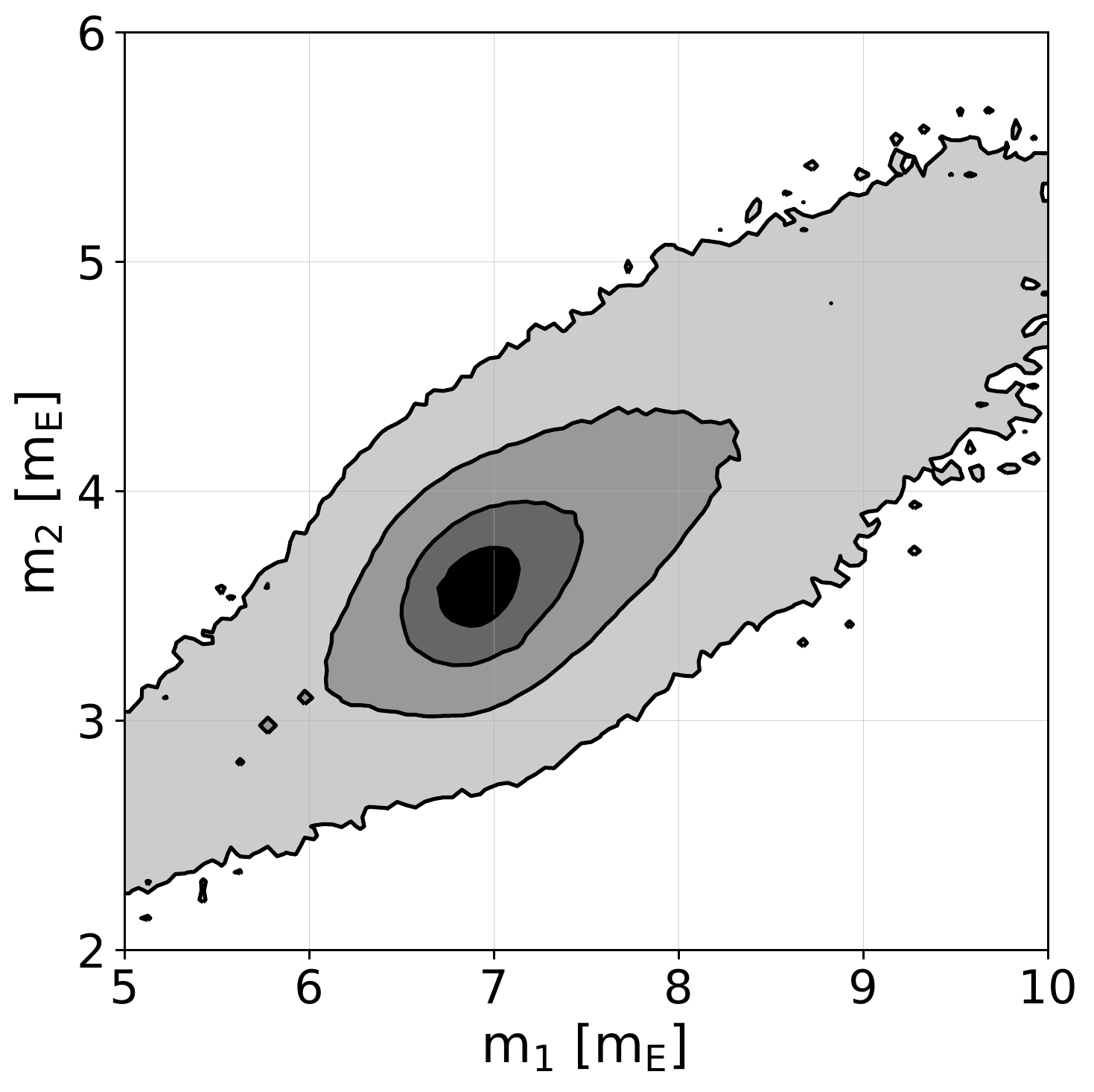}
\includegraphics[height=5cm]{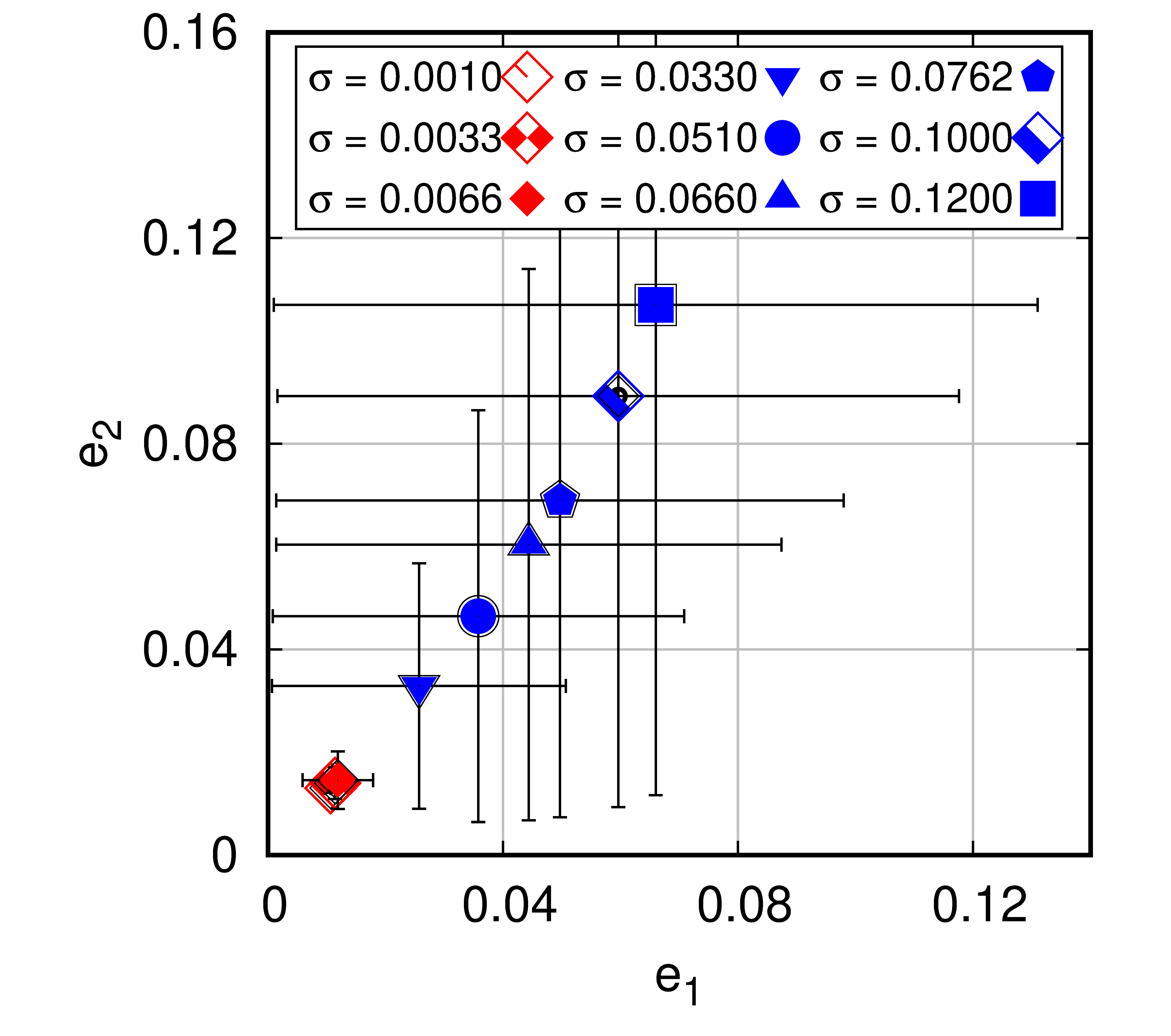}
}}
\centerline{
\hbox{
\includegraphics[height=5cm]{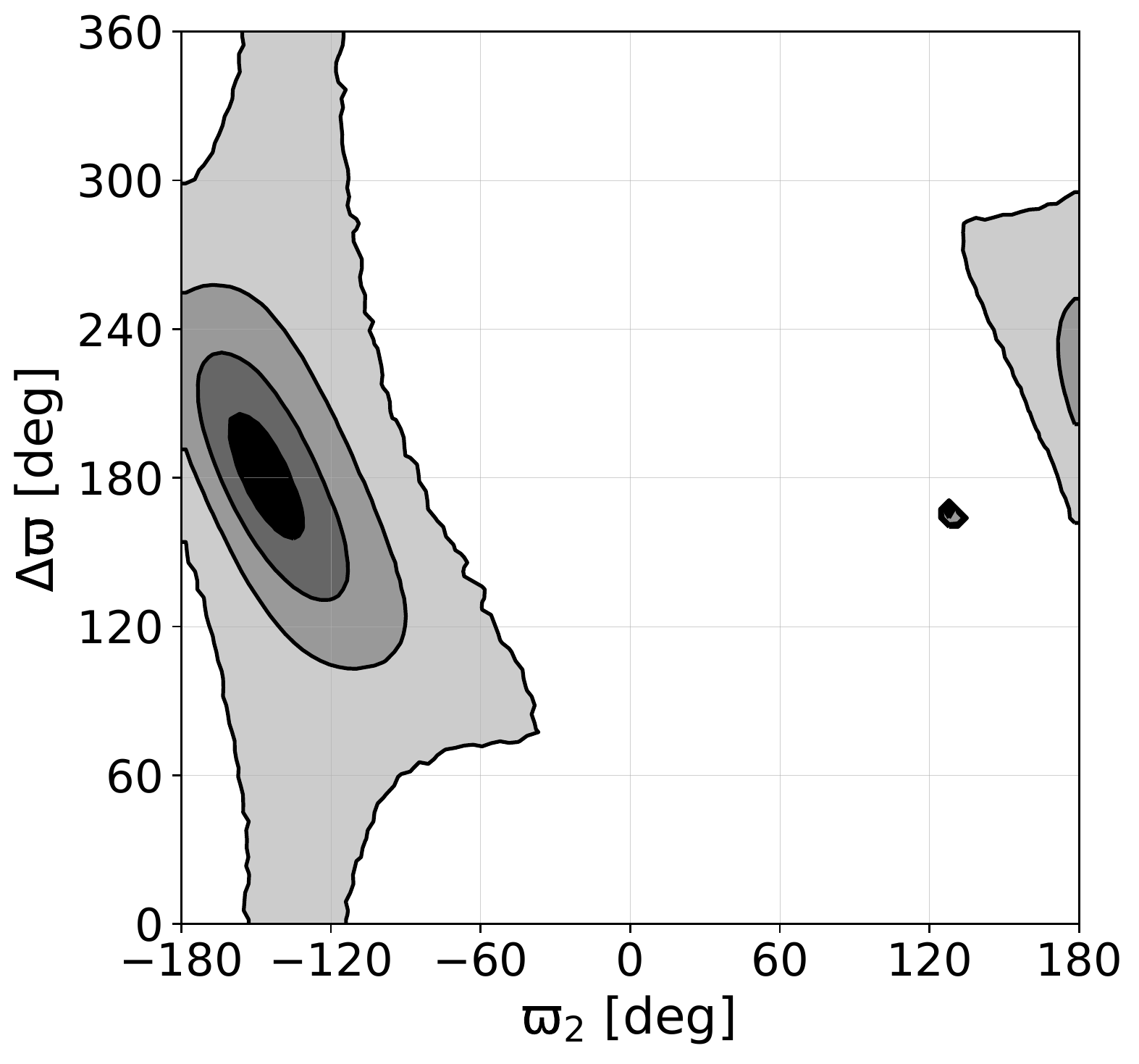}
\includegraphics[height=5cm]{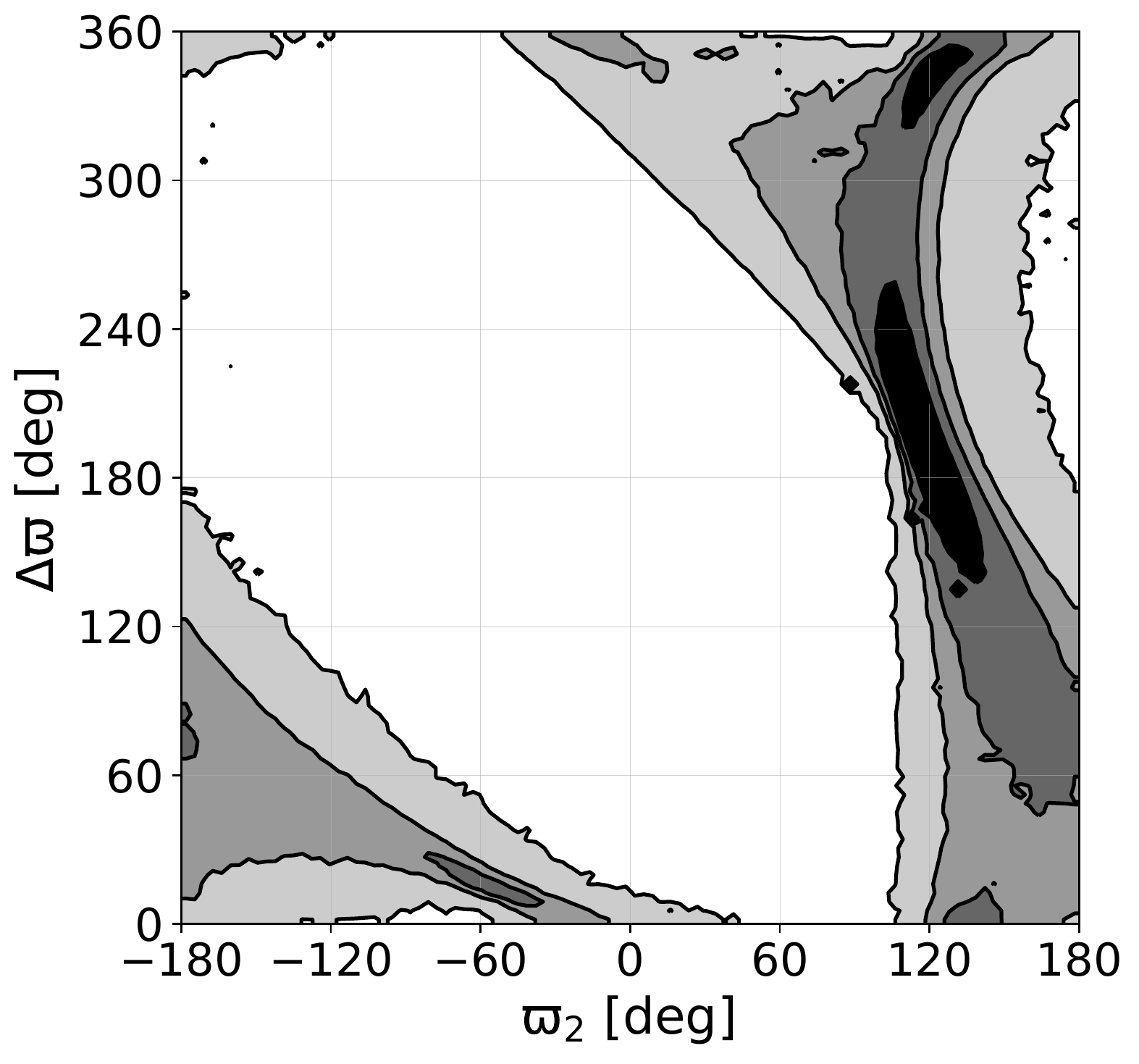}
\quad\includegraphics[height=5cm]{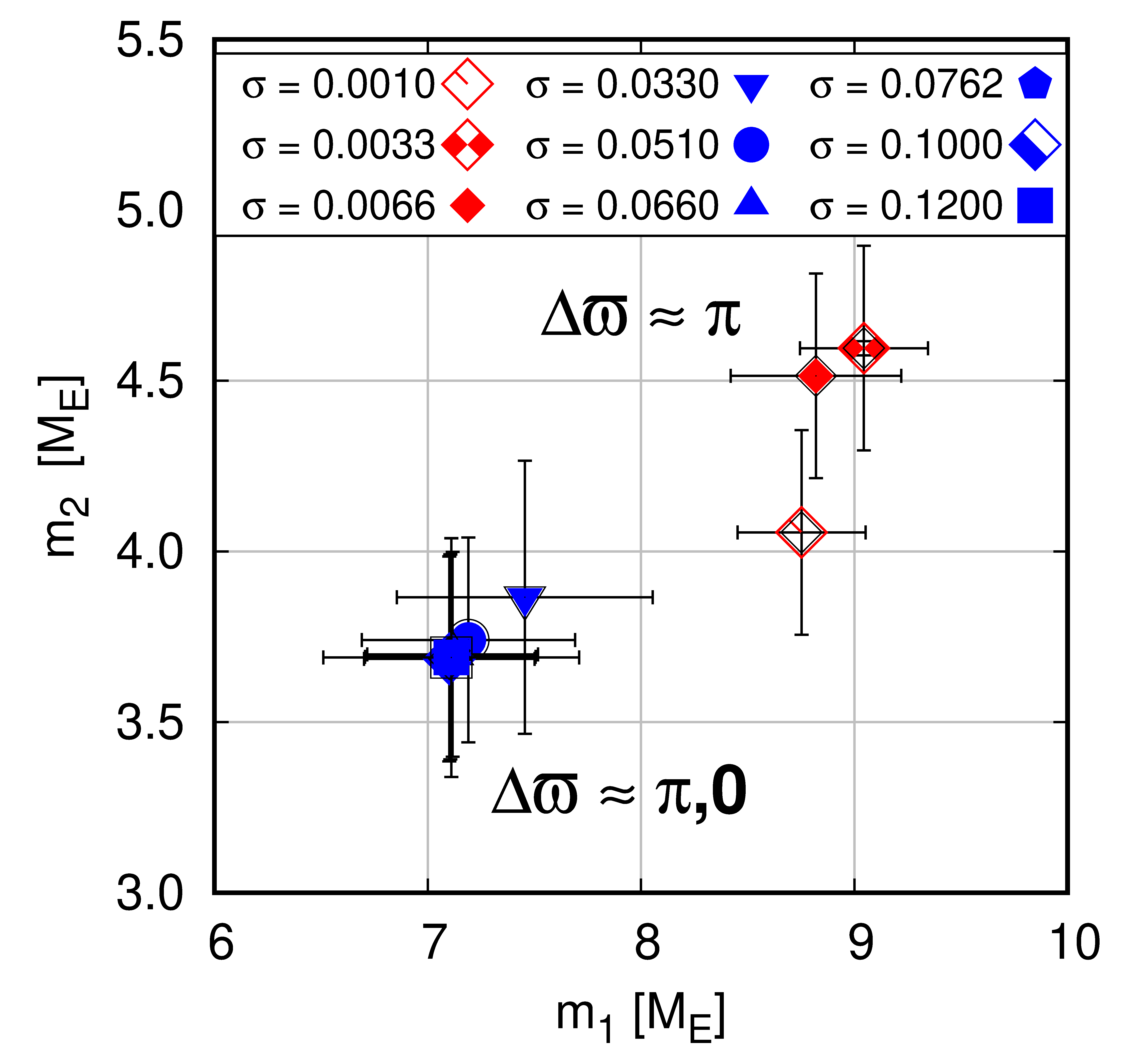}
}
}
\caption{Two--dimensional projections of the MCMC-derived posterior: the left column is for the  $(m_{\rm 1},m_{\rm 2})$- and $(\varpi_2,\Delta\varpi)$-plane, where $\Delta\varpi\equiv\varpi_2-\varpi_1$, respectively for $\sigma_{x_i,y_i}=0.0066$, representative of a single $\Delta\varpi=180^{\circ}$ mode, and  the middle column is for $\sigma_{x_i,y_i}$=0.12, with two-modal $\Delta\varpi=180^{\circ},0^{\circ}$ posterior. Contours illustrate the 14--, 50--, 86-- and 99.9--th percentile of the posterior samples. The right column is for the median values of eccentricities and masses derived as medians of the posterior samples, for different eccentricity priors $\sigma\equiv \sigma_{x_i,y_i}$. Their formal uncertainties are marked with cross-hairs. We distinguish between clear single-mode $\Delta\varpi=180^{\circ}$ solutions yielding $\chi^2_{\nu}\simeq 1.1$ (red symbols) and other models with asymmetric or dual-mode posterior in $\Delta\varpi$ (blue symbols). 
}
\label{figure:fig3}
\end{figure*}
In Fig.~\ref{figure:fig3}, we show the posterior distributions in the $(m_{\rm 1},m_{\rm 2})$- and $(\varpi_2,\Delta\varpi)$-plane,  $\Delta\varpi\equiv\varpi_2-\varpi_1$, for two representative $\sigma_{x_i,y_i} =$ 0.0066 and 0.12, respectively.  For eccentricity priors smaller than the critical one, $\sigma^C_{x_i,y_i} \simeq 0.03$  (tentatively), the posterior is single-modal (the left column of Fig.~\ref{figure:fig3}), with  $\Delta\varpi\simeq 180^{\circ}$. We found the posterior more and more asymmetric in $\Delta\varpi$ for $\sigma_{x_i,y_i}>\sigma^C_{x_i,y_i}$, and with two clear local extrema,  $\Delta\varpi \simeq 180^{\circ},0^{\circ}$ for $\sigma_{x_i,y_i}> 0.06$. Close to $\sigma^C_{x_i,y_i}$, the $\Delta\varpi=180^{\circ}$ mode bifurcates, and a second mode $\Delta\varpi=0^{\circ}$ (aligned orbits) emerges. 
As the best-fitting TTV model, we report in Tab.~\ref{table:tab3} a low-eccentric ($e_i\simeq 0.01$), $\Delta\varpi=180^{\circ}$  solution for $\sigma_{x_i,y_i}=0.0033$.
It yields similarly small  $\chi_{\nu}^2\simeq1.1$ as the best GEA models, $\chi^2_{\nu}\sim 1$,  close to a local minimum, when compared with models for $\sigma_{x_i,y_i}=0.001$ ($\chi_{\nu}^2=1.22$) and $\sigma_{x_i,y_i}=0.0066$ ($\chi_{\nu}^2=1.27$). 

Plots in the right column of Fig.~\ref{figure:fig3} illustrate the median values of eccentricities and masses derived from the posterior samples,  for a number of runs with $\sigma_{x_i,y_i}\in [0.001,0.12]$. There is a strong correlation of the median eccentricity with the priors, while mass estimates seem to be clustered, yet in two different regions, relative to  $\sigma^C_{x_i,y_i}\simeq 0.03$. Beyond that value, $(x_i,x_j)$ and/or $(y_i,y_j)$  are correlated, and $\Delta\varpi=180^{\circ},0^{\circ}$ modes appear clearly for $\sigma_{x_i,y_i}>0.06$, a value likely dependent on the MCMC sampling strategy.

This experiment shows that it is not possible to distinguish between solutions exhibiting the two $\Delta\varpi$ modes and different eccentricities with only TTV observations. Also, the mass-eccentricity degeneracy \citep[e.g.,][]{Hadden2014,JontofHutter2016} cannot be fully removed with a ``reasonable''  selection of the eccentricity priors. The masses and eccentricities are globally weakly constrained, mostly due to the two-modal $\Delta\varpi=0^{\circ},180^{\circ}$ posterior.  Additional constraints, like a particular type of periodic configurations \citep{Migaszewski2018}, or flowing from the migration history of the system, could be helpful for resolving this issue.  Indeed, as described in detail in Sect.~\ref{section6}, for reasonable disk decay and migration time-scales $(\tau_{a},\tau_{e})$, we obtain an agreement between the numerical simulations and our best-fitting solutions, having synthetic systems with $\Delta\varpi=180^{\circ}$ and moderate eccentricities.
\begin{table}
\centering
\caption{
The TTV and light-curve model parameters and their uncertainties from the MCMC sampling. For the TTV model, the eccentricity priors $\sigma_{x_i,y_i}=0.0033$, resulting in $\Delta\varpi=180^{\circ}$ and $\chi^2_{\nu}\simeq 1.1$.
For all parameters, we estimate the uncertainties as the 16--th, and 84--th percentile of the samples. The $T_0$ epoch is BKJD$-139$~days. The star mass is $m_{\star}= 1.02\,\msun$, its radius $R_{\star}=0.972~\Rsun$ \citep{Rowe2015}. We inferred the orbital elements $a$, $e$, $\varpi$ and ${\cal M}$ (the mean anomaly at the epoch) from the primary parameters.
}
\label{table:tab3}
\begin{tabular}{l r r}
\hline
Planet & KOI-1599.02 & KOI-1599.01  \\
\hline
\hline
 $P\,$[d] & 13.6088 $\pm$0.0006 & 20.4415 $\pm$ 0.0013  \\
 $x \equiv e \cos\varpi$ & 0.007 $\pm$ 0.003 & -0.009 $\pm$ 0.003 \\
 $y \equiv e \sin\varpi$  & 0.009 $\pm$ 0.003 & -0.011 $\pm$ 0.003 \\
$T\,$[d] & 74.012 $\pm$ 0.006 & 72.946 $\pm$ 0.008 \\
mass $m \,[\mE]$  & 9.0 $\pm$ 0.3 & 4.6 $\pm$ 0.3\\
\hline
$a\,[\au]$ &  0.112293  & 0.147280 \\
$e$ &   0.0114   & 0.0140  \\
$\varpi$ [deg] & 49.951 & 230.175 \\
${\cal M}\,$[deg] & 167.660 & 22.141 \\
\hline
$R\,[\RE]$ & 1.9$\pm$0.2 & 1.9$\pm$0.3 \\
\hline
\end{tabular}
\end{table}
\section{The 3:2 MMR dynamics}
\label{section4}
Although both the GEA and MCMC-sampling experiments make it not possible to constrain eccentricities without additional assumptions, a striking feature of the TTV models is their clustering in the anti-aligned libration mode $\Delta\varpi=180^{\circ}$. This feature is expected as a natural outcome of inward and convergent migration of two-planet systems \citep{Lee2002,Batygin2013}. Combinations of low---moderate eccentricity and  aligned---anti-aligned configurations, which fit the TTV observations are also possible. Also the best-fitting, low $\Chi \simeq 1$ systems exhibit period ratios close to 3:2. In order to assess whether these systems are dynamically resonant, in spite of possible large-amplitude libration of the critical angles, or only their coherence (such as shown in Fig.~\ref{figure:fig4}), we performed additional numerical experiments regarding four classes of configurations, illustrated in Fig.~\ref{figure:fig2} and listed in Tab.~\ref{table:tab2}.

A clear libration of the critical angles may not be the decisive factor for identifying the MMR dynamics \citep[e.g.,][]{Migaszewski2017a,Petrovich2013,Delisle2012,Henrard1983}. Here, we follow the strictly dynamical understanding of the resonance (MMR), as the 3:2 commensurability region in the parameter space, in which the proper (fundamental) frequencies are closely commensurate and dynamically bordered by separatrices (boundaries between different modes of orbital evolution). The presence of separatrices in multi-dimensional planetary systems usually leads to zones of chaotic motions.  In order to detect such structures in the orbital elements space, we computed dynamical maps in terms of  the Maximal Lyapunov Characteristic Exponent (MLCE) expressed by the MEGNO indicator \citep{Cincotta2000}. We also computed the proper mean motions $n_i$, $i=1,2$ (fundamental frequencies associated with the orbital periods) and their ratio ($f_2 / f_1 \equiv n_2 /n_1$).

\subsection{Numerical mapping of the 3:2 MMR}

{A proper parametrization of the dynamical maps is required for detecting the resonance structure. Besides the common $(a_i,e_i)$--representation, which crosses all MMRs \citep{Laskar2001}, we considered the so called representative plane of initial conditions \citep{Michtchenko2001,Callegari2006}.  The equations of motion of the system are governed by the Hamiltonian expressed in Poincar\'e coordinates \citep[e.g.,][]{Michtchenko2001}, 
\begin{equation}
{\cal H}  = \sum_{i=1,2} 
\left( \frac{\vec{p}^2}{2\beta_i} -\frac{\mu_i \beta_i}{|\vec{r}_i|} \right)
- k^2 \frac{m_1 m_2}{|\vec{r}_1-\vec{r}_2|}
+ \frac{\vec{p}_1 \cdot \vec{p}_2}{m_{\star}},
\label{eq:HamP}
\end{equation}
where $\mu_i=k^2({m_{\star}}+m_i)$, $\beta_i=k^2 m_{\star} m_i/\mu_i$. Besides the total energy integral ${\cal H}$, the total angular momentum $C =G_1 + G_2$ is preserved, where $G_i=L_i \sqrt{1-e_i^2}$ and $L_i=\beta_i \sqrt{\mu_i a_i}$. Moreover, the system Eq.~\ref{eq:HamP} averaged near a particular MMR $(p+q)/p$, where $p,q$ are integers (here $p=2$, $q=1$), exhibits a particular integral $K= (p+q) L_1 + p L_2$ that bounds variability of semi-major axes in resonant motion.

The $C$ and $K$ integrals depend on four elements $e_1, e_2, a_1, a_2$, and if their values are fixed, two linear equations may be solved against the two remaining variables.  Since the resonant and non-resonant dynamics are governed by librations of the critical angles $\phi_{3:2,1}$ and $\phi_{3:2,2}$ around $0^{\circ}$ or $180^{\circ}$, or circulations, still involving these two critical values, the structure of the phase space restricted to the same $C$ and $K$ levels may be illustrated in a plane of eccentricities  $\Sigma(e_1,e_2)$. This plane is composed of four quadrants with $\phi_1,\phi_2$ fixed at $0^{\circ}$ or $180^{\circ}$, encoded by variables $x \equiv e_1 \cos\phi_1$ and $y \equiv e_2 \cos\phi_2$. Here, we fix $\varpi_1=0^{\circ}, {\cal M}_1=0^{\circ}$, and then the remaining angles are  $(\varpi_2=0^{\circ}, {\cal M}_2=0^{\circ})$, 
$(\varpi_2=180^{\circ}, {\cal M}_2=180^{\circ})$, 
$(\varpi_2=0^{\circ}, {\cal M}_2=180^{\circ})$, and 
$(\varpi_2=180^{\circ}, {\cal M}_2=0^{\circ})$, respectively. These pairs of angles define quadrants I, II, III and IV of the representative plane $\Sigma$.
}

We derived the proper mean motions with the frequency modified Fourier transform \citep[Frequency Modified Fourier Transform aka Numerical Analysis of Fundamental Frequencies, NAFF,][]{Laskar1990,Laskar1993,Nesvorny1996} of the time series $\left\{ a_i(t_k) \exp[ \mbox{i} \psi_i(t_k)] \right\}$,  where $\psi_i$ are appropriate angles forming the conjugate action-angle pairs with the osculating semi-major axes $a_i(t)$ (equivalent to rescaled actions $L_i$), sampled at discrete moments $t_k$,  $k=1,2, 3,\ldots,2^K$ (here $K=18$), inferred from the Poincar\'e coordinates \citep[e.g.,][]{book:Morbidelli2002,Gozdziewski2008b}. As the conjugate angles, we may choose the mean longitudes $\psi_i\equiv \lambda_i(t) = \Mmean_i(t) + \varpi_i(t)$ or the mean anomalies $\psi_i \equiv \Mmean_i(t)$. The meaning of the NAFF-derived fundamental frequencies $n_i$ is then subtly different. In the $\psi_i=\lambda_i(t)$ settings, the proper frequencies $n_i$ are related to the inertial frame, while for $\psi_i=\Mmean_i(t)$, they represent the orbital motion in particular rotating reference frames related to the rotations of the longitudes of pericenter of each orbit. If the orbital configuration of both planets is periodic, as explained below, then their apsides rotate with the same frequency, thus there is a common reference frame corresponding to the rotation of the planetary system as a whole.

We found the conjugate angles distinction as important for detecting regions close to the periodic orbits associated with the 3:2 MMR, or rather the 3:2 commensurability. The periodic orbits, meaning a repetitive, relative configuration of planets in the rotating reference frame with one of the planets, are associated with centers of the mean motion resonances \citep{Hadjidemetriou2006}. These solutions, when related to symmetric periodic orbits, may be characterised by $\Delta\varpi=0^{\circ},180^{\circ}$ and $\phi_{3:2,1}=0^{\circ},180^{\circ}$ in the averaged system \citep[also,][]{Voyatzis2008} or low-amplitude librations around these values in the original (full) system. Regarding the 3:2 MMR, these conditions correspond to a constant value of $2\Mmean_1-3\Mmean_2 \simeq 0^{\circ},180^{\circ}$ for the periodic orbits. Therefore, by using the NAFF we may check whether $f_1/f_2 \simeq3/2$ in selected orbital parameter planes, thus detecting the regions associated with the resonant periodic orbits, besides illustrating the width and structure of the resonance.

\subsection{The 3:2 MMR structure for selected models}

\begin{figure*}
\centerline{
\hbox{
\includegraphics[width=0.49\textwidth]{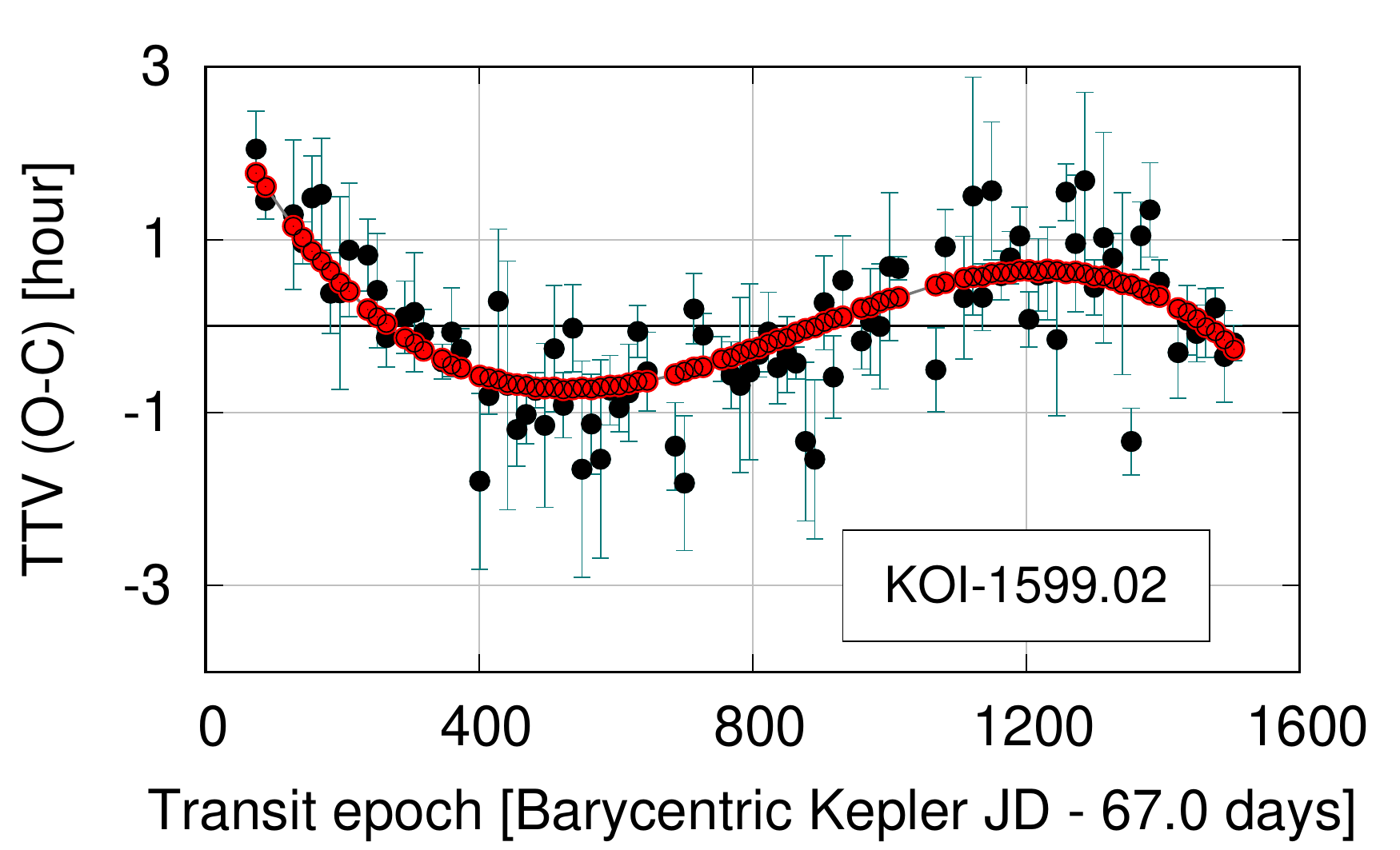}
\includegraphics[width=0.49\textwidth]{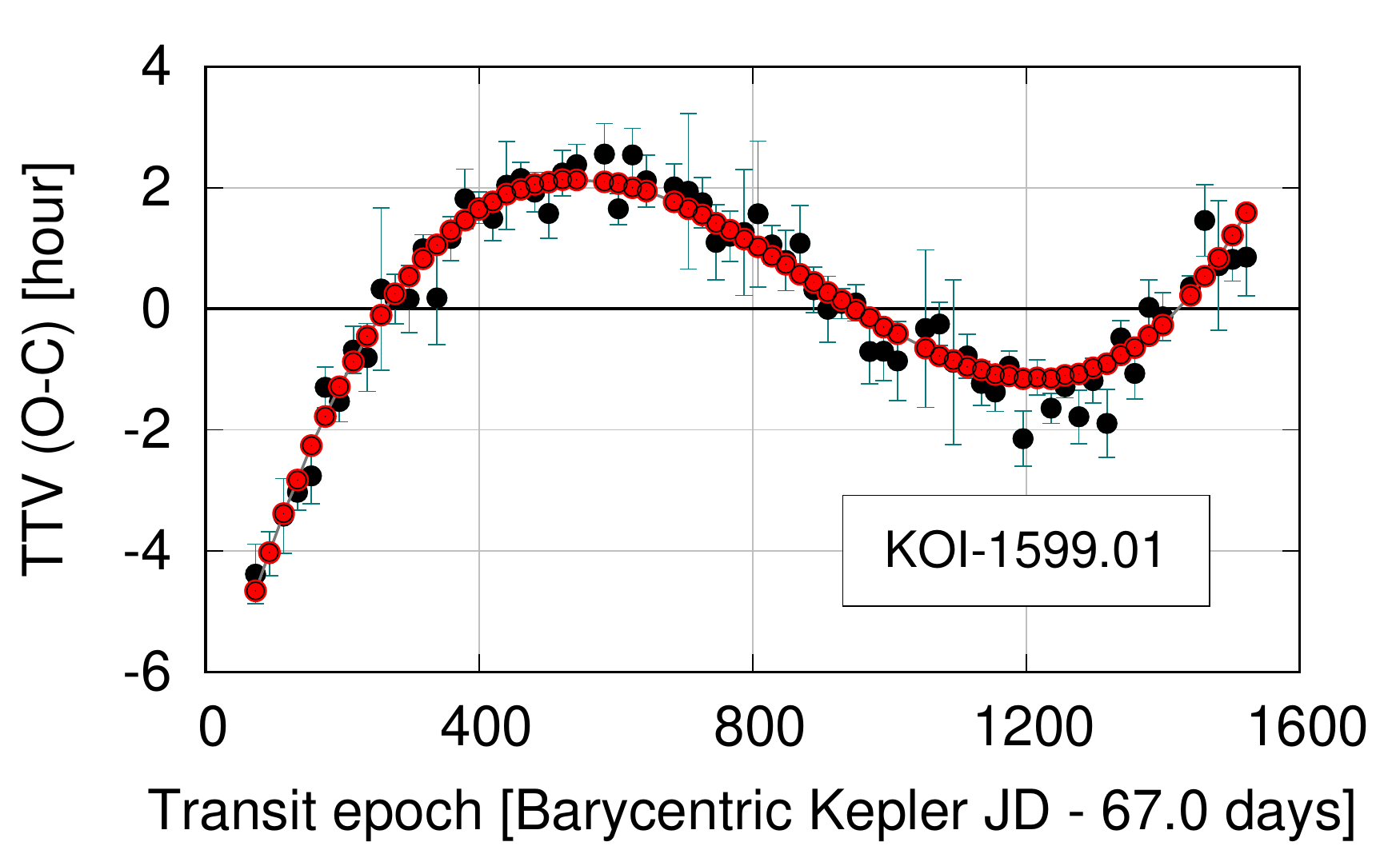}
}
}
\centerline{
\hbox{
\includegraphics[width=0.49\textwidth]{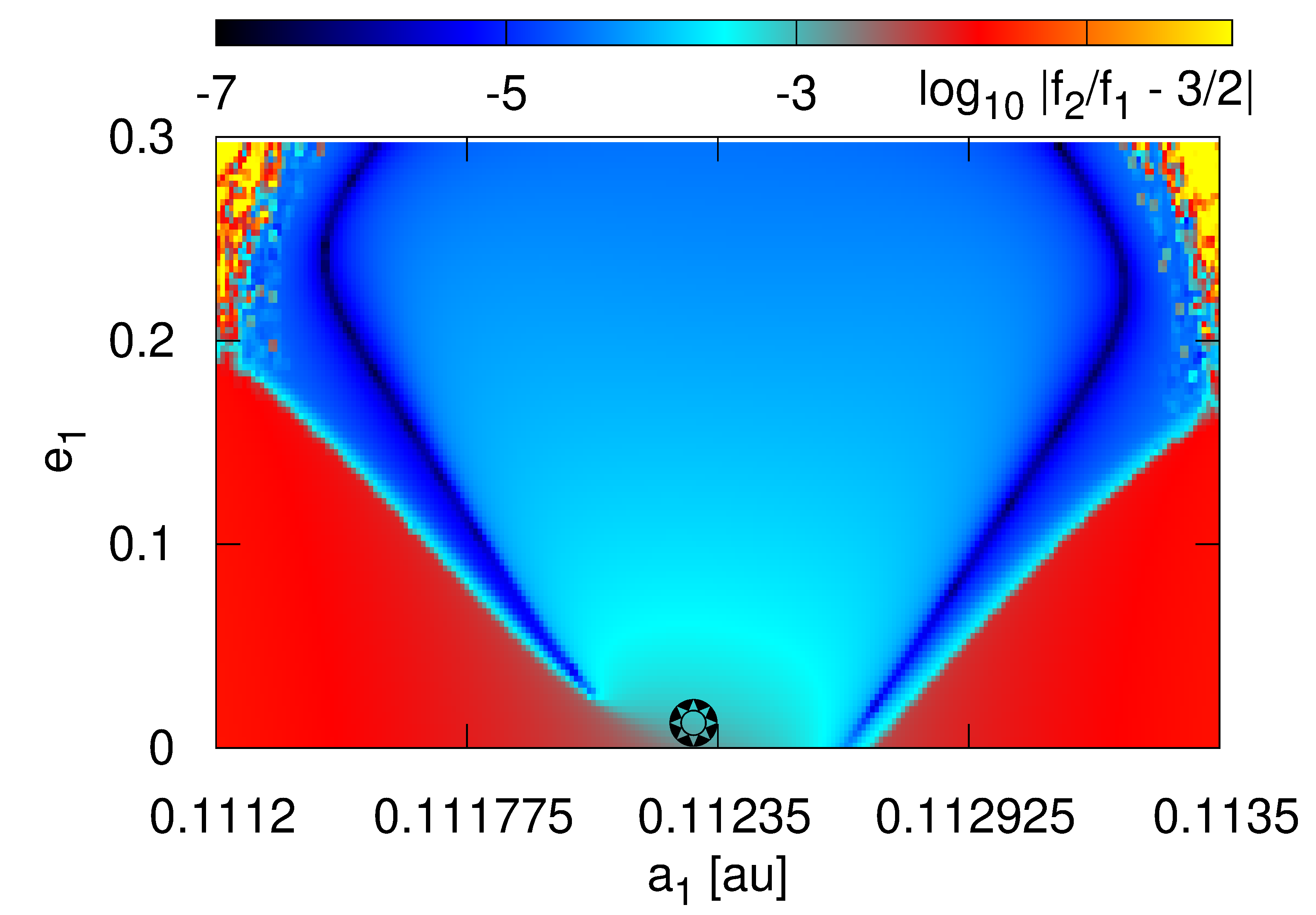}
\includegraphics[width=0.49\textwidth]{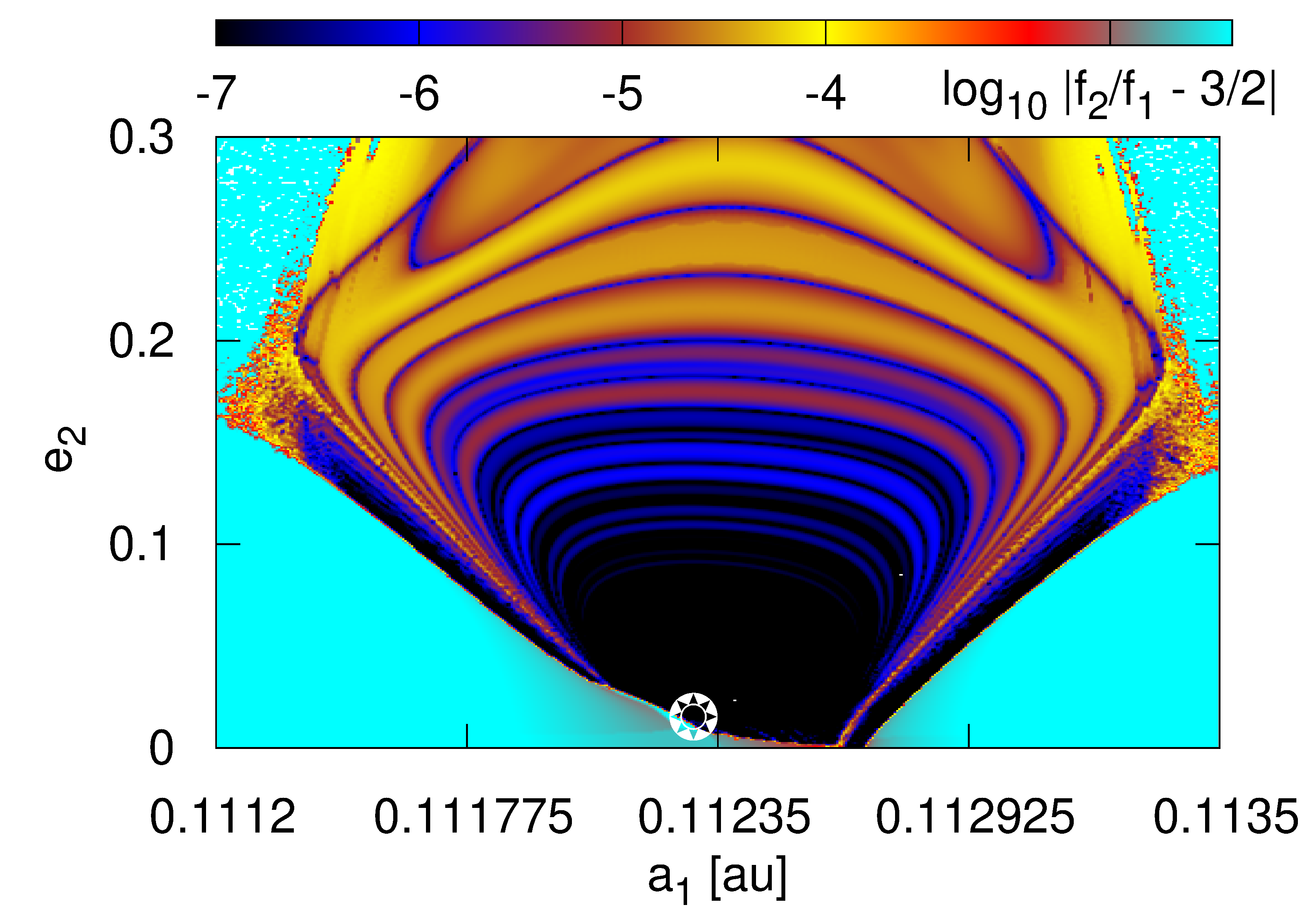}
}
}
\centerline{
\hbox{
\includegraphics[width=0.49\textwidth]{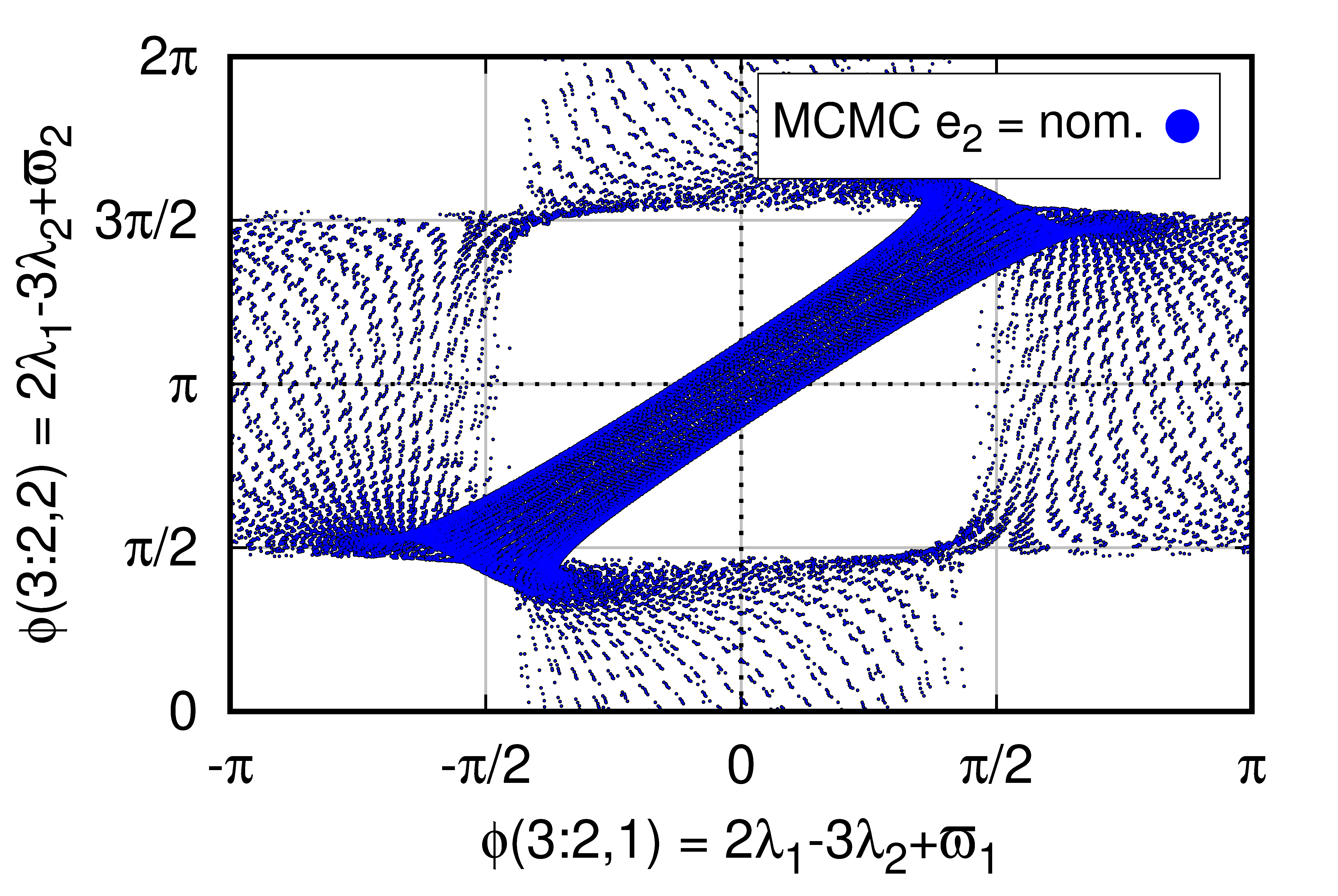}
\includegraphics[width=0.49\textwidth]{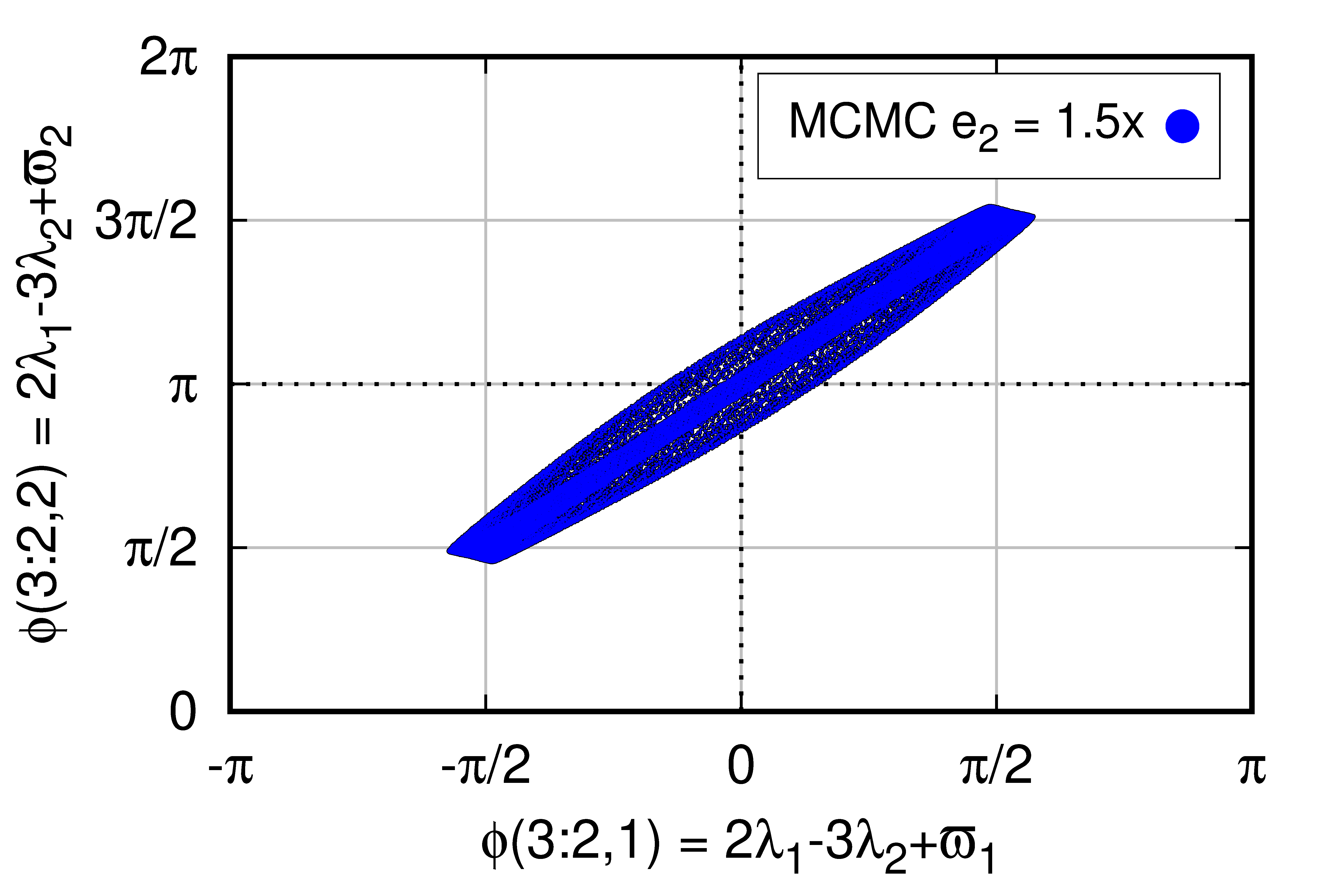}
}
}
\caption{
{\em Upper row}: 
Synthetic TTV signals and the measurements for the representative MCMC solution with anti-aligned apsides (see Tab.~\ref{table:tab3}). This model yields $\chi^2_{\nu}\simeq 1.1$.  {\em Middle row}: Dynamical FMFT maps of the 3:2~MMR, obtained with $\lambda$-NAFF (left panel) and $\Mmean$-NAFF (right panel), respectively for calculating the proper mean motions $f_1$, $f_2$ and their ratio. The star marks the nominal initial condition. The grid has the resolution of $512\times 360$ points, each integrated for $2^{18}$ time-steps of $0.4$~d, for constructing the time series and resolving the proper frequencies $f_{1,2}$.
{\em Bottom row}: 
Evolution of the critical angles, for the nominal  MCMC solution (the left panel) and for a model with eccentricity $e_2\sim 0.02$ (the nominal value increased by $50\%$, the right panel).
}
\label{figure:fig4}
\end{figure*}

In the top row of Fig.~\ref{figure:fig4}, we over-plotted the observed TTVs for both planets and synthetic TTV signals obtained from the best-fitting MCMC solution (Tab.~\ref{table:tab3}). In the middle row, we show two dynamical maps in the $(a_i,e_i)$--plane for this model illustrating the 3:2 commensurability region. The left-hand map shows a deviation of the proper mean motions ratio relative to the exact 3:2~MMR value, computed by using the mean longitudes (from herein, $\lambda$-NAFF). In this map, the 3:2 commensurability region may be indicated by small values of $|f_1/f_2-3/2|$. 
We obtained the right-hand map with the FMFT analysis of time series involving the mean anomalies ($\Mmean$-NAFF). In this plot, the family of symmetric periodic orbits is associated with the 3:2~MMR and close solutions appear as dark-blue and dark regions in the $(a_1,e_2)$--plane. Note that this plane is different from the one presented in the left panel. In the black regions in this map, the $f_2/f_1$ deviation from 3/2 may be as small as $10^{-9}$. We note that due to a complex projection of the multi-dimensional elements space onto the selected two--dimensional plane, the family of periodic orbits generally does not appear as a simple curve or an isolated region. We choose the $(a_1,e_2)$--plane after some experiments with the aim of crossing the phase-space of the 3:2 MMR in a representative way (yet the choice of the crossing plane is not unique).

In the bottom row  of Fig.~\ref{figure:fig4}, we show the evolution of two resonant solutions, the nominal MCMC one (left), and the one with eccentricity for the second planet $\simeq 0.02$, i.e., the nominal value increased by $50\%$ (right). A different proximity to the periodic orbit is related to different behaviour of the critical angles.
The NAFF dynamical maps reveal the MMR structure and mark the separatrices. However, while the right separatrix is clearly visible in the whole $e_{1,2}$-range, the left separatrix apparently ``diffuses'' and vanishes at small eccentricities.
{In fact, in the small-eccentricity regime the separatrix may not exist \citep{Henrard1983,Delisle2015}, and the $\Sigma$-plane is more useful for resolving the resonance structure.}

\begin{figure*}
\centerline{
\hbox{
\includegraphics[width=0.45\textwidth]{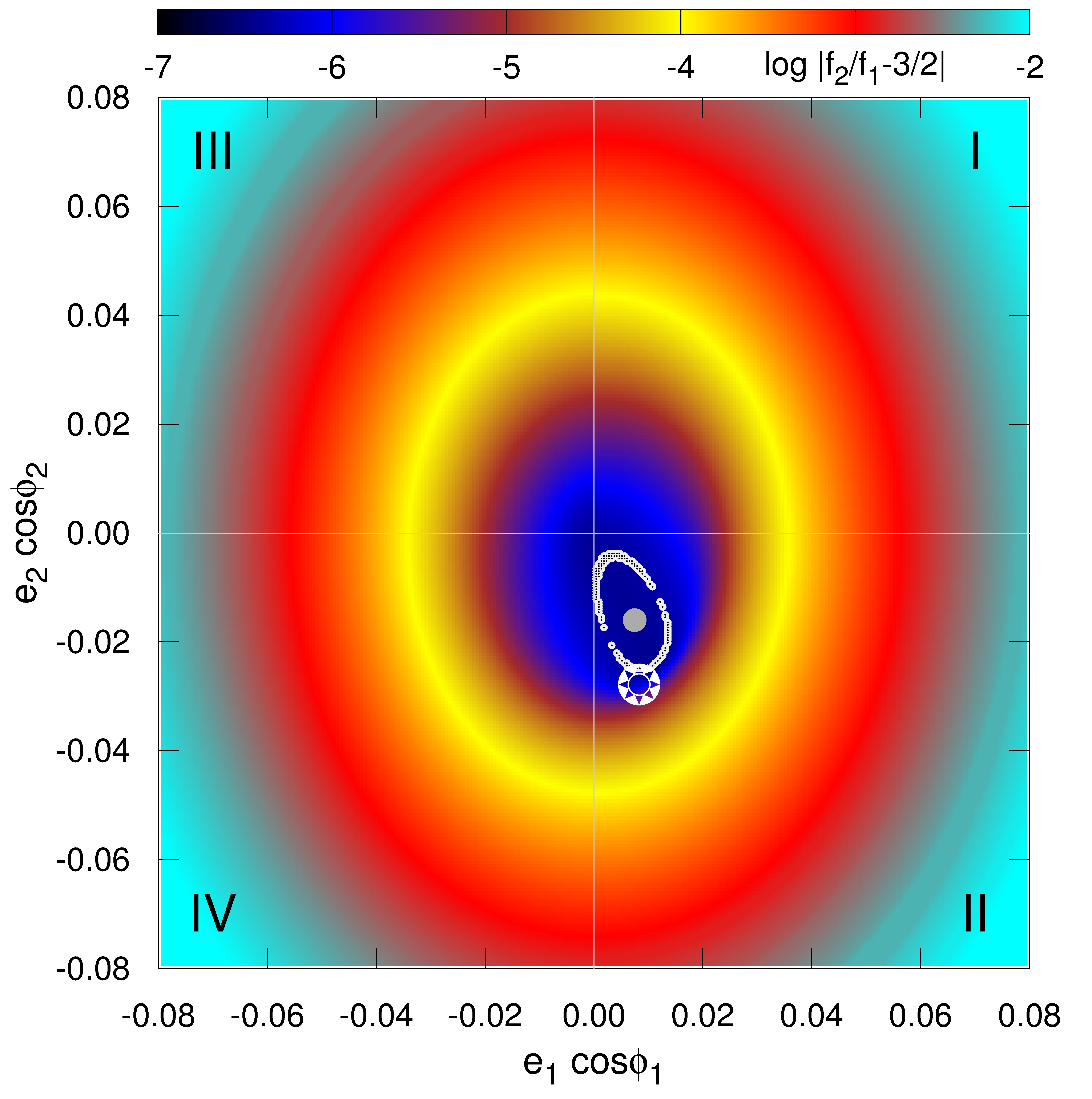}
\includegraphics[width=0.45\textwidth]{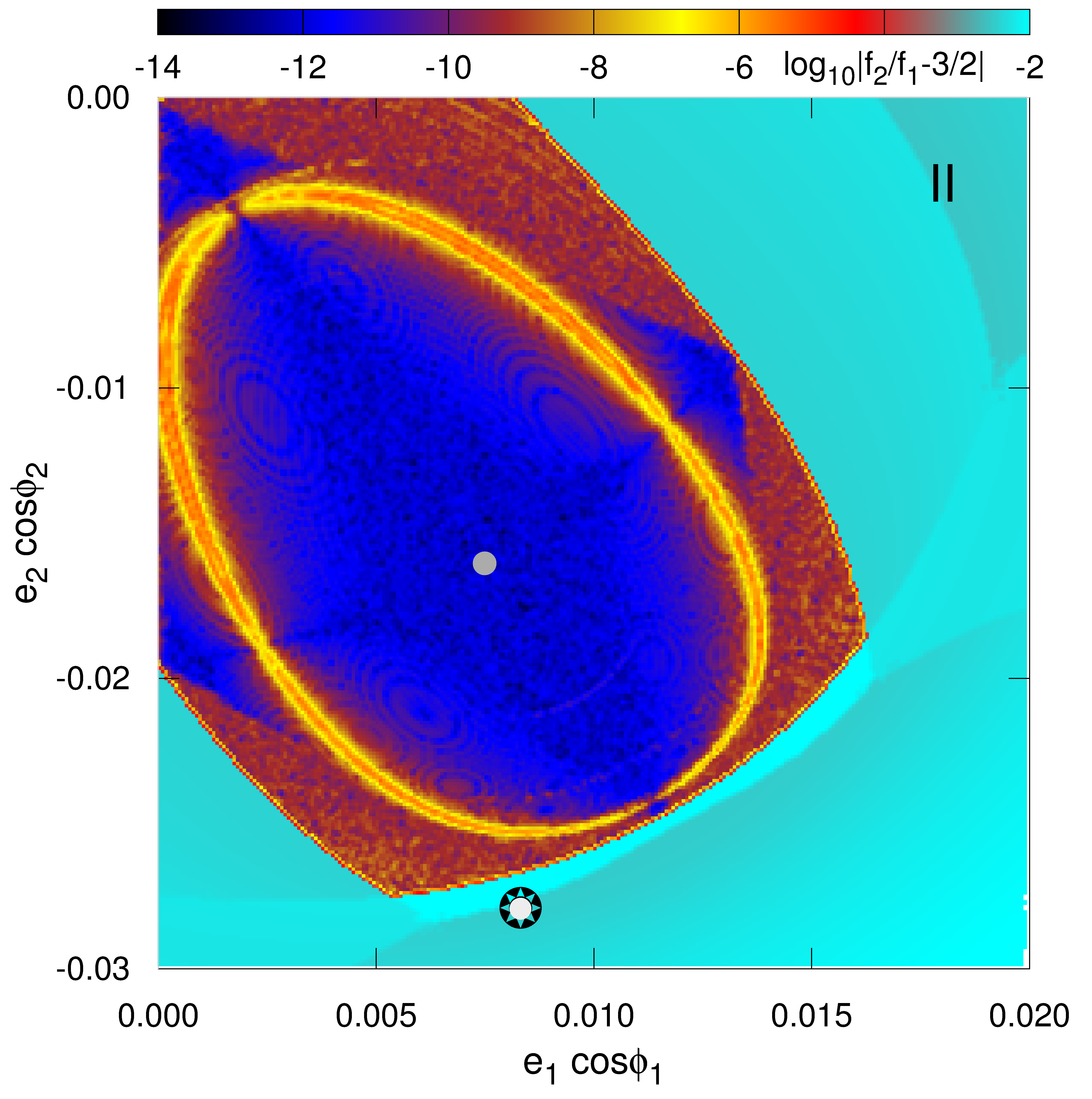}
}
}
\caption{
{Dynamical maps in terms of $|f_2/f_1-3/2|$ in the representative plane of initial conditions $\Sigma$ for the $C$ and $K$ integrals fixed at their values computed for the best-fitting MCMC model. The left panel is for the $\lambda$-NAFF and the right panel is for a close-up of quadrant~II, involving the nominal solution (the star symbol), derived with the $\Mmean$-NAFF. Small white filled circles in the left panel are for unstable solutions detected with the MEGNO indicator $\Y>5$ 
in a grid of $256\times256$ initial conditions. The MEGNO indicator was integrated for 72,000 outermost periods ($\sim 2\times 10^6$ days), for each point in the grid. The filled circle marks stable periodic orbit associated with the 3:2 MMR. The resolution of the maps is $256\times256$ points. See the text for more details.}
}
\label{figure:fig5}
\end{figure*}

{Figure~\ref{figure:fig5} shows two dynamical maps in the $\Sigma$-plane computed in terms of $|f_2/f_1-3/2|$. The left-hand panel is for $\lambda$-NAFF. In this plot we also marked unstable solutions detected with the symplectic MEGNO indicator, computed for 72,000 years with a time-step of 0.5~days. They form a loop, with a filled circle inside, which marks a stable periodic orbit associated with the 3:2~MMR. The nominal model (marked with a star symbol) lies outside the loop. The right-hand panel shows this region in terms of the $\Mmean$-NAFF (only the quadrant~II is shown). The loop of unstable solutions overlaps with an increase of $|f_2/f_1-3/2|$, and the libration zone has a sharp border.}

\begin{figure*}
\centering
\centerline{
\hbox{
\includegraphics[width=0.33\textwidth]{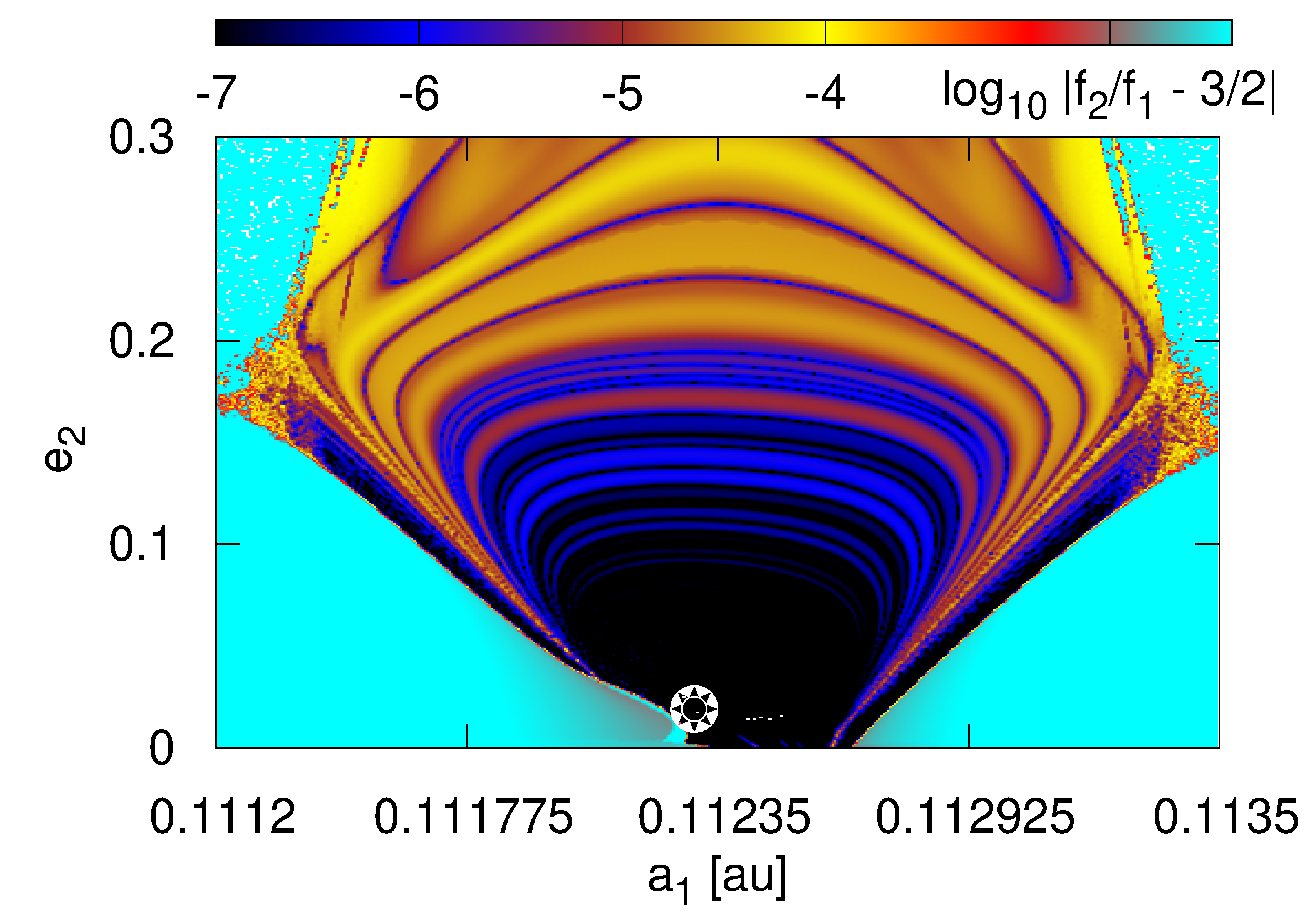}
\includegraphics[width=0.33\textwidth]{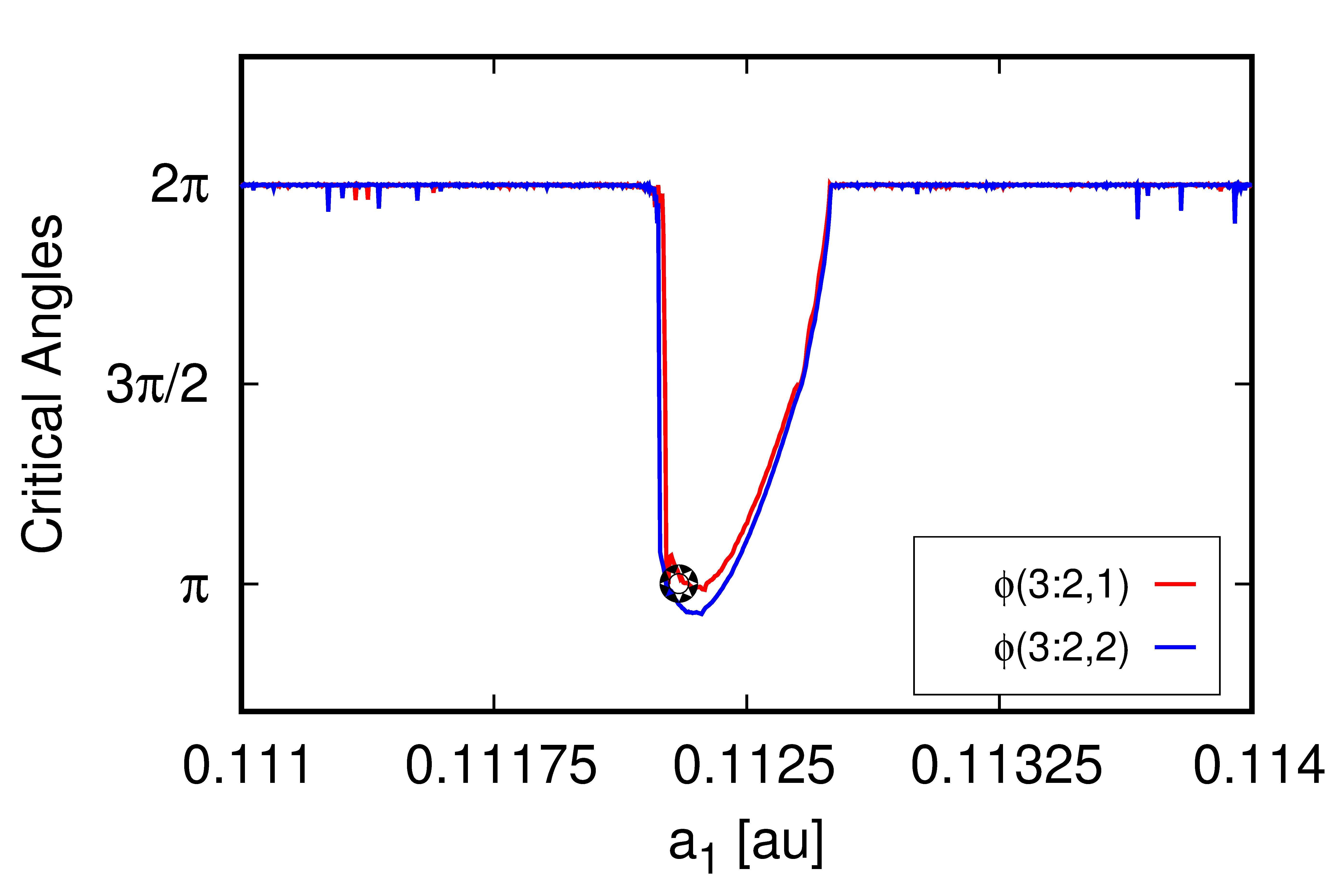}
\includegraphics[width=0.33\textwidth]{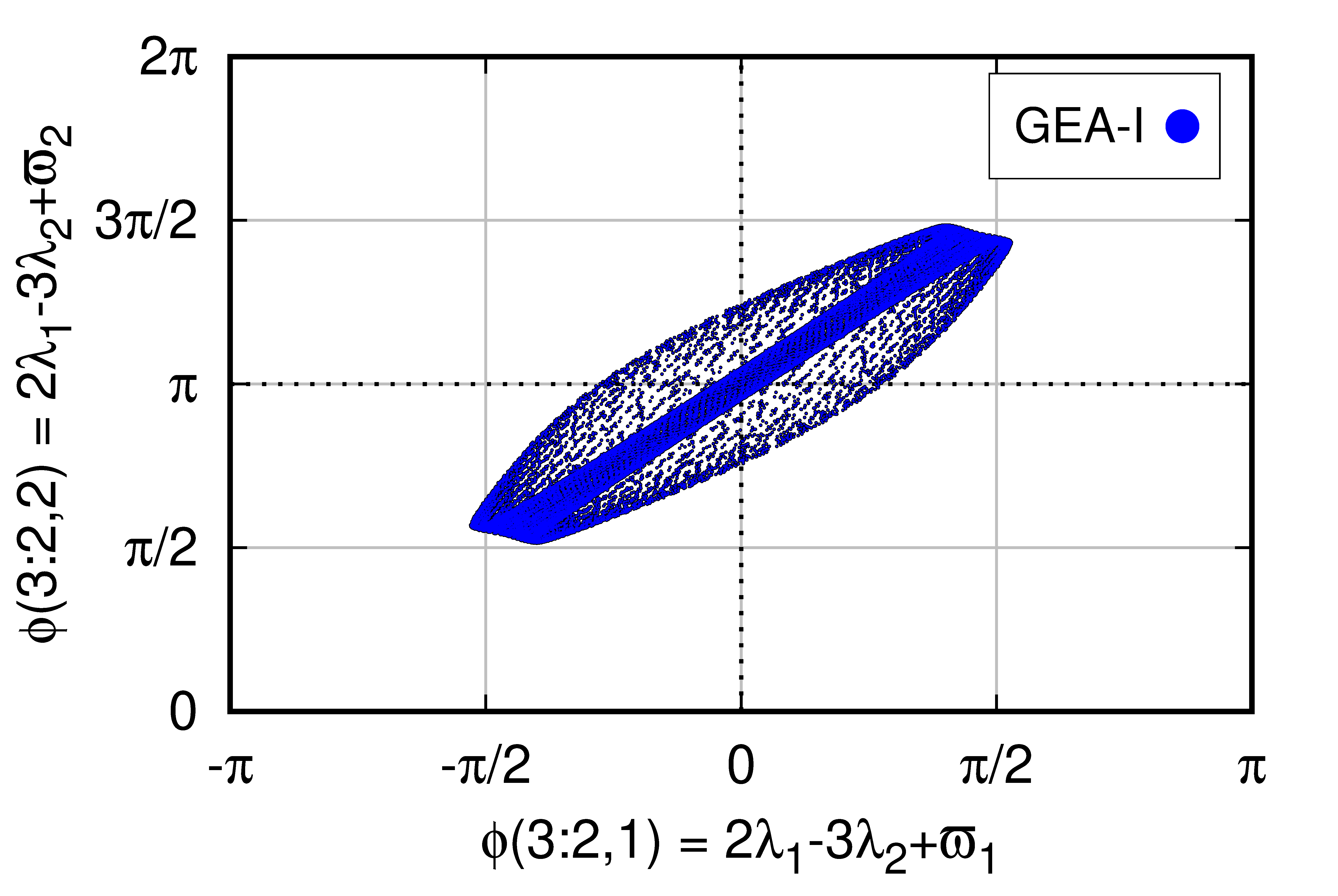}
}
}
\centerline{
\hbox{
\includegraphics[width=0.33\textwidth]{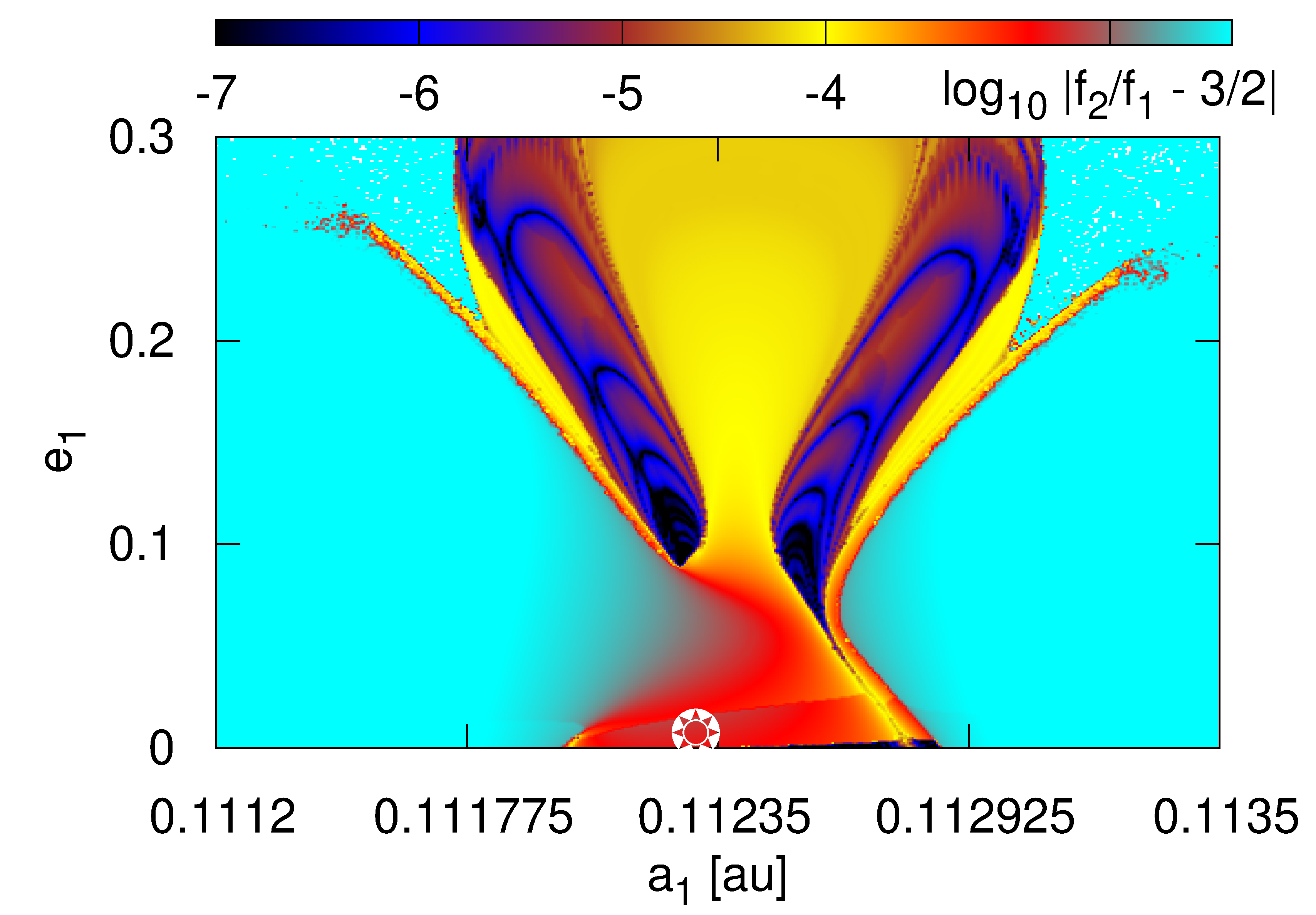}
\includegraphics[width=0.33\textwidth]{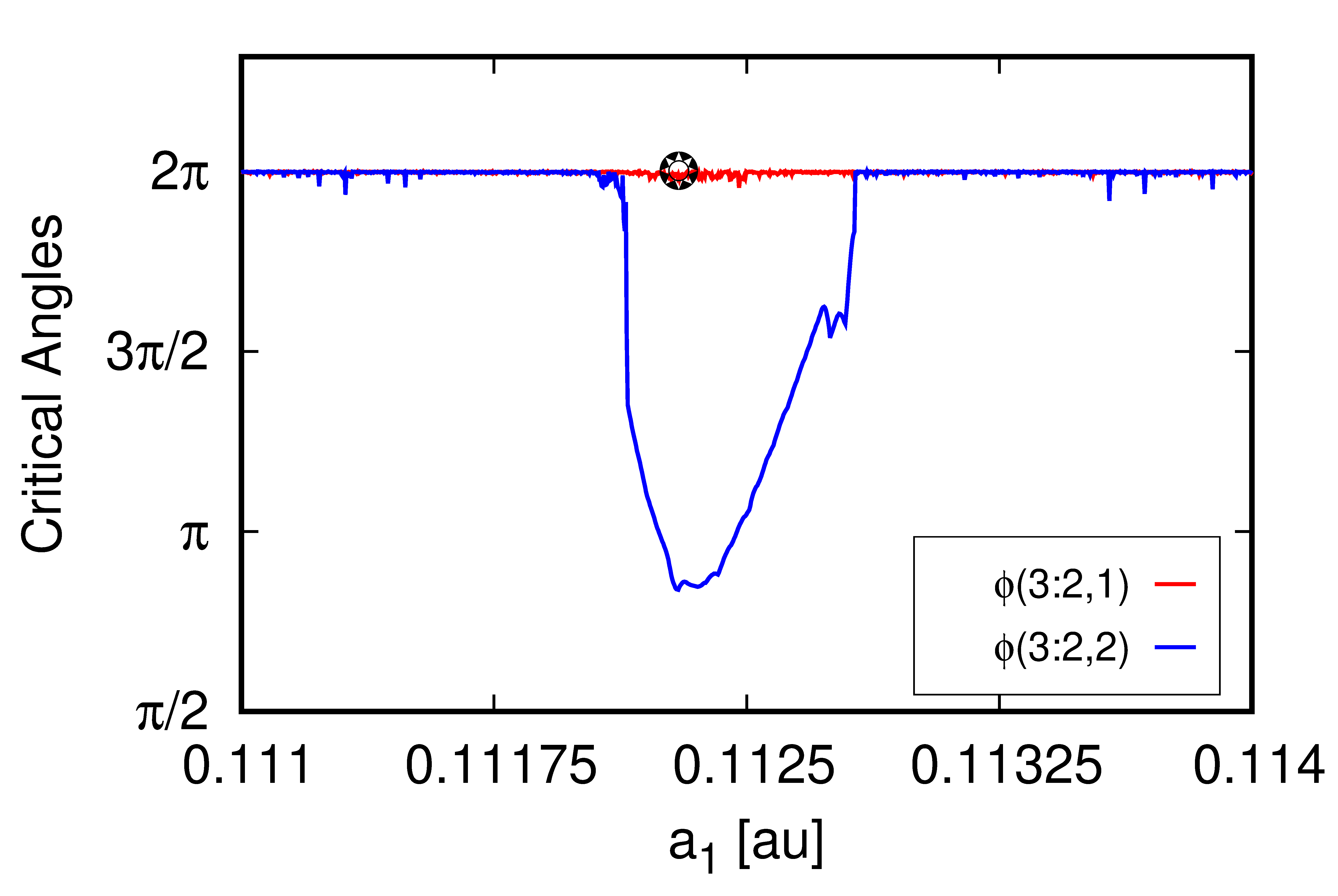}
\includegraphics[width=0.33\textwidth]{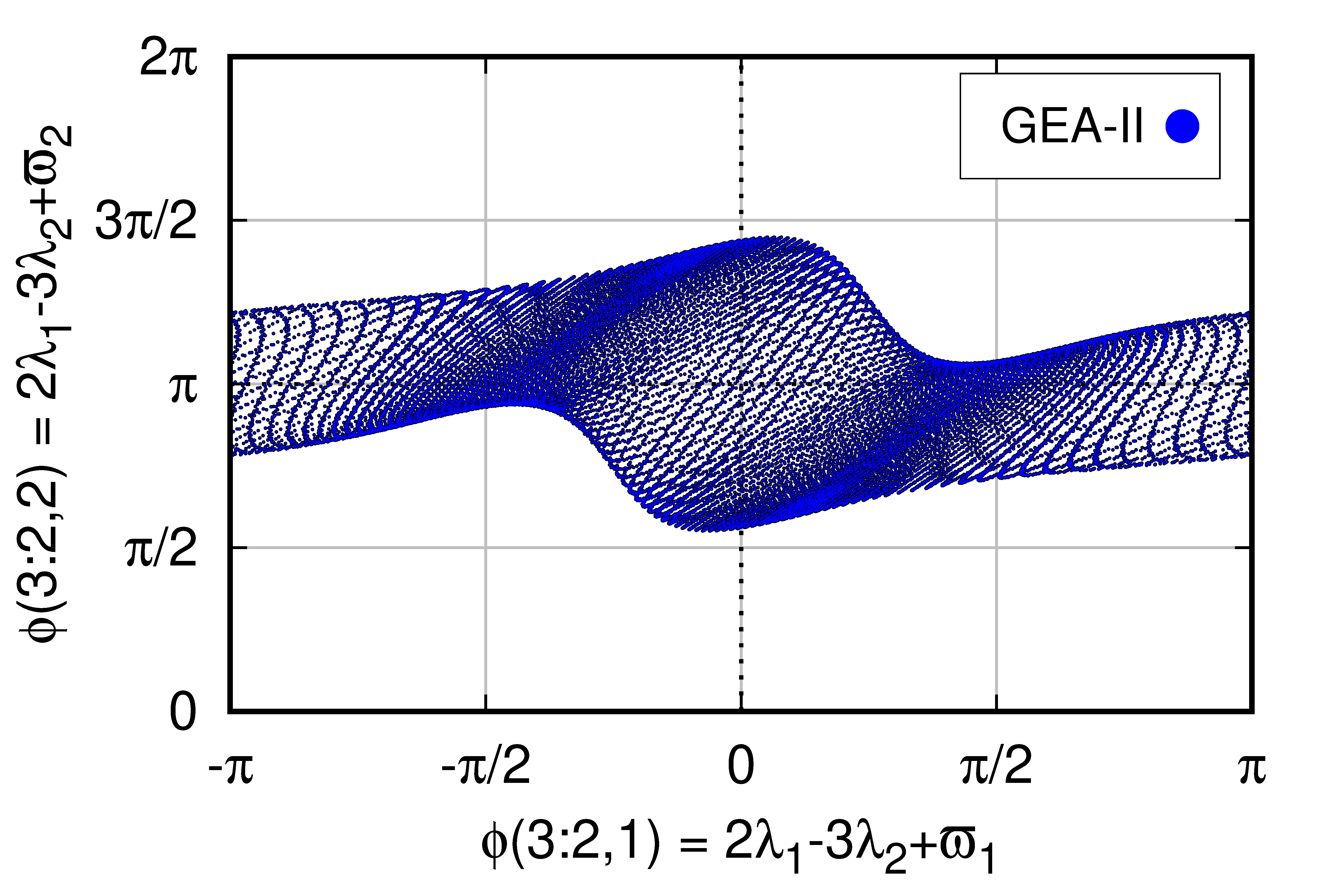}
}
}
\centerline{
\hbox{
\includegraphics[width=0.33\textwidth]{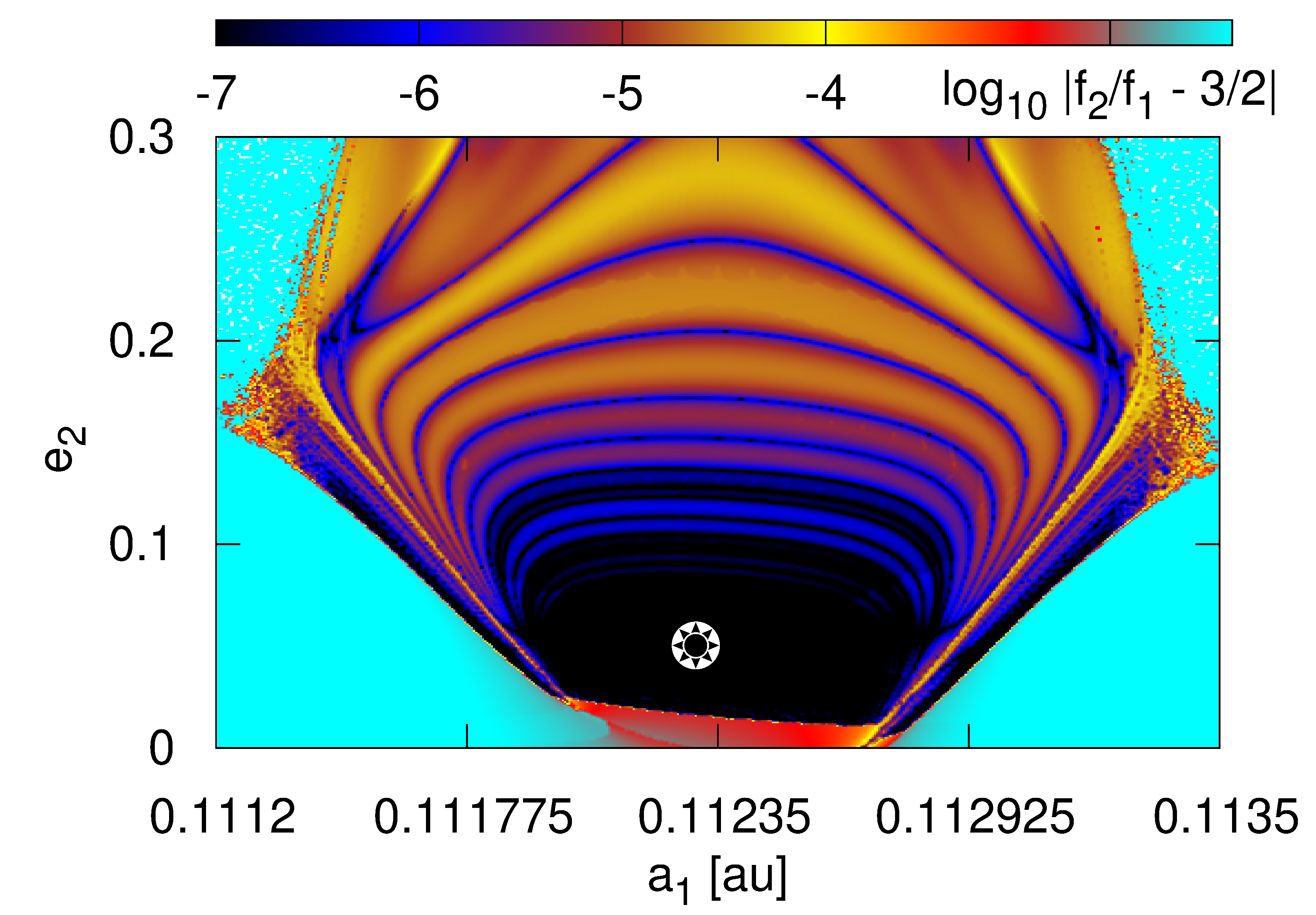}
\includegraphics[width=0.33\textwidth]{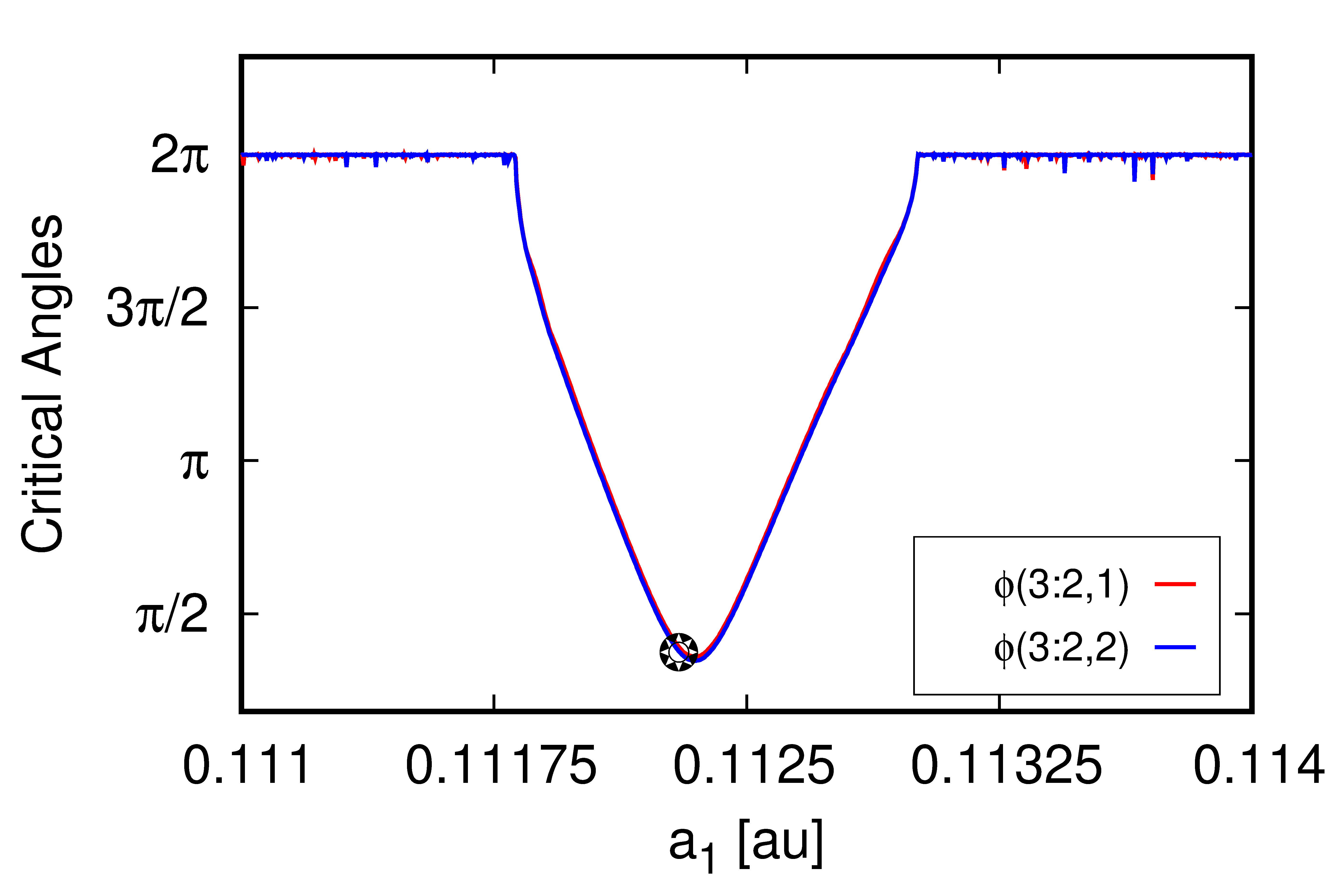}
\includegraphics[width=0.33\textwidth]{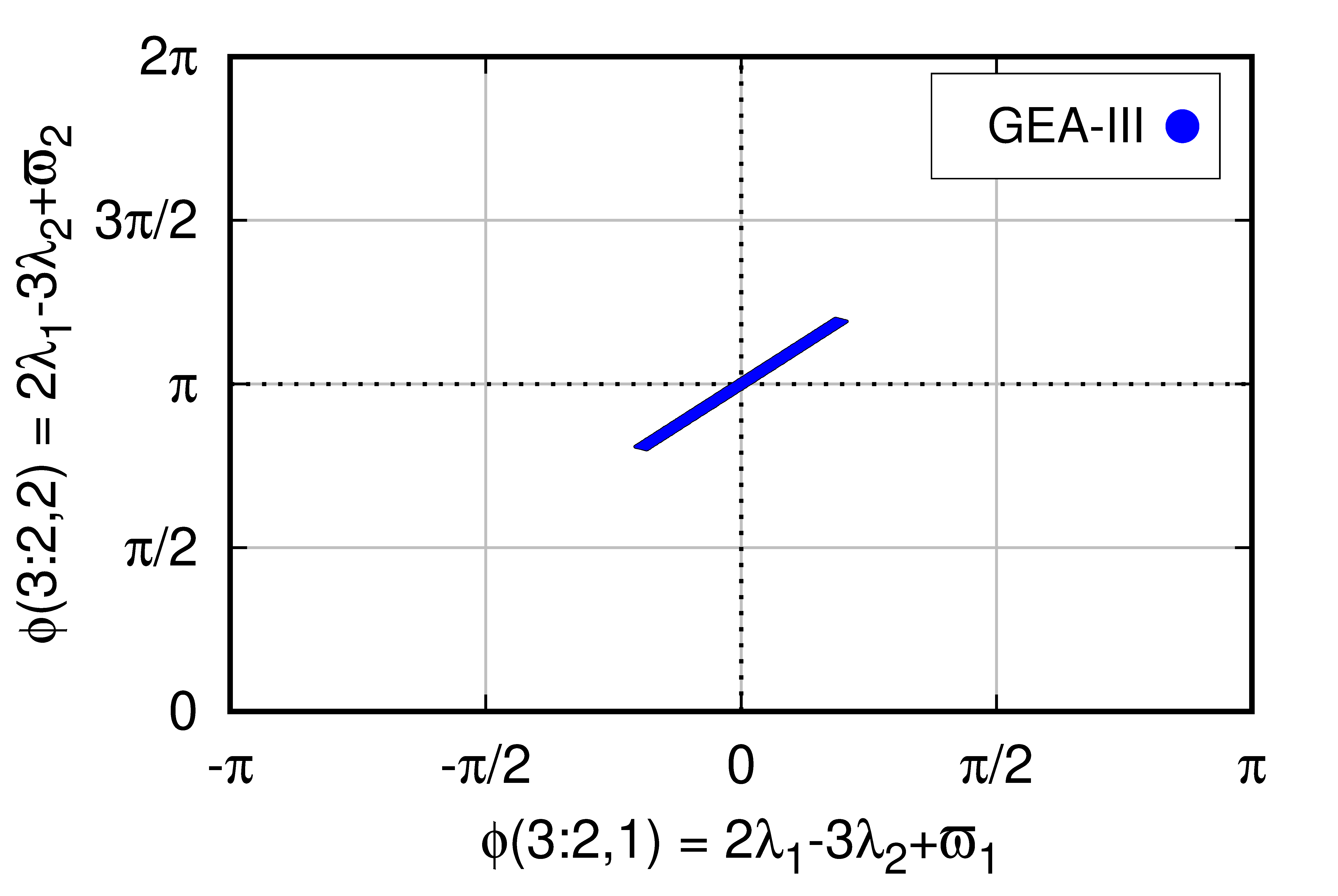}
}
}
\centerline{
\hbox{
\includegraphics[width=0.33\textwidth]{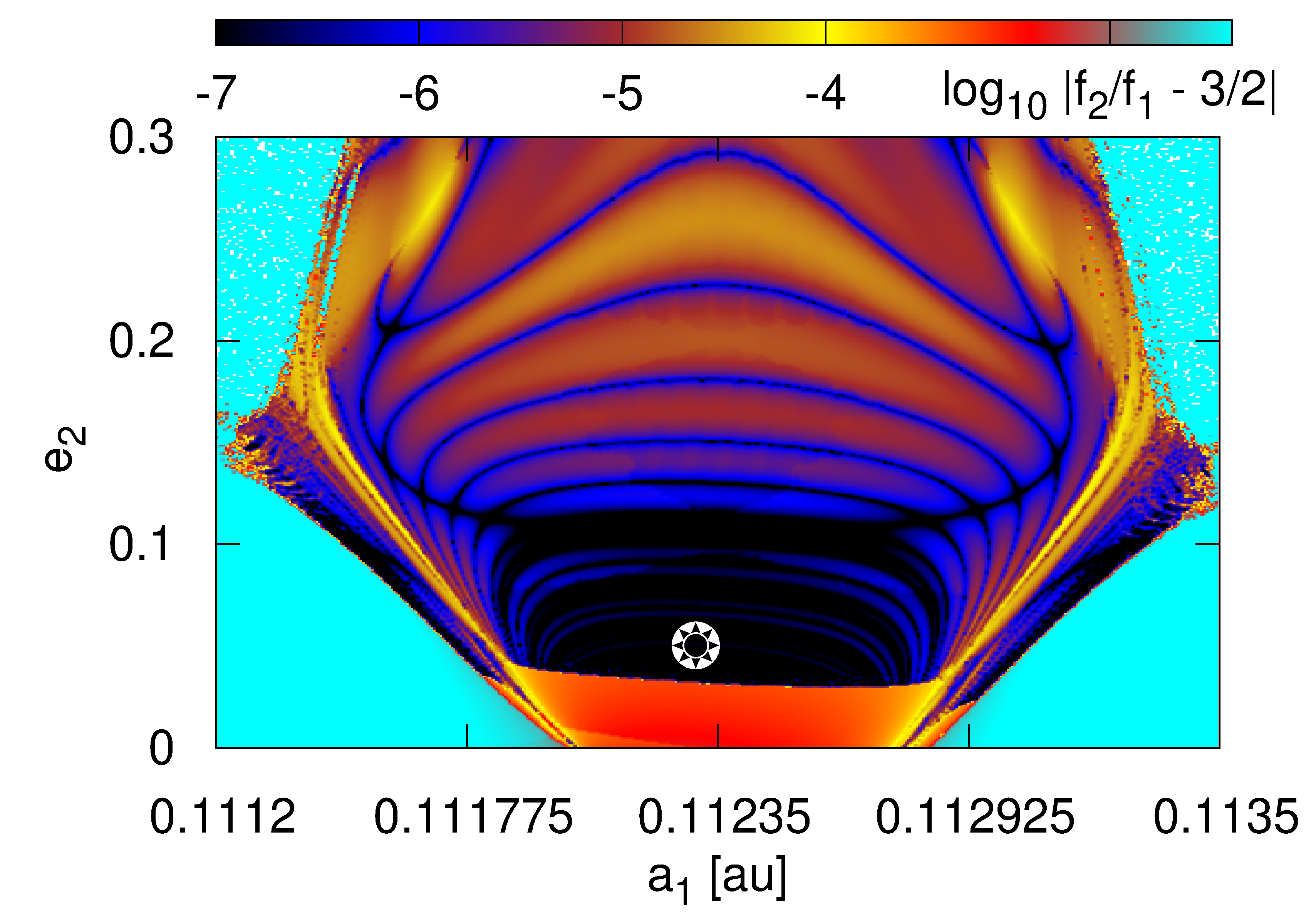}
\includegraphics[width=0.33\textwidth]{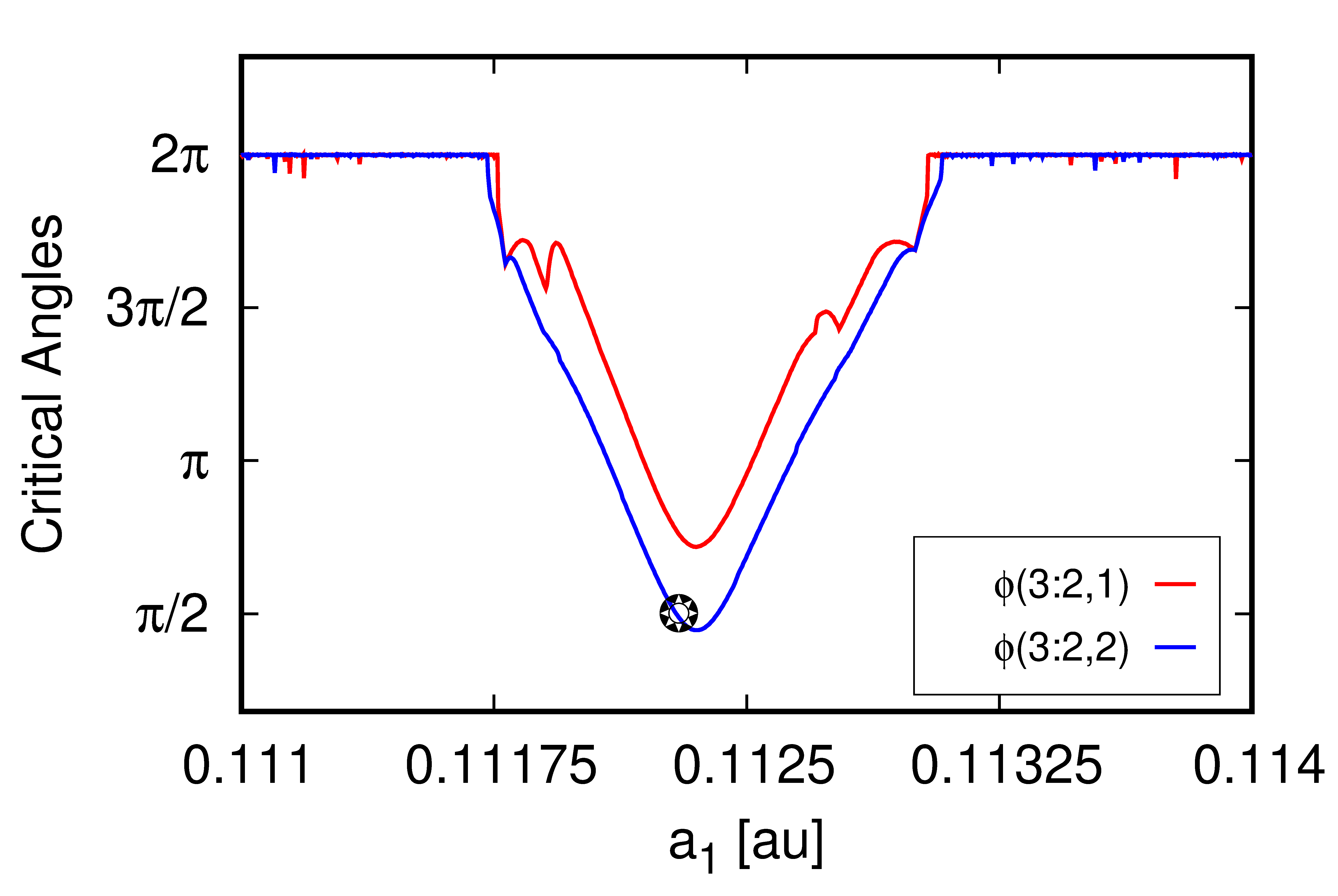}
\includegraphics[width=0.33\textwidth]{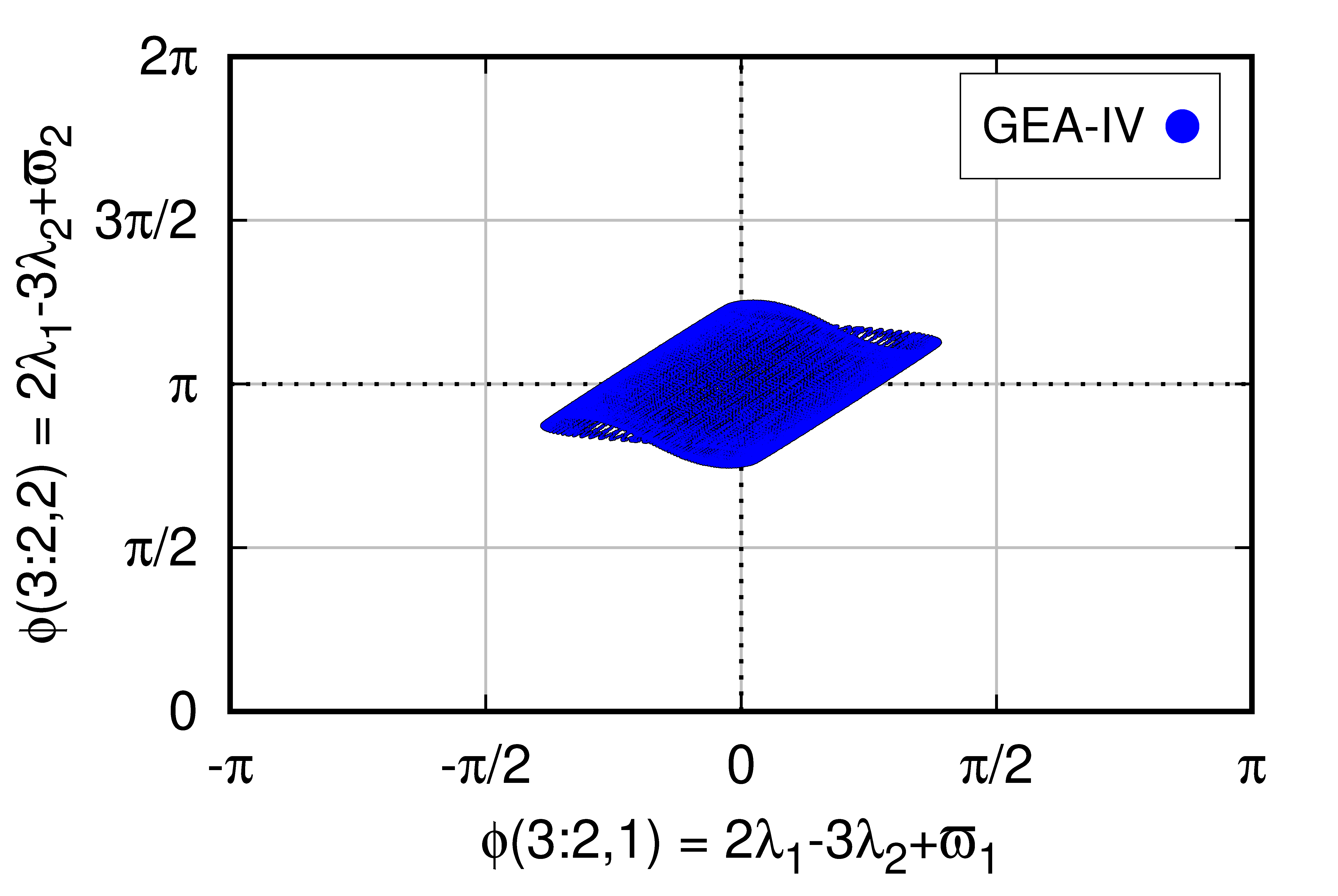}
}
}
\caption{
\textit{Left column:} Two--dimensional $\Mmean$-NAFF maps in the
$(a_{\rm 1},e_{\rm 2})$-- and $(a_{\rm 1},e_{\rm 1})$--plane (second panel from the top). The resolution is $512\times360$ points, the integration time is $\sim 5\times 10^3$ outermost orbital periods ($2^18$ samples with the time-step of 0.5~d). We used the 4-th order symplectic integrator SABA$_4$ scheme. The GEA best-fitting configurations are marked with black star symbol (see Tab.~\ref{table:tab2} for their orbital elements).
\textit{Middle column:} One--dimensional amplitude scan of the critical angles ($\phi_{\rm 3:2,1}$ and $\phi_{\rm 3:2,2}$). The grid has $2\times10^3$ initial conditions, each of them integrated for $1$~kyr with the SABA$_4$ scheme. The best-fitting GEA initial conditions are marked by a black star-like symbol.
\textit{Right column:} Time evolution of the critical angles. The integration time is equal to $10^5$~yrs. Initial conditions are marked with GEA-I, anti-aligned, low eccentricity (LE) configuration; GEA-II, aligned, low eccentricity (LE) configuration; GEA-Model~III, anti-aligned, moderate eccentricity (ME) configuration, and GEA-Model~IV, anti-aligned, high eccentricity (HE) configuration, respectively.
}
\label{figure:fig6}
\end{figure*}

In the left column of Fig.~\ref{figure:fig6}, for a reference as well as for a better representation of the TTV models, we report two--dimensional $(a, e)$ $\Mmean$-NAFF dynamical maps for the four representative GEA models in Tab.~\ref{table:tab2}. For the anti-aligned, low eccentric (GEA~I), moderate eccentric (GEA-III), and high eccentric (GEA-IV) models, respectively, the resonant structures are generally similar. Some differences may be observed between fine structures presented in the maps, which indicate a small dependence on eccentricities. As explained above, the left separatrix does not clearly appear at low eccentricities due to its very narrow width. Large regions of strong dynamical instability appear at moderate eccentricities beyond the resonance borders, yet the 3:2 resonance persists for eccentricities as large as 0.3. All the GEA solutions are found inside the 3:2 MMR dynamical structure bounded by two separatrices in the $(a_1,e_2)$-- and $(a_1,e_1)$--planes. We recall that in the $\Mmean$-NAFF maps, dark-blue zones and strips, with the lowest values of $|f_1/f_2-3/2|$ correspond to the periodic orbits in the rotating frame, and are associated with the 3:2 MMR centers for the given map coordinates.

In the middle column of Fig.~\ref{figure:fig6}, we calculate the amplitude of the two critical angles in  one--dimensional $a_1$ scans across fixed, nominal eccentricities. As for the NAFF structures, these scans are similar to each other, showing critical angles librating with low amplitude close to the centre of the 3:2 commensurability region. Significant differences appear for the aligned case, in which the first critical angle ($\phi_{\rm 3:2,1}$) circulates in all the 1-dim domain. In the right column of Fig.~\ref{figure:fig6}, the time evolution of the critical angles is represented in the ($\phi_{\rm 3:2,1}$,$\phi_{\rm 3:2,2}$)-plane.

\begin{figure*}
\centerline{
\vbox{
\hbox{
\includegraphics[width=0.48\textwidth]{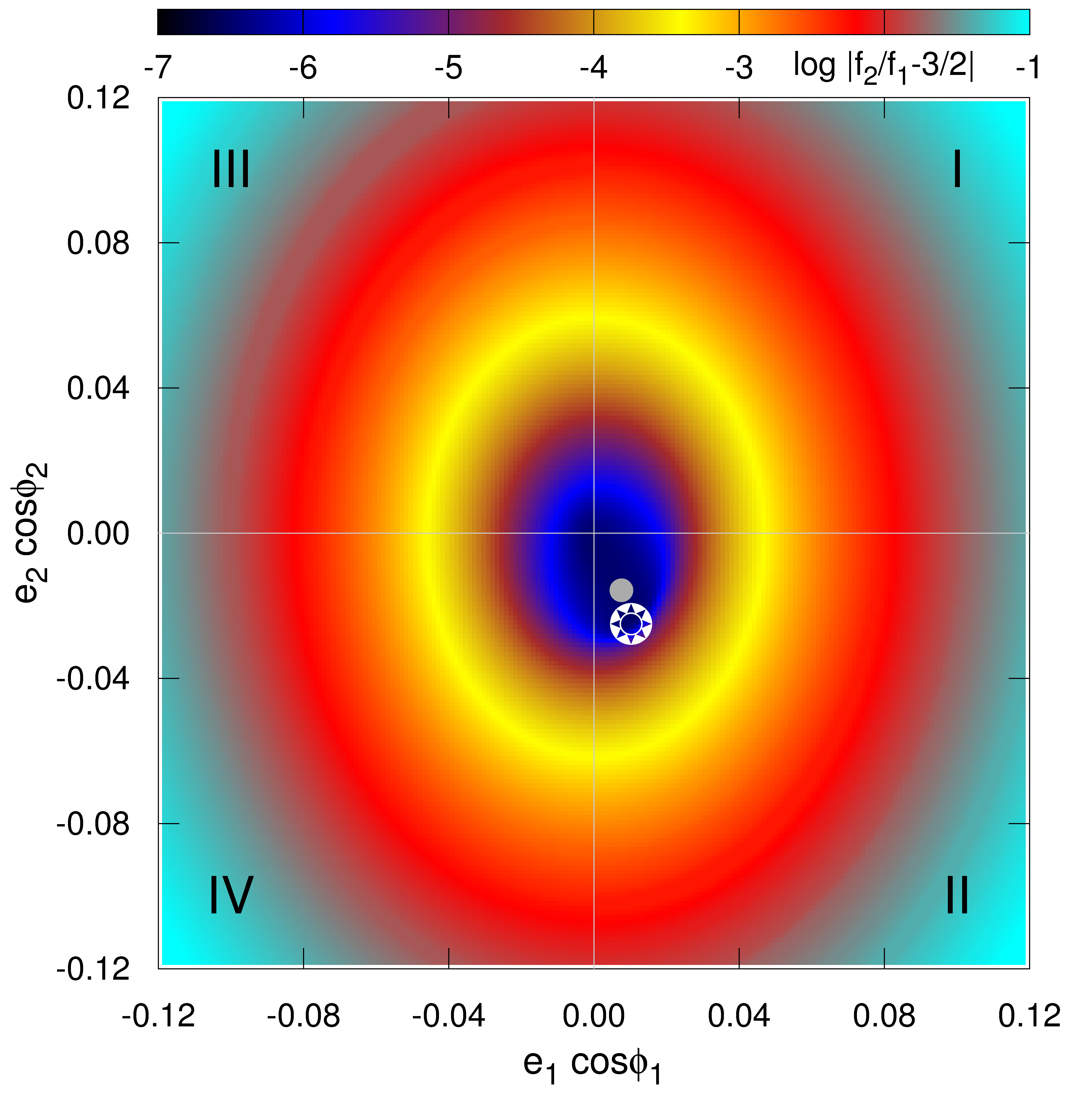}
\quad
\includegraphics[width=0.48\textwidth]{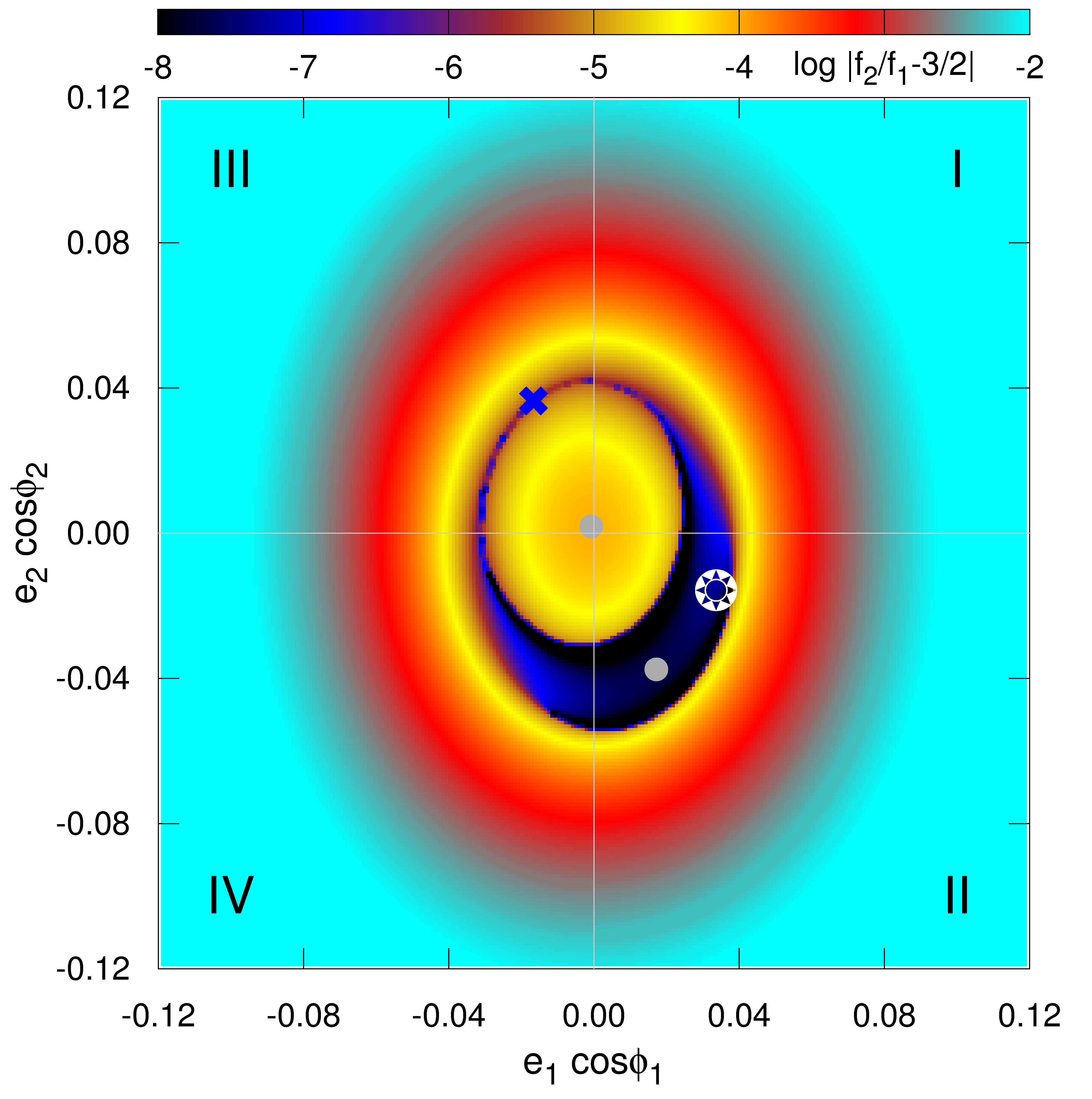}
}
\medskip
\hbox{
\includegraphics[width=0.48\textwidth]{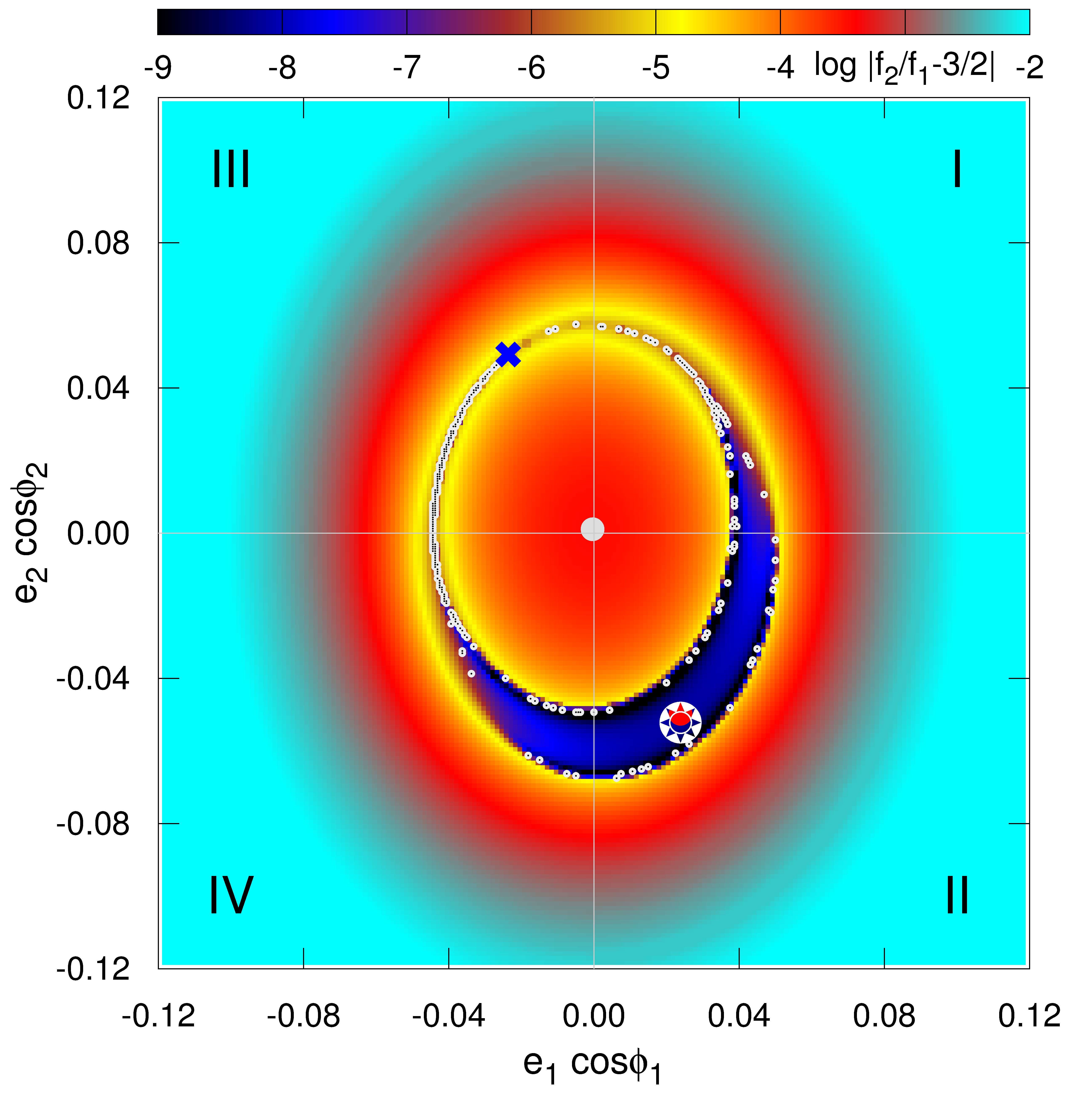}
\quad
\includegraphics[width=0.48\textwidth]{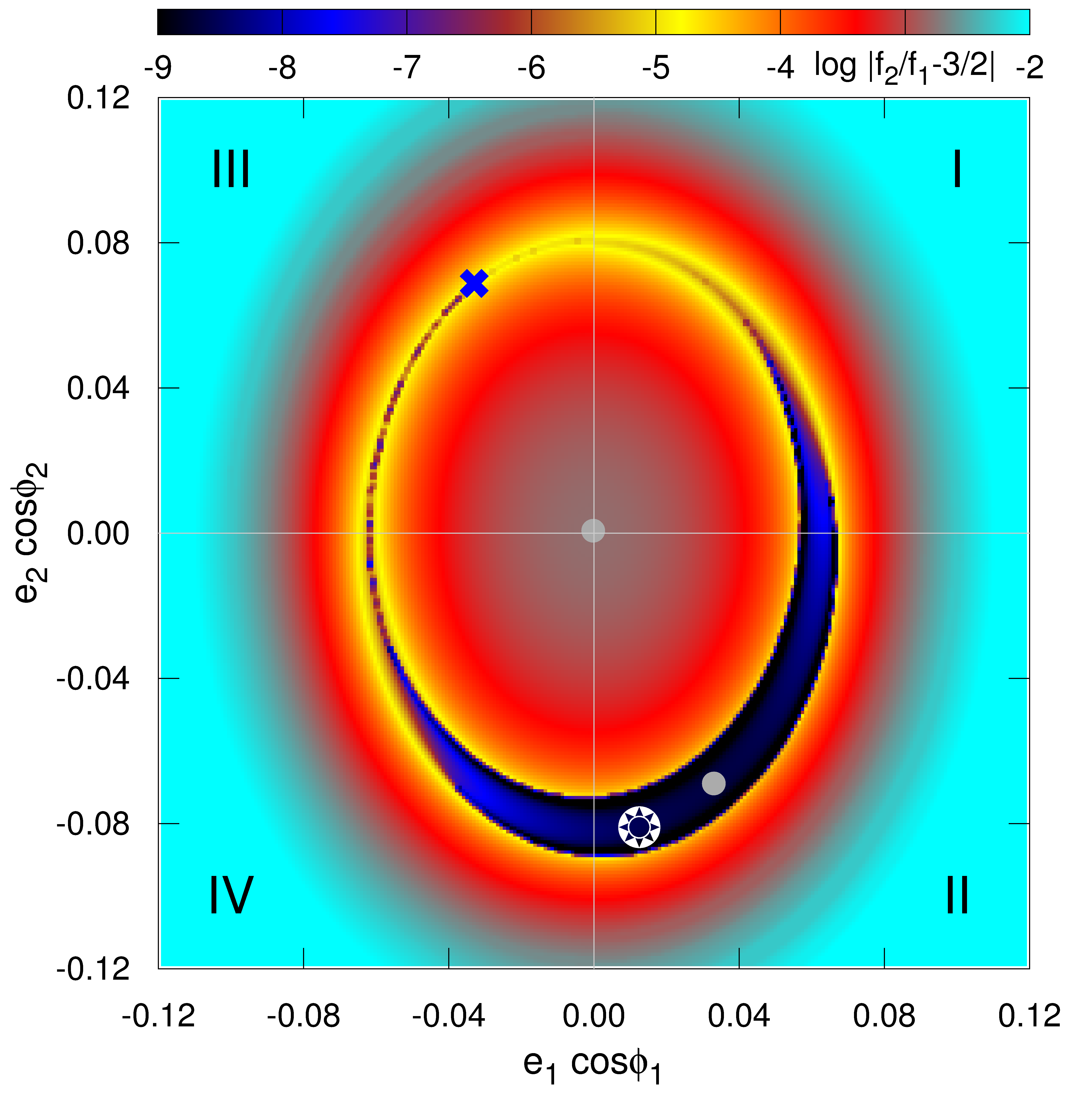}
}
}
}
\caption{
{Dynamical maps in terms of $|f_2/f_1-3/2|$ in the representative plane of initial conditions $\Sigma$ for the $C$ and $K$ integrals fixed at their values computed for the best-fitting, representative GEA models. All maps were computed by using the $\lambda$-NAFF with the resolution of $256\times256$ initial conditions, integrated for $2^{20}$ time-steps of 1~day with the SABA$_4$ symplectic scheme. Subsequent panels from the top-left to the bottom-right are for the GEA~I, GEA~II, GEA~III and GEA~IV solutions, respectively. The nominal solution is marked with the star symbol at each map. Small filled circles for the GEA~III (the bottom-left panel) are for unstable solutions detected with the MEGNO indicator $\Y>5$  in a grid of $256\times256$ initial conditions. The MEGNO indicator was integrated for 72,000 outermost periods ($\sim 2\times 10^6$ days), for each point in the grid. The large filled circles mark stable periodic orbits associated with the 3:2 MMR, and the blue crosses are for unstable periodic orbits. See the text for more details.}
}
\label{figure:fig5a}
\end{figure*}

{In Fig.~\ref{figure:fig5a}, we show the $\lambda$-NAFF dynamical maps for the four GEA best-fitting solutions, computed at the $\Sigma$-plane. The GEA~I solution
is topologically similar to the MCMC one (see Fig.~\ref{figure:fig5}), being close to the stable periodic orbit related to the 3:2 MMR. 
The GEA~II, GEA~III, and GEA~IV are topologically similar to each other. All of them are inside the dynamically resonant region determined by the very low values of $|f_2/f_1-3/2|$. For GEA~III (which we chose as a representative for this group of configurations), the MEGNO unstable regions overlap with the $\lambda$-NAFF resonant borders. Interestingly, the location of the GEA~III best-fitting solution coincides almost exactly with the position of the periodic orbit.}

%
\section{Migration}
\label{section6}
%
As we demonstrated above, the TTV fitting cannot give any unique best-fitting configuration. There are observationally permitted systems with eccentricities ranging from very small values up to 
$\sim 0.1$ and possibly beyond. The relative orientation of the apsidal lines cannot be constrained either, since both aligned and anti-aligned orbits are possible. We have shown also that the MCMC best-fitting configurations depend strongly on the eccentricities priors.  

It is well known that convergent migration results in resonant systems with anti-aligned apsides \citep[e.g.,][]{Batygin2013}. Moreover, if the migration is smooth and acts long enough the final systems are periodic configurations \citep{Migaszewski2015}. Recently, \cite{Migaszewski2018} showed that anti-aligned orbits can be misinterpreted as aligned ones due to the TTV model degeneracy. This is why we limit our analysis to anti-aligned systems and we will not make any attempts to reconstruct the aligned configurations on the way of migration.  However, even with this restriction, a migration-based reconstruction of the KOI-1599 system formation is a non-trivial task due to complex and non-deterministic constraints. 

Inward migration of planets due to their tidal interaction with the protoplanetary disc is a widely accepted formation scenario for short-period planets. It is also well known that the convergent migration of two planets results in locking them into mean motion resonance \citep[e.g.,][]{Snellgrove2001,Lee2002}. In this section, we study the planetary migration as a way in which the KOI-1599 system was formed. We use the parametric model of migration \citep{Papaloizou2000,Beauge2006,Moore2013,Voyatzis2016}, in which the $N$-body astrocentric Newtonian equations of motion of the $i$-th planet are completed with the following acceleration which mimics the planet-disc interaction:
\[
\vec{f}_i = -\frac{\vec{v}_i}{2 \tau_i} - \frac{\vec{v}_i - \vec{v}_{c,i}}{\tau_i \kappa_i^{-1}},
\]
where $\vec{v}_i$ is the $i$-th planet's astrocentric velocity, $\vec{v}_{c,i}$ is its Keplerian velocity for a circular orbit of radius $r_i$ (the astrocentric distance of planet~$i$). The time scale of migration is denoted with $\tau_i$, while the circularization time scale is given by $\tau_i$ divided by a factor $\kappa_i$. Within the model, we can choose the parameters freely in order to obtain the observed configuration, without considering particular disc properties.

Since, as shown above, the observational system appears as dynamically resonant, and likely has evolved into the 3:2~MMR via migration, we may {\em a priori} consistently impose the TTV signals characteristics, like their amplitudes and periodicities, as well as the period ratio. In this way, we avoid using the best-fitting  Keplerian orbital elements as a target for the migration simulations, since they are dependent on additional assumptions (like eccentricity priors).

Before constrain the observables mentioned above, we ask how the TTV signals of a system which was formed on the way of migration should look like. The TTV signals of a periodic configuration should be sinusoidal with a period that equals the so-called super-period \citep{Lithwick2012}, which is related to the rotation of the system as a whole, i.e., $\Tsp = |q/\Pinobs - (q+1)/\Poutobs|$ for the (q+1):q resonance, where $\Pinobs$ and $\Poutobs$ denote the mean periods of the transit times (TT) series computed over the time interval longer than the super-period. Additionally, the signals are in anti-phase. 

\begin{figure}
\centerline{
\vbox{
\hbox{
\includegraphics[width=0.47\textwidth]{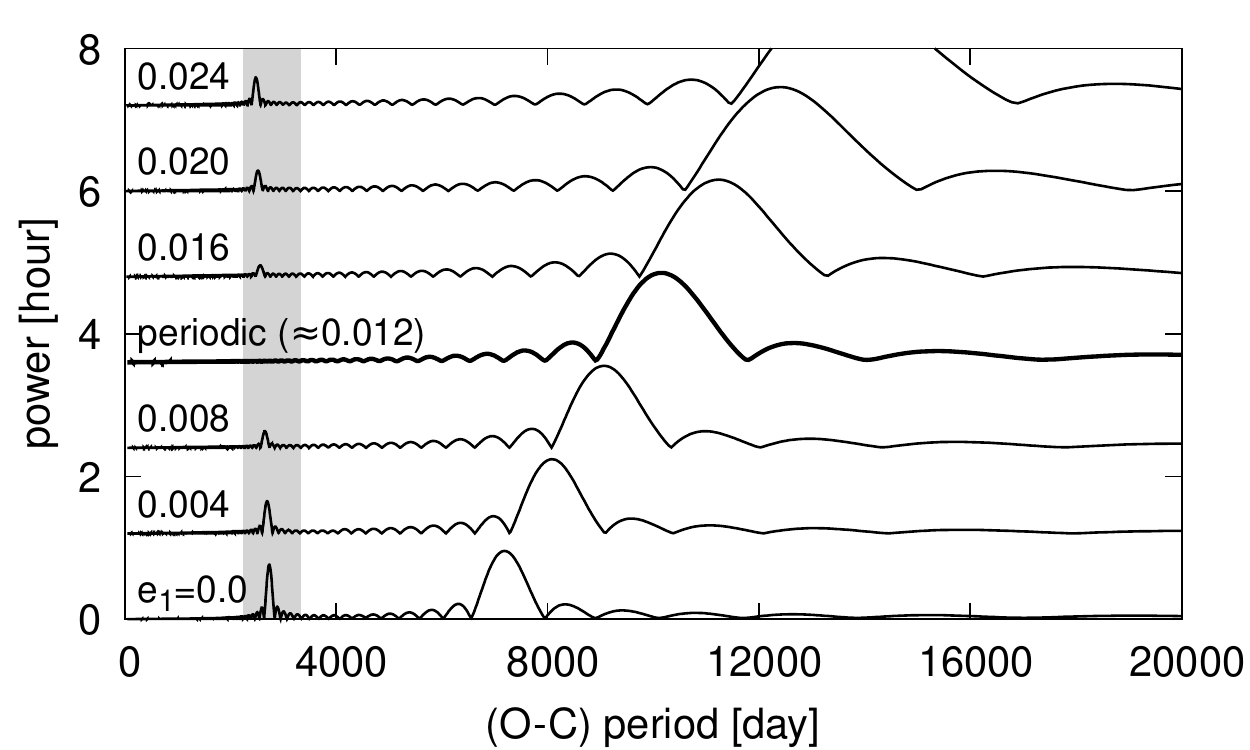}
}
}
}
\caption{
Lomb-Scargle periodograms of the inner planet of an example periodic configuration (the thick curve) and configurations with different $e_1$ (labelled). The planets' masses equal the best-fitting configuration listed in Tab.~\ref{table:tab3}, i.e., $m_1 = 9\,\mE$, $m_2 = 4.6\,\mE$. The semi-major axes are chosen in order to have $\ratio = 1.501$ for the periodic configuration. The grey area indicates roughly the periodicity in the TTVs. See the text for details.
}
\label{figure:fig7}
\end{figure}

If the system is shifted from the periodic configuration, there appears another periodicity in the TTVs, which is related to the resonant modulations of the semi-major axes. Figure~\ref{figure:fig7} shows the Lomb-Scargle periodograms of example systems of masses like in the KOI-1599 representative best-fitting configuration (Tab.~\ref{table:tab3}) and the $\ratio = 1.501$ (close to the value for the best-fitting system). The thick curve in the centre of the plot corresponds to the periodic system. There is only one periodicity $\approx 10000\,$days, i.e., much longer than the observing window. The remaining six curves are for the configurations shifted from periodic, i.e., the eccentricity of the inner planet is being varied between $0$ and $0.024$ (for the periodic system $e_1 \approx 0.012$), while all other parameters are left unchanged. The super-period changes from one system to another, because when changing $e_1$, $\ratio$ changes as well, even though $P_2/P_1$ (the osculating period ratio) is the same. But more importantly, an additional periodicity $\sim 3000\,$days appears. The grey area indicates roughly the observed periodicity of the TTVs (see the Sect.~\ref{subsection7.1}). The amplitude of this additional signal increases with the distance from the periodic configuration. Therefore, we conclude that the KOI-1599 system is shifted from periodic, the observed TTVs are due to the resonant modulations of $a_1, a_2$ and the observing window is much narrower than the super-period.

\subsection{TTV signals characteristics}
\label{subsection7.1}
In the next step, we constrain the TTVs characteristics. As we mentioned, the TTV signals can be modelled with two sine (or cosine) functions. In other words, the transit times model $t_n$ should consist of three parts:
\begin{equation}
t_n = t_n^{\idm{(lin)}} + \Delta t_n^{\idm{(s-p)}} + \Delta t_n^{\idm{(res)}},
\label{eq:model_tt}
\end{equation}
where 
\[
t_n^{\idm{(lin)}} = n \Pobss + T
\]
is the linear model of transit times, usually used in order to find the observed period $\Pobss$, i.e., the mean time distance between subsequent transits and the epoch of the first transit, $T$. For a system of two interacting planets those quantities, in general, depend on the width of the observing window. Therefore, the values obtained for the window of $\sim 4$~years (as it is for the \kepler{} candidates) usually differ from the ones computed over time longer than the longest periodicity in TTV. This means that for a wide enough observing window (in the example illustrated in Fig.~\ref{figure:fig7} the window should be wider than $\sim 10000\,$days) $\Pobss$ and $T$ stop depending on the observing window width. Further in this section, $\Pobss$ and $T$ have the sense given above.

The model of the transit times series is completed with two cosine functions  
\[
\Delta t_n^{\idm{(s-p)}} = \Asp \cos(\Nsp t_n^{\idm{(lin)}} + \phisp),
\]
and
\[
\Delta t_n^{\idm{(res)}} = \Ares \cos(\Nres t_n^{\idm{(lin)}} + \phires).
\]
The first one describes the TTV caused by the rotation of the system as a whole, with a characteristic frequency related to the super-period, $|\Nsp| \equiv 2\pi/\Tsp$, i.e., $\Nsp = (q+1) n_2 - q n_1$, where $n_1 \equiv 2\pi/\Pinobs$, $n_2 \equiv 2\pi/\Poutobs$. The amplitude of this signal $\Asp$ is related to the eccentricity in the following way:
\[
\frac{\Asp}{\Pobss} = \frac{\emean}{\pi},
\]
where $\emean$ is the mean value of the eccentricity. The above relation stems from the fact that for the Keplerian motion (for small $e$) the true anomaly can be expressed through the mean anomaly as $\nu \approx \Mmean + 2 e \sin \Mmean$, therefore the maximal difference between anomalies with respect to the full angle is $(2e)/(2\pi)$, what gives the amplitude with respect to the period. The phase of the signal for the inner planet can be computed when $\Pinobs$, $\Poutobs$, $T_1$ and $T_2$ are known as
\[
\phiinsp = \frac{3}{2} \pi - \phieq + q n_1 T_1 - (q+1) n_2 T_2,
\]
where $\phieq$ is the equilibrium value of the resonant angle $\phi_1 \equiv q \lambda_1 - (q+1) \lambda_2 + \varpi_1$, i.e., $\phieq = 0$ for $\ratio > 1.5$. The phase for the outer planet $\phioutsp = \phiinsp + \pi$.

The second signal $\Delta t_n^{\idm{(res)}}$ stems from the resonant evolution of the system, i.e., the semi-major axes modulations. Unlike $\Nsp$ (which depends on $\ratio$ only), the frequency of the second signal $\Nres$ depends on the planets' masses and eccentricities. 

The model of the transit times, Eq.~\ref{eq:model_tt} may be now used in order to find the best-fitting parameters $\Pinobs$, $\Poutobs$, $T_1$, $T_2$, $\Ainsp$, $\Aoutsp$, $\Nsp$, $\phiinsp$, $\phioutsp$, $\Ainres$, $\Aoutres$, $\Nres$, $\phiinres$, $\phioutres$. In general, that would make $14$ free parameters to be found if there were no dependencies between them. However, as already mentioned $\Nsp$, $\phiinsp$ and $\phioutsp$ are functions of $\Pinobs, \Poutobs, T_1, T_2$. The amplitudes $\Ainsp, \Aoutsp$ depend on $\ratio$ when the masses are given. Moreover, the phases of the resonant part of the model are in anti-phase, i.e., $\phioutres = \phiinres + \pi$. As a result there are only $8$ free parameters which have to be found. The dependence of $\Ainsp$ and $\Aoutsp$ on $m_1, m_2$ may seem a problem, since the masses are known within $\sim 50\,\%$ uncertainties. However, $\Nsp$ and the phases are independent of the masses and for $\ratio$ very close to $3/2$ the super-period is at least an order of magnitude longer than the observing window  while for $\ratio$ shifted significantly from $3/2$ the amplitude $\Asp$ is below the TTs uncertainties. As a result, the best-fitting free parameters of the model are almost independent of the masses, assuming the super-Earth mass range.
\begin{figure}
\centerline{
\vbox{
\hbox{
\includegraphics[width=0.23\textwidth]{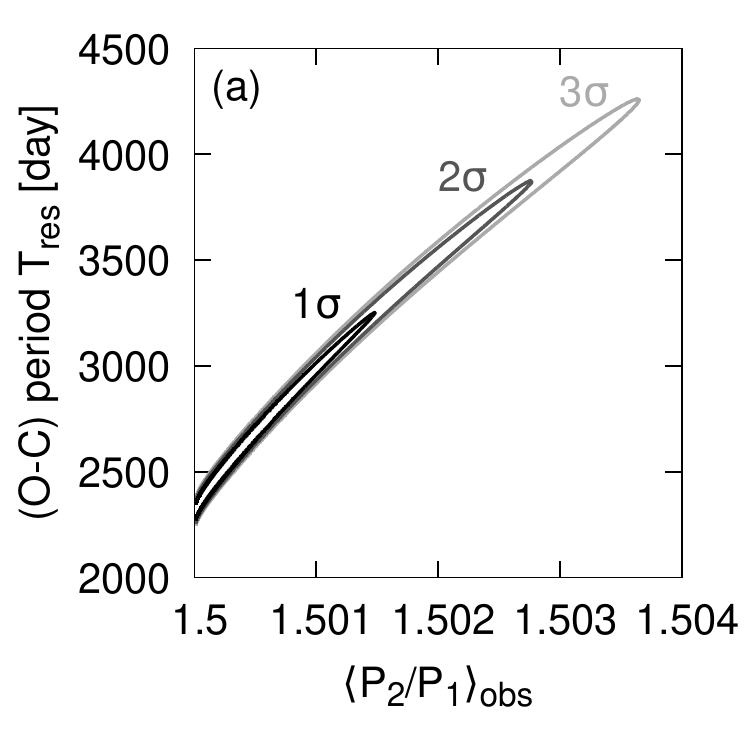}
\includegraphics[width=0.23\textwidth]{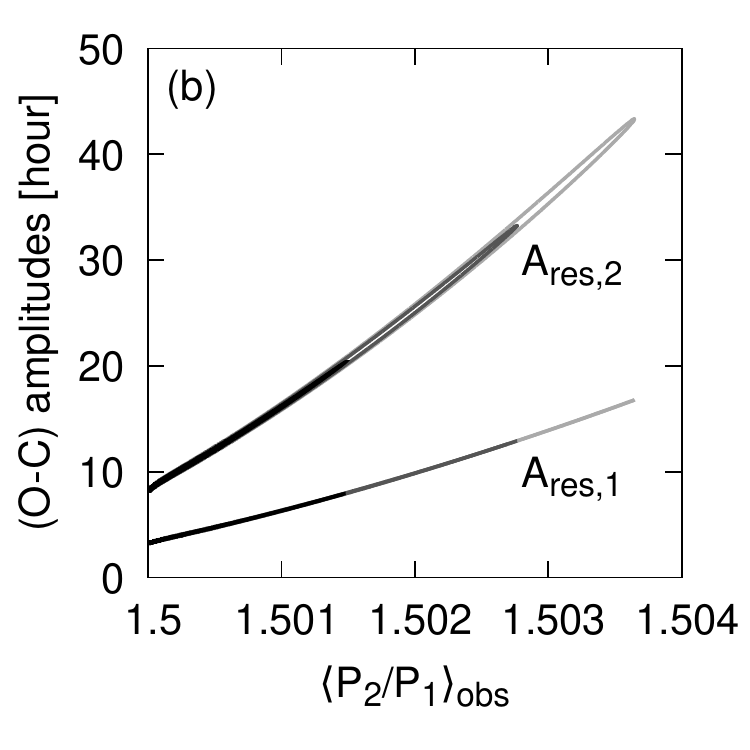}
}
}
}
\caption{
Results of the geometric fit to the KOI-1599 transit times data presented as $1 \sigma , 2 \sigma $ and $3 \sigma$ confidence levels at two parameter planes. The assumed planets' masses are $m_1 = 7.0\,\mE, m_2 = 3.6\,\mE$ and the stellar mass $m_{\star} = 1.02\,\msun$.}
\label{figure:fig8}
\end{figure}

The results of fitting the model given by Eq.~\ref{eq:model_tt} to the transit times series of the KOI-1599 system are presented in Fig.~\ref{figure:fig8}. There is no minimum of the $\chi^2$ function. In order to illustrate the shape of $\chi^2$ we choose and fix $\ratio$ and $\Nres$ from a grid and find a minimum of $\chi^2$ in the space of the remaining parameters. Next, we compute the confidence levels and plot them in Fig.~\ref{figure:fig8}a. The (O-C) period, $\Tres \equiv 2\pi/\Nres$ is correlated with $\ratio$, although there is maximal allowed $\ratio \approx 1.5037$, and thus also the period can be constrained to be in a range of $(\sim 2200, \sim 4250)\,$days. For the representative configurations listed in Tab.~\ref{table:tab3} and Tab.~\ref{table:tab2} $\ratio \lesssim 1.5016$.

We did not check for $\ratio < 1.5$, because the periodic configurations for $\ratio$ close to, but lower than $1.5$ are unstable. As we study the migration-induced formation of the KOI-1599 system, $\ratio$ could be lower than $1.5$ if i) the migration was divergent and stopped when $\ratio$ was just below $1.5$, ii) the convergent migration was very fast so the system passed through the resonance and stopped just below it, or iii) the convergent migration--induced  resonance capture was only temporary. The two former situations are in principle possible, but the formation of this particular configuration could have happened only by pure coincidence. The latter case would require $m_1 < m_2$ in order to make the capture over-stable \citep[e.g.,][]{Delisle2015,Xu2016}, i.e., the equilibrium should be unstable against migration, and we know from the MCMC TTV modelling that $m_1 \approx 2 m_2$.

\begin{figure*}
\centerline{
\vbox{
\hbox{
\includegraphics[width=0.4\textwidth]{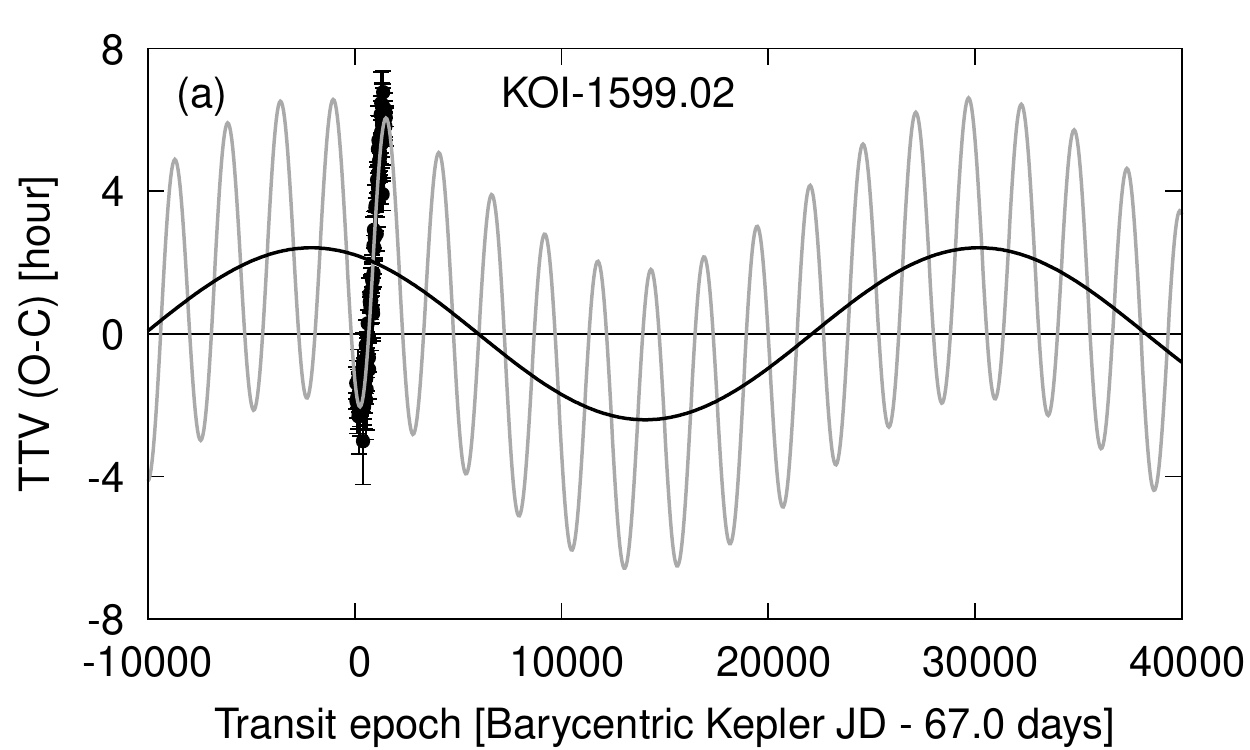}
\includegraphics[width=0.4\textwidth]{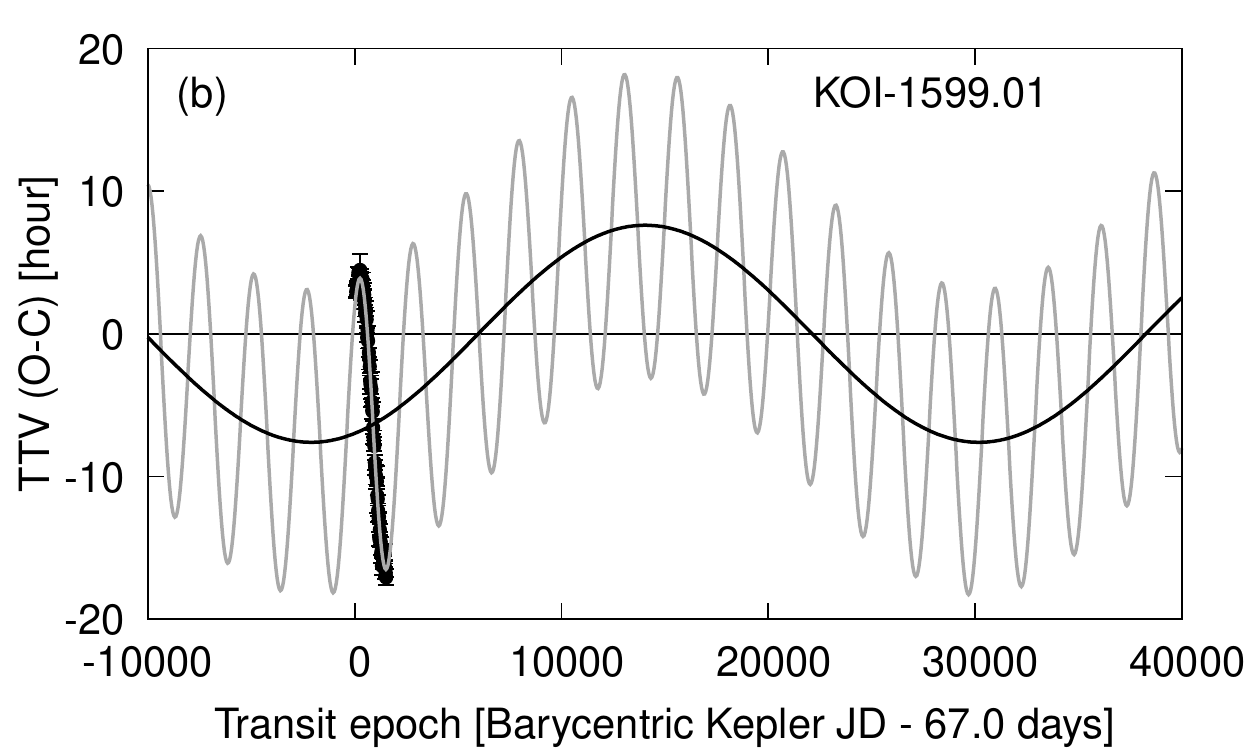}
}
\hbox{
\includegraphics[width=0.4\textwidth]{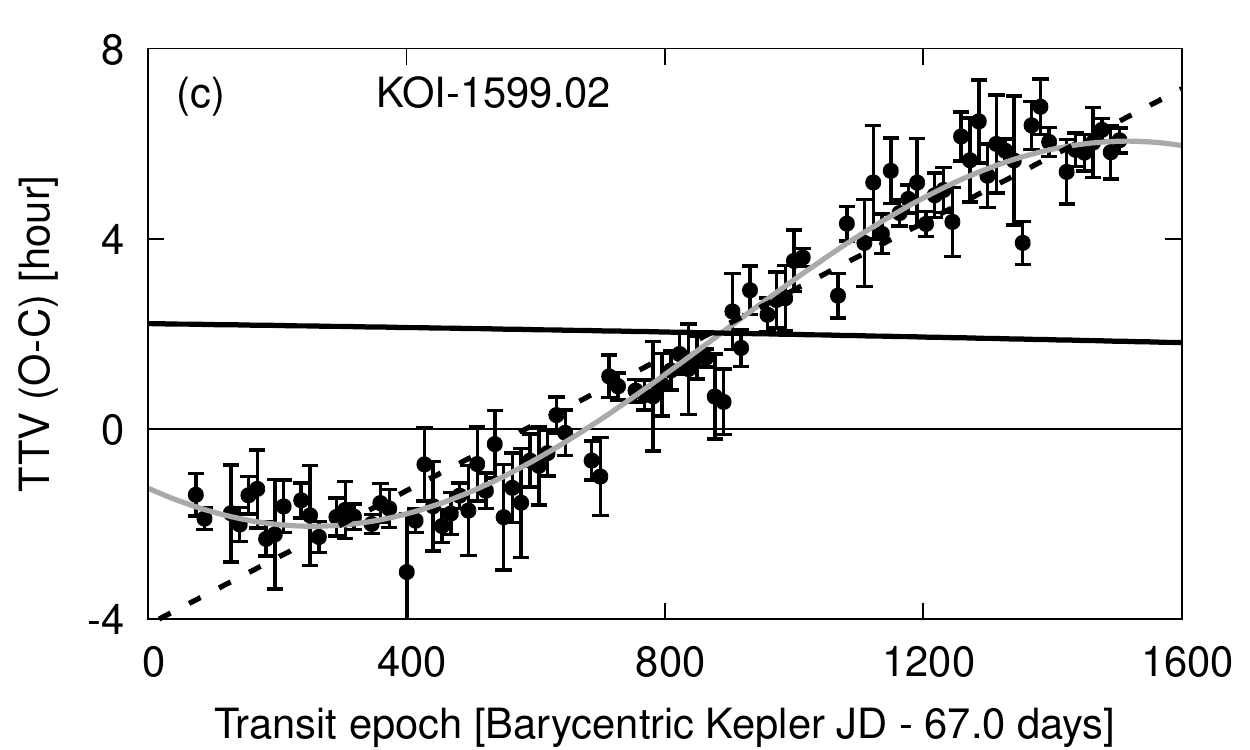}
\includegraphics[width=0.4\textwidth]{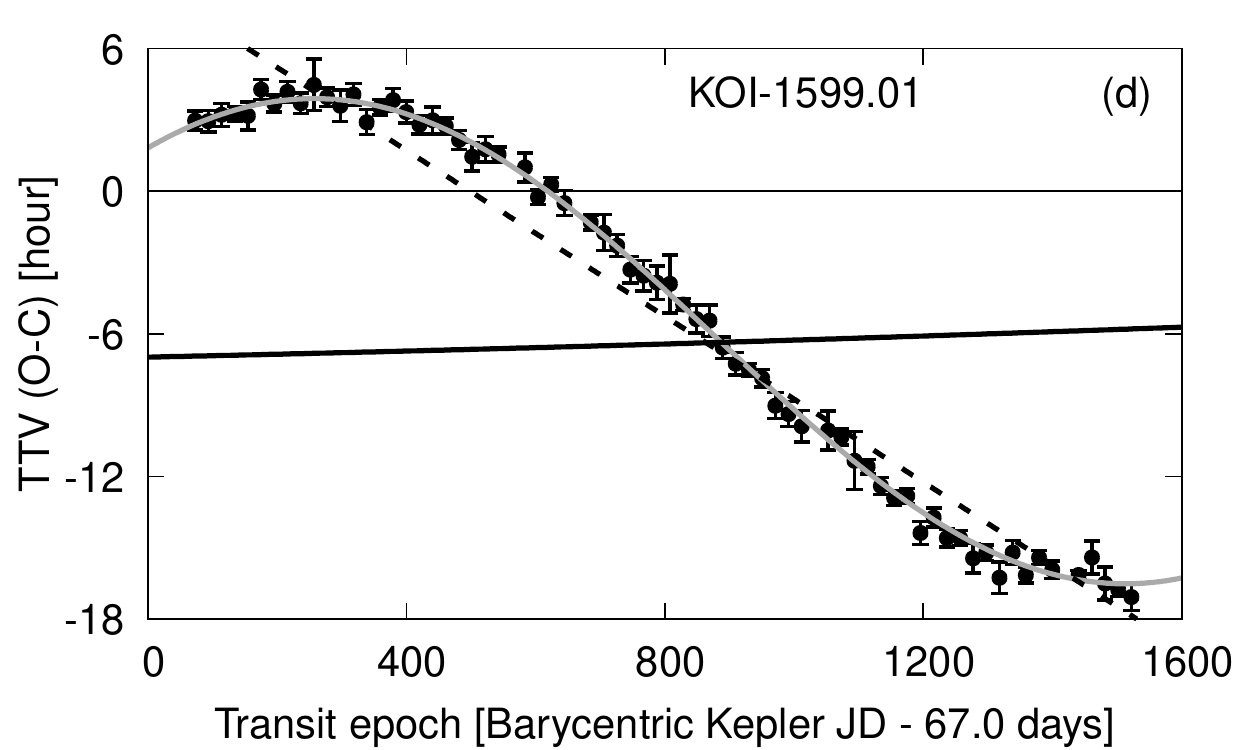}
}
}
}
\caption{The TTV data with the TTs model (Eq.~\ref{eq:model_tt}) over-plotted. The black curves correspond to the TTV modulations due to the system rotation, while the grey curves indicate the TTV due to the resonant evolution of the system slightly shifted from the periodic configuration. The dashed lines in the two bottom panels indicate the transit times resulting from the linear model of transits fitted to the data. See the text for details.}
\label{figure:fig9}
\end{figure*}

Figure~\ref{figure:fig8}b illustrates the results of the fitting procedure for the amplitudes of the resonant modulations in TTV. They are even more strongly correlated with $\ratio$ and the allowed values range from $\sim 3$ to $\sim 18$ hours for the inner planet's signal and from $\sim 8$ to $\sim 43$ hours for the outer planet. The amplitudes seem to be significantly overestimated when looking at the standard (O-C)~diagrams presented in Fig.~\ref{figure:fig4}. Nevertheless, the situation becomes clear from Fig.~\ref{figure:fig9}. The two top panels show the (O-C)~diagrams, but instead of taking $P,T$ values from the standard linear fit to the TTs series, we use $\Pobss$ and $T$ stemming from the model given by Eq.~\ref{eq:model_tt}. Because, as mentioned earlier, there is no single best-fitting model, we chose arbitrarily $\ratio$ from the allowed range, i.e., $\ratio \approx 1.50032$ (more precisely, $\log_{10}(\ratio - 1.5) = -3.5$), for which the best-fitting $\Tres \approx 2560\,$d, which is almost twice as long as the observing window. The super-period for this value of $\ratio$ equals $\approx 40000\,$d. The two components of the TTV model are drawn with grey and black curves, respectively. While the top panels show the model and the data in a time interval long enough to encompass the super-period, the bottom panels of Fig.~\ref{figure:fig9} focus at the observing window only. The dashed lines in the bottom panels indicate the standard linear model of TTs. We see that not only the mean periods computed for the observing window differ from $\Pinobs$ and $\Poutobs$, but also the TTV amplitudes one could infer from the standard (O-C)~diagrams are much smaller than the actual amplitudes of the resonant modulations of the TTV.

\subsection{Constraining the migration parameters}

In the previous subsection, we put constraints on the observables $\ratio, \Tres, \Ainres$ and $\Aoutres$. The next step is to find the migration parameters and possibly the planet's masses which could reconstruct the observed quantities of the TTV data. As we mentioned, in order to fit the model, Eq.~\ref{eq:model_tt}, to the data, we needed to fix the planets' masses, because $\Ainsp, \Aoutsp$ depend on $m_1, m_2$ and we could not treat the amplitudes as free parameters due to the too narrow observing window. Luckily, the resulting constraints on the TTV signal observables do not depend on the masses in the  mass range expected for the KOI-1599 planets. The amplitudes $\Ainres, \Aoutres$, on the other hand, are free parameters of the TTs model, as is the period $\Tres$. All those three quantities depend on the planets masses as well as on the eccentricities. As illustrated in Fig.~\ref{figure:fig7}, $\Ainres$ (and similarly $\Aoutres$) increases when the system shifts away from the periodic configuration. It is known, though, that very slow convergent migration into a first order MMR, like 3:2, results in a configuration that is very close to periodic \citep{Migaszewski2015}. In such a case one could expect small amplitudes $\Ainres$ and $\Aoutres$. Therefore, in order to obtain significant amplitudes of the TTV signal, the migration must be fast enough. Not too fast, though, since the system would pass through the resonance.

We expect from this qualitative analysis that $\Ainres$ and $\Aoutres$ will help us to constrain the migration rate, while the planets' masses could be constrained with $\Tres$, since smaller masses mean slower resonant modulations of $a_1, a_2$. The forth observable, $\ratio$ will be useful to constrain the $\kappa$ parameter since it governs the equilibrium period ratio, i.e., too high values of $\kappa$ would not let the system evolve towards $\ratio$ very close to $1.5$, and we know that $\ratio \lesssim 1.5037$.

\begin{figure*}
\centerline{
\hbox{
\includegraphics[width=0.33\textwidth]{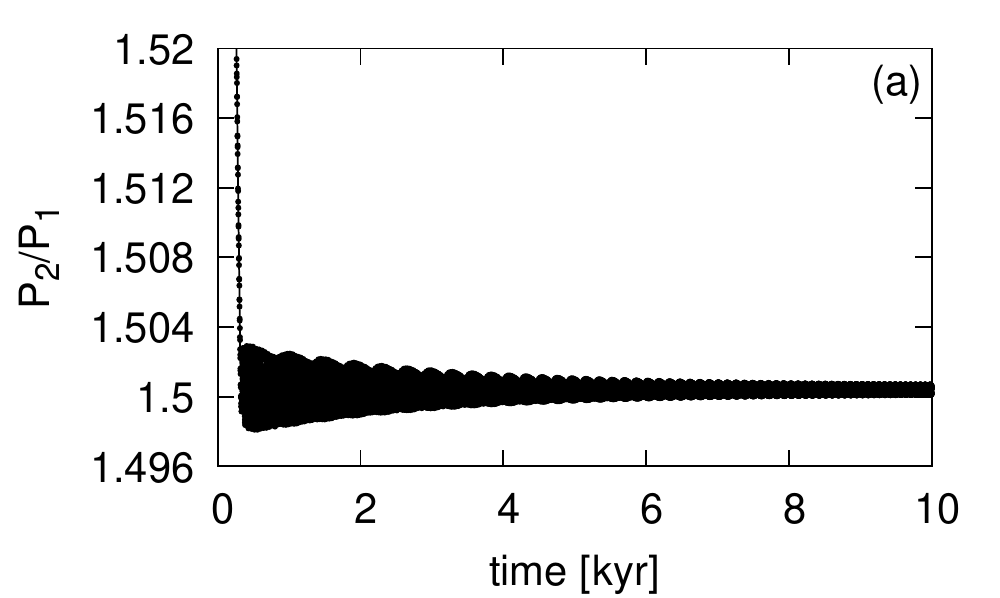}
\includegraphics[width=0.33\textwidth]{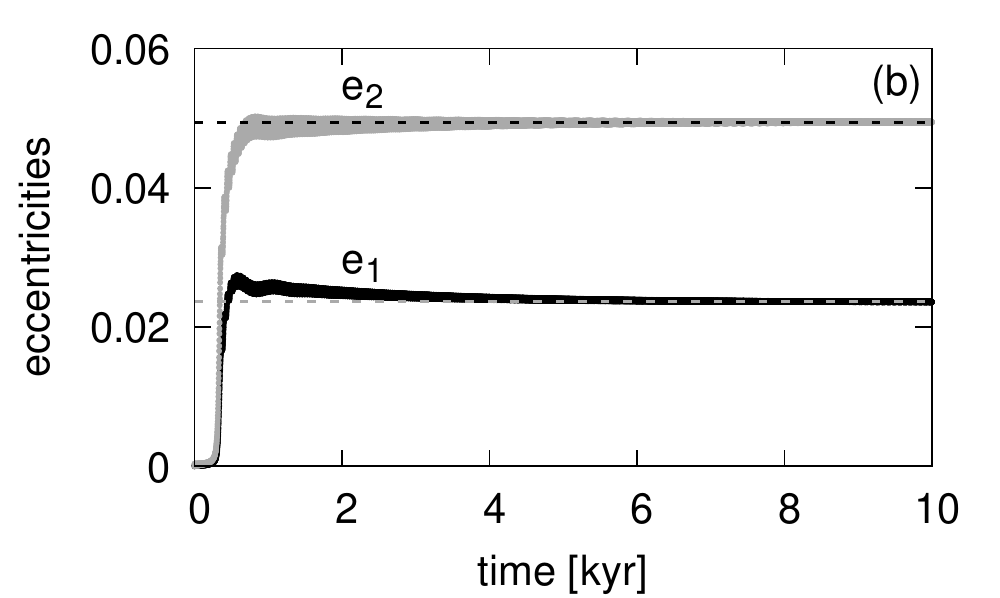}
\includegraphics[width=0.33\textwidth]{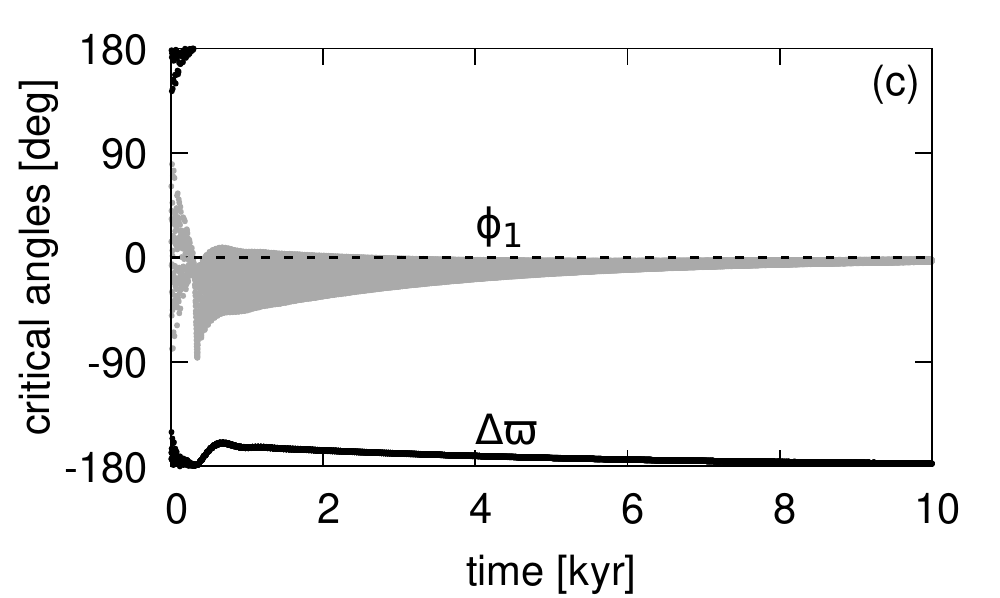}
}
}
\caption{The migration evolution of an example configuration (with $m_1 = 7.0\,\mE$, $m_2 = 3.6\,\mE$) illustrated as time-variation of the osculating period ratio (panel~a), eccentricities (panel~b) and the critical angles (panel~c).}
\label{figure:fig10}
\end{figure*}

After running a series of the migration simulations for different $\tau_1, \tau_2, \kappa_1 = \kappa_2 = \kappa$ and $m_1, m_2$ we found a very good agreement with the observables for $m_1 = 7.0\,\mE$, $m_2 = 3.6\,\mE$, $\tau_1 = 10\,$kyr, $\tau_2 = 4\,$kyr and $\kappa = 27$. The evolution of the osculating period ratio, the eccentricities as well as the critical angles is shown in subsequent panels of Fig.~\ref{figure:fig10}. The period ratio reaches the value of $1.5$ and librates in its vicinity with a decreasing amplitude. The eccentricities increase up to the values close to their equilibria (dashed lines), i.e., the so-called forced eccentricities \citep[e.g.,][]{Lithwick2012}. Note that the eccentricities as well as the masses are close to the GEA-III model (see Tab.~\ref{table:tab2}).
During further migration both $e_1$ and $e_2$ tend towards the dashed lines. The resonant angle $\phi_1$ and the difference between the longitudes of the apsidal lines $\Delta\varpi$ are slightly shifted from their equilibria, respectively $0$ and $\pi$, and similarly to the eccentricities, they tend towards them during the migration. 

\begin{figure*}
\centerline{
\vbox{
\hbox{
\includegraphics[width=0.4\textwidth]{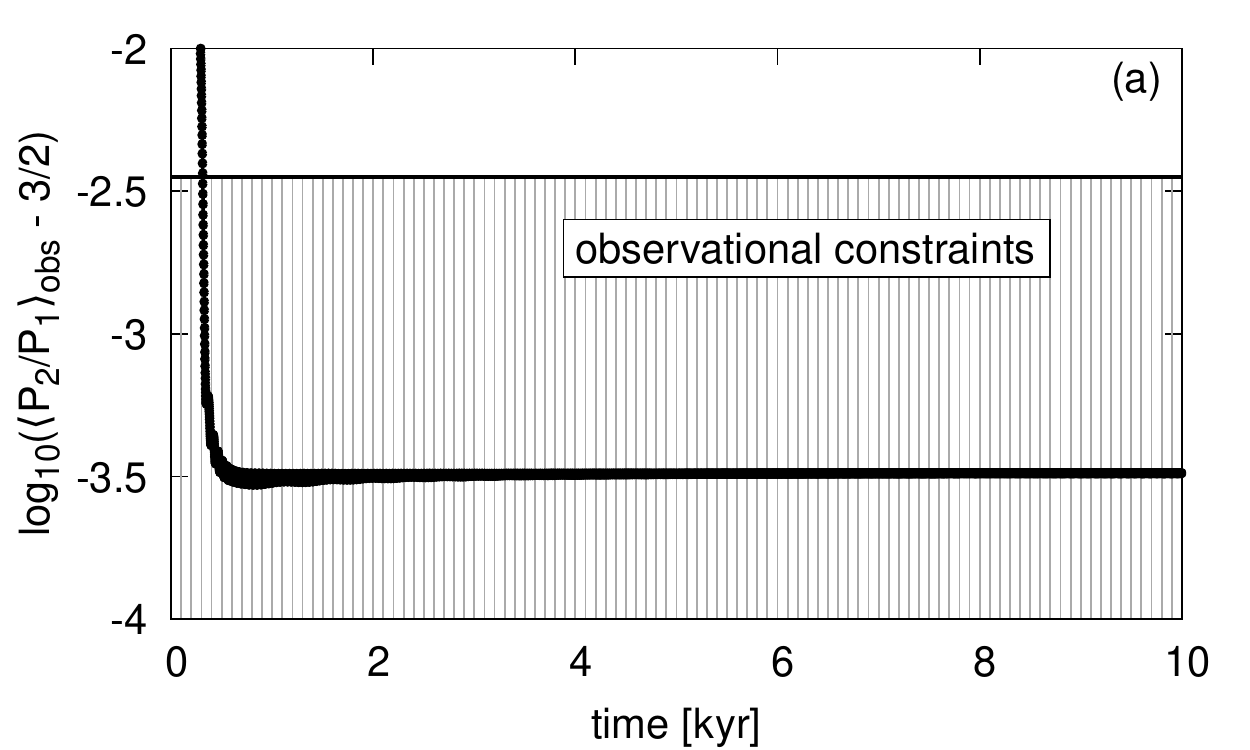}
\includegraphics[width=0.4\textwidth]{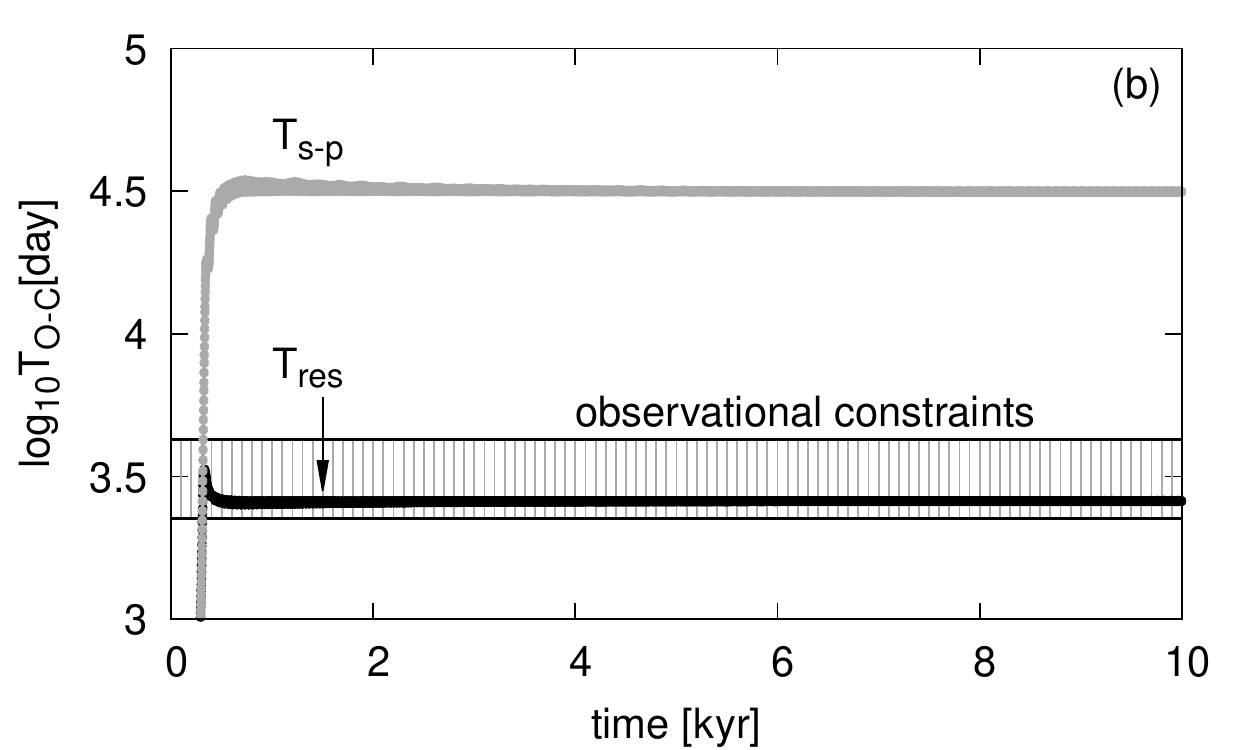}
}
\hbox{
\includegraphics[width=0.4\textwidth]{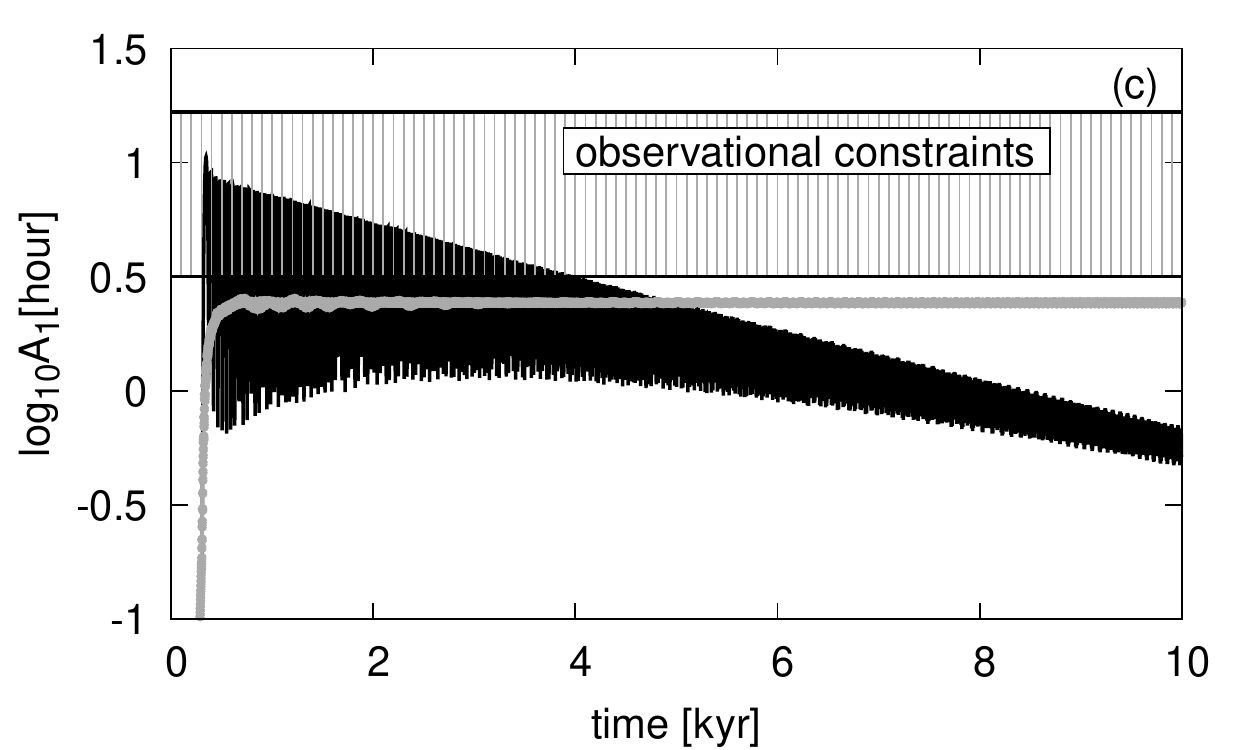}
\includegraphics[width=0.4\textwidth]{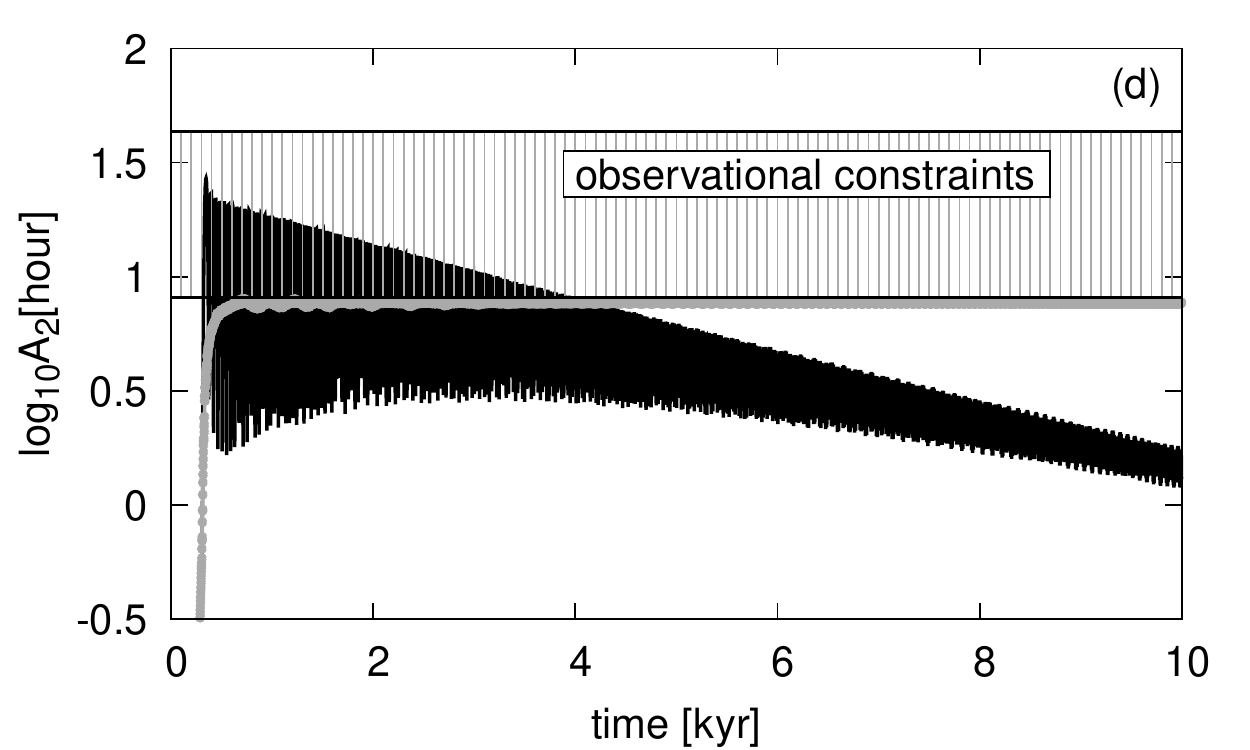}
}
}
}
\caption{The migration evolution of the observables for the same system as presented in Fig.~\ref{figure:fig10}, i.e., the observed period ratio (panel~a), the two main periodicities of the (O-C)~signal (panel~b) and the amplitudes corresponding to these periodicities, $\Ares$ is shown with the black, while $\Asp$ with the grey colour (panels~c and~d for the inner and outer planets, respectively). See the text for details.}
\label{figure:fig11}
\end{figure*}

Every $\sim 2.5\,$yr of the migration simulation we checked how would the (O-C)~signal look like, by integrating the $N$-body equations of motion (without the dissipative terms) over $\sim 300\,$yr (in order to encompass the super-period), evaluating TTs and computing the observables $\ratio$, $\Tres$, $\Ainres$ and $\Aoutres$. At each time the system was scaled to the size of KOI-1599. The results of the analysis are presented in Fig.~\ref{figure:fig11}. Despite large oscillations of $P_2/P_1$ after entering the resonance (Fig.~\ref{figure:fig10}a), $\ratio$ is almost constant -- see Fig.~\ref{figure:fig11}a, note that the period ratio is presented as $\log_{10}(\ratio - 3/2)$ -- and is kept at the value well inside the observational limits, actually very close to the example illustrated in Fig.~\ref{figure:fig9} (and close to the GEA-III model). Looking at this solution globally, we may notice that it belongs to a family of $(\sim 7, \sim 3.6)$~Earth mass solutions which permit both modes of librations, as illustrated in Fig.~\ref{figure:fig3}.

Figure~\ref{figure:fig11}b illustrates the evolution of both $\Tres$ and $\Tsp$. The period of the resonant modulations (the black curve) reaches a value within the observational constraints shortly after entering the resonance and stays unchanged. On the contrary, the amplitudes (shown in Fig.~\ref{figure:fig11}c,d) oscillate between high values (at the beginning of the resonance capture they are even higher than the observational ranges) and very low values, lower than the TTs uncertainties. Such large oscillations result from the fact that the migration is being switched-off every time we compute the synthetic TTV, and depending on the phase in which it is switched-off, we obtain higher or lower amplitudes of the resonant modulations. Therefore, one should look at the behaviour of the amplitudes in a mean sense. The ranges in which $\Ainres$ and $\Aoutres$ oscillate decrease while the system reaches the periodic configuration. Therefore, the consistency between the synthetic system stemming from the migration and the observations is temporary. Nevertheless, the consistency holds for up to $\sim 4\,$kyr after entering the resonance. If the migration, which is very fast at the beginning, slowed down, the system could remain consistent with the TTVs even longer.

\begin{figure*}
\centerline{
\hbox{
\includegraphics[width=0.33\textwidth]{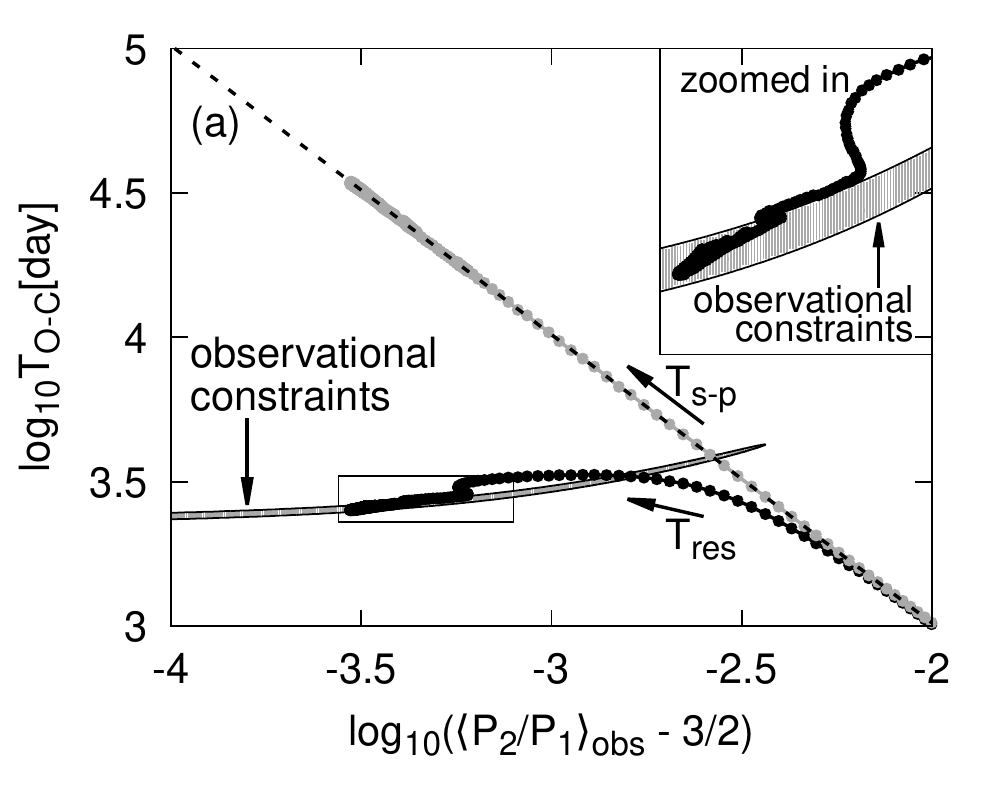}
\includegraphics[width=0.33\textwidth]{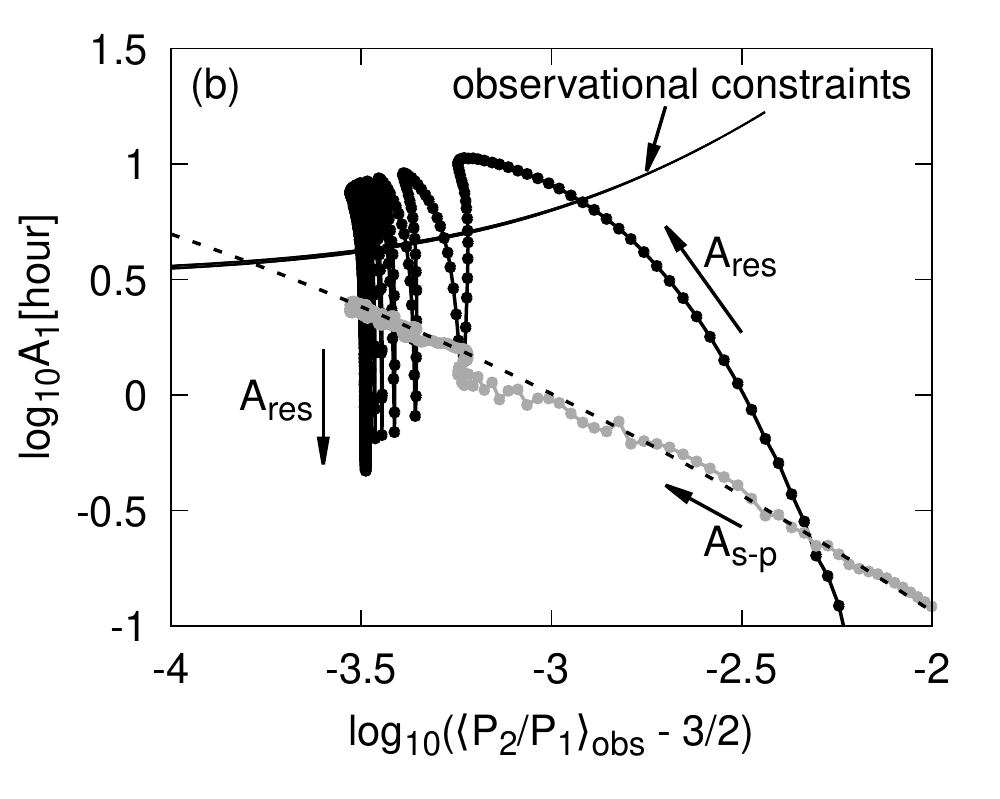}
\includegraphics[width=0.33\textwidth]{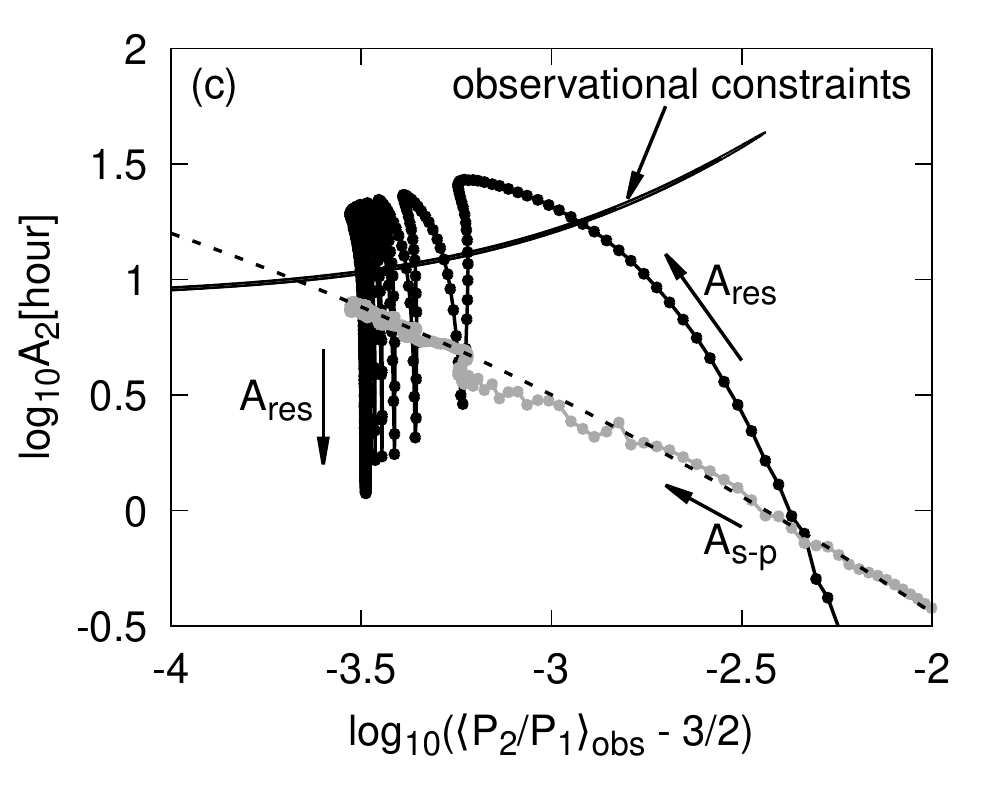}
}
}
\caption{The evolution shown in Fig.~\ref{figure:fig11} presented in different planes. The narrow slightly curved grey areas are the observational constraints. The direction of the evolution is marked with the arrows. The grey dots/curves denote the observables related to the system rotation, while the black dots/curves are for the resonant part of the TTs model. The dashed curves indicate the $\Tsp, \Ainsp$ and $\Aoutsp$ as functions of $\ratio$ for the periodic configurations.}
\label{figure:fig12}
\end{figure*}

The evolution of the observables can be also presented in a different manner, see Fig.~\ref{figure:fig12}. The observational constrains, illustrated first in Fig.~\ref{figure:fig8}, look here as very narrow, slightly curved areas. Instead of studying the time evolution of the four observables, in Fig.~\ref{figure:fig12} we show the evolution at the planes with $\ratio$ at the $x$-axis. The correspondence between the synthetic system and the data is clear, especially in Fig.~\ref{figure:fig12}a, in which the observables fits the observational constraints.

The general conclusion from this section is that the KOI-1599 system could have been formed on the way of migration. The necessary conditions are that the migration is fast enough, the eccentricity damping should not be too effective and the masses should be in the lower limits of the MCMC TTV analysis. The time-scales of the migration $\tau_1 = 10\,$kyr and $\tau_2 = 4\,$kyr, means that the time-scale of the period ratio variation, $\tau_X$ ($\tau_X = -X/\dot{X}$, where $X = P_2/P_1$), equals $10\,$kyr. The initial semi-major axis of the inner planet equals $0.4\,\au$. Therefore, at the beginning of the simulation, i.e., shortly before entering the resonance, $\tau_X$ is as long as $4 \times 10^4$ of the inner planet's orbital period. For the migration slowed down by a factor of $2$, the amplitudes $\Ares$ for the synthetic system fits the observational constrains only marginally. On the other hand, fastening the migration by a factor of $2$ makes the system pass through the resonance. 

The $\Tres$ dependence on the planets' masses constrains them relatively well. In order to fit the observational limits for all the four observables, the masses should equal $7.0$ and $3.6\,\mE$ within $\sim 10\,\%$ ranges. The observational constraints of $\kappa$ are also quite strong. For $\kappa \gtrsim 100$ the system only marginally reaches the observational upper limit of $\ratio \sim 1.5037$. For lower $\kappa$ the resonant modulations are being damped slower, and the synthetic system stays consistent with the observations over a longer time interval. The $\kappa$ parameter can be therefore limited only from the top, to be $\lesssim 100$.

\section{The density dichotomy of KOI-1599 planets}
\label{section5}
As we found with the photometric analysis of the LCs, the KOI-1599 Super-Earths have very similar radii of $\simeq 1.9~\RE$ Surprisingly,  our estimates differ from data in \cite{Rowe2015}. They report a large, likely incorrect radius $\sim 64 \RE$ of the inner planet, implying its density $10^4$ times smaller than the density of the Earth. 

The mass estimates in the TTV model in Tab.~\ref{table:tab3} have small uncertainties of $\simeq 20\%$. Their values are consistent with predictions from the mass--radius relation $M = 10^C \times R^E$, where $C=0.50 \pm 0.03$ and $E=0.64\pm 0.06$ \citep{Mills2017}.  

\begin{figure}
\centering
\includegraphics[width=0.47\textwidth]{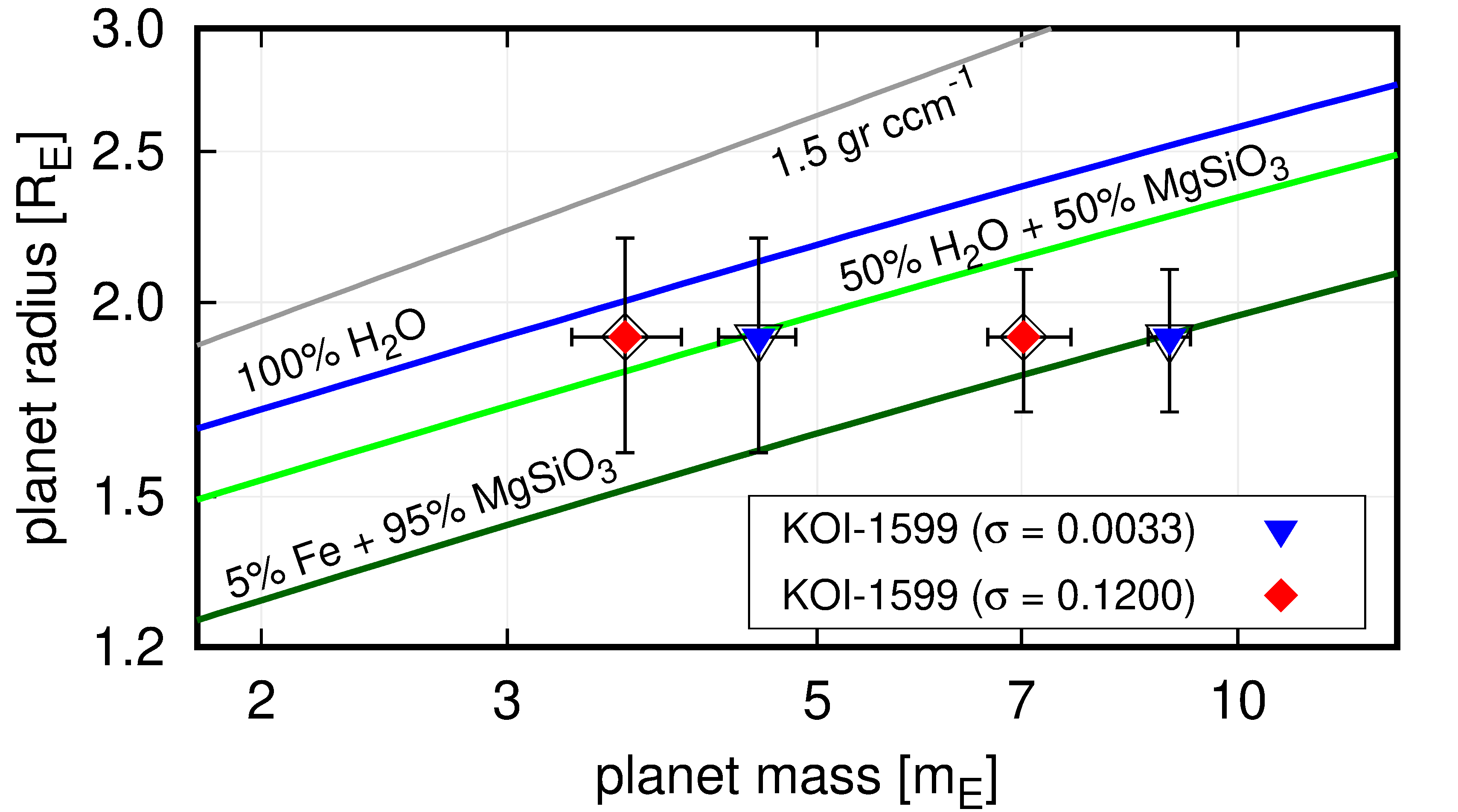}
\caption{
The mass-radius diagram for mass estimates of the single-mode $\sigma_{x_i,y_i}=0.0033$ (blue squares) and dual-mode $\sigma_{x_i,y_i}=0.1200$ (red triangles) MCMC classes of orbital configurations.
}
\label{figure:fig13}
\end{figure}

Having the relatively accurate mass and radii estimates, we may infer the densities of the two planets. Given the similar radii and the inner planet twice as massive as the outer one, the internal compositions of the planets are likely different.  We used two-layer theoretical models by \cite{Zeng2016} in order to localise the planets in the mass-radius diagram in Fig.~\ref{figure:fig13}. 

We report two different internal composition models obtained for two MCMC classes of solutions. For the MCMC-model with $\Delta\varpi=180^{\circ}$ (Tab.~\ref{table:tab3}), the inner planet has a bulk density of ($7.2 \pm 0.3$)\,\gccm{}, roughly 1.5 times the density of the Earth. The density of the second planet is $(3.7 \pm 0.3 $)\,\gccm{}, roughly the density of Mars (i.e. 3.9\,\gccm{}). As for the GEA search, also in the case of MCMC search, we have found different local best-fitting solutions, for the dual-mode solutions ($\Delta\varpi=0^{\circ},180^{\circ}$), the masses (Fig.~\ref{figure:fig3}) as well as  the derived densities are significantly smaller, (5.6 $\pm$ 0.5)\,\gccm{} and (2.9 $\pm$ 0.6)\,\gccm{}, respectively. 

Both classes of the internal structure  models predict the inner planet as a Super-Earth with a rocky envelope and a small iron-rich core. The outer planet is found in the region between rocky and icy-giant planets, populated by a new class of objects. They are called  ``Super-Venus'' planets \citep[e.g.,][]{Kane2013} if they are, as KOI-1599.02, close to the host star.
Other multi-planet systems with planets having dissimilar densities or very different radii and in a very compact orbital configuration are known and well studied, such as Kepler-29 \citep{Migaszewski2017a}, Kepler-30 \citep{Panichi2018}, and Kepler-36 \citep{Carter2012}.
Among others, Kepler-36 is the most similar system to KOI-1599, models of the structure of such systems predict planets having water in the form of a thick hot atmosphere, as molecular fluid, or ionic fluid, depending on the planet's mass, composition and UV flux \citep[e.g.,][]{Lopez2012,Lopez2013,Nettelmann2008}. 
Recently, \cite{Bodenheimer2018} propose a new scenario for the formation of Kepler-36, likely applicable also to KOI-1599. By using this new approach, which includes the accretion and dissolution of planetesimals into the gaseous envelope of the planets, the effect of in-situ formation and migration, as well as the influence of the mass loss due to XUV radiation it may be possible to understand the difference in the internal structure of the two planets in the KOI-1599 system.

%
\section{Conclusions}
\label{conclusion}
%
With the dynamical photometry, we aim to validate two transiting, candidate planets attributed to the KOI-1599 \kepler{} target.

From the Q1-Q17 DR-25 LCs, we derived the complete series of TTVs with a few hours amplitude. They exhibit a clear anti-correlation trend, indicating that the candidate planets mutually interact. Dynamical experiments make it possible to determine the upper limits of the candidate masses, roughly below 3 Jupiter masses that provide dynamically stable configurations. 
The photometric analysis of the LCs of KOI-1599 reveals that the planets have very similar radii, $\simeq 1.9$ times the Earth's radius.

The orbital model of the TTVs makes it possible to constrain the masses in the Super-Earth range.
The uncertainties may be as small as $\simeq 20\%$. However, we assumed that the orbits are low-eccentric ($e_i \simeq 0.01$) and anti-aligned, consistent with predictions of the planetary migration theory \citep[e.g.,][]{Batygin2013}. Given the similar radii, the internal compositions of the planets may be different.

Remarkably, despite of the eccentricities and $\Delta\varpi=0^{\circ},180^{\circ}$ libration modes, the best-fitting orbital models are dynamically ``easily'' involved in the 3:2 MMR.  Dynamical maps reveal that the best-fitting configurations are inside the separatrix region of the 3:2 MMR, although the critical angles can circulate or librate with amplitudes reaching $360^{\circ}$ (so only are coherent). The apparently natural libration--circulation criterion of the critical angles may lead to an incorrect identification of the MMR.

We determine if a particular configuration of the system is dynamically resonant by comparing its position in the phase space w.r.t. the location of the MMR separatrix. Additionally, by calculating the proper mean motions we reinforce the conclusion that a system may be dynamically resonant, having critical angles circulate.

The results of our statistical and dynamical experiments are suggestive of two real planets. 
By using a simple model for planet-disk interaction, we tried to reproduce the present orbital configuration of the KOI-1599 two-planet system. 
We find that for fast migration and slow eccentricity damping (with the migration--to--damping time scale ratio $\kappa \lesssim 100$), the synthetic systems well reproduce the best-fitting configurations characterised by librating critical angles and by low values of planetary masses equal to $7.0 \mE$ for the inner and $3.6 \mE$ for the outer planet, respectively.
The fast migration ($\tau \sim 10^4$ orbital periods) means rapid resonance entrance, a result of which is that the system deviates from the branch of periodic configurations. The deviation is responsible for the observed TTV signal (i.e., TTVs result from the semi-major axes modulation instead of the system rotation). The requirement of $\kappa \lesssim 100$ is necessary so the system could reach an appropriate value of the period ratio, i.e., very close to the nominal value of 3:2 MMR. Additionally, the synthetic systems are dynamically resonant, have anti-aligned orbits and small eccentricities, consistent with the ones obtained by our best-fitting configurations.

%
\section{Acknowledgements}
%
We would like to thank the anonymous referees for their review and fruitful comments that improved the quality of our manuscript. K.~G. thanks the staff of the Pozna\'n Supercomputer and Network Centre (PCSS) for the support and CPU resources (grant No.~313). This work has been supported by Polish National Science Centre MAESTRO grant DEC-2012/06/A/ST9/00276.

\bibliographystyle{mn2e}
\bibliography{ms}
\label{lastpage}

\section*{On-line material}
\setcounter{figure}{0}
\setcounter{table}{0}
\renewcommand{\thefigure}{SM\arabic{figure}}
\renewcommand{\thetable}{SM\arabic{table}}
\section{Transit time variations (this work)}

\noindent\begin{minipage}{\linewidth}
\centering
\captionof{table}{Midpoint transit times, TTVs and relative errors obtained with light-curves analysis of KOI-1599. {See Sect.~2 for details about the calculation of transit times, TTVs and relative uncertainties}.}
\label{tab:koi199ttvA}
\begin{tabular}{clllll}
\hline\hline
Planet & Transit  & Transit time & TTV & $+1\sigma$ & $-1\sigma$  \\
\smallskip
[KOI] & number &  [BJD-2454900] & [min] & [min] & [min] \\
\hline\hline
1599.02   &  0 &    73.93730     &   122.717    &   26.32     &   27.52     \\
 &           1 &    87.55370     &   87.178     &   13.15     &   14.01     \\
 &           4 &   128.40290     &   77.299     &   51.90     &   70.57     \\
 &           5 &   142.01930     &   57.686     &   14.66     &   28.18     \\
 &           6 &   155.63570     &   88.877     &   29.03     &   18.07     \\
 &           7 &   169.25211     &   91.397     &   38.87     &   59.50     \\
 &           8 &   182.86850     &   22.838     &   28.12     &   14.79     \\
 &           9 &   196.48490     &   22.925     &   66.80     &   71.63     \\
 &          10 &   210.10130     &   52.747     &   46.43     &   19.96     \\
 &          12 &   237.33411     &   49.190     &   24.77     &   19.80     \\
 &          13 &   250.95050     &   24.624     &   39.51     &   86.60     \\
 &          14 &   264.56690     &  -8.122      &   20.26     &   18.53     \\
 &          16 &   291.79971     &   6.163      &   25.17     &   23.04     \\
 &          17 &   305.41611     &   9.518      &   41.43     &   30.37     \\
 &          18 &   319.03250     &  -4.637      &   16.06     &   15.55     \\
 &          20 &   346.26529     &  -24.710     &   12.05     &   12.25     \\
 &          21 &   359.88171     &  -4.190      &   30.79     &   18.62     \\
 &          22 &   373.49811     &  -16.402     &   14.67     &   32.50     \\
 &          24 &   400.73090     &  -107.683    &   61.19     &   85.05     \\
 &          25 &   414.34729     &  -48.326     &   12.69     &   16.89     \\
 &          26 &   427.96372     &   17.122     &   50.10     &   42.39     \\
 &          27 &   441.58011     &  -41.213     &   86.34     &   26.55     \\
 &          28 &   455.19650     &  -71.885     &   25.59     &   15.96     \\
 &          29 &   468.81290     &  -61.488     &   20.35     &   33.16     \\
 &          30 &   482.42930     &  -44.597     &   12.11     &   19.99     \\
 &          31 &   496.04568     &  -68.904     &   56.76     &   56.88     \\
 &          32 &   509.66211     &  -15.768     &   43.91     &   49.32     \\
 &          33 &   523.27850     &  -54.950     &   22.65     &   22.39     \\
 &          34 &   536.89490     &  -1.541      &   30.28     &   55.09     \\
 &          35 &   550.51129     &  -99.346     &   75.11     &   58.06     \\
 &          36 &   564.12769     &  -67.982     &   34.75     &   53.27     \\
 &          37 &   577.74408     &  -92.362     &   68.90     &   69.06     \\
 &          38 &   591.36047     &  -44.496     &   24.21     &   42.57     \\
 &          39 &   604.97693     &  -56.808     &   16.75     &   82.67     \\
 &          40 &   618.59332     &  -46.340     &   33.71     &   24.55     \\
 &          41 &   632.20972     &  -3.830      &   17.99     &   27.39     \\
 &          42 &   645.82611     &  -31.666     &   27.32     &   29.84     \\
 &          45 &   686.67529     &  -83.376     &   30.34     &   19.15     \\
 &          46 &   700.29169     &  -109.051    &   46.68     &   52.10     \\
 &          47 &   713.90808     &   11.880     &   24.32     &   29.71     \\
 &          48 &   727.52448     &  -6.322      &   15.13     &   18.85     \\
 &          50 &   754.75732     &  -22.896     &   15.61     &   12.67     \\
 &          51 &   768.37372     &  -34.099     &   23.23     &   14.49     \\
 &          52 &   781.99011     &  -41.098     &   60.81     &   77.79     \\
 &          53 &   795.60651     &  -31.853     &   61.23     &   17.74     \\
 &          54 &   809.22290     &  -19.642     &   15.77     &   33.50     \\
 &          55 &   822.83929     &  -4.219      &   27.62     &   23.92     \\
 &          56 &   836.45569     &  -28.886     &   25.16     &   89.21     \\
 \hline
&  & & & & Continue \\
\hline\hline
\end{tabular}
\end{minipage}
\begin{table}
\centering
\caption{Transit midpoint times, TTVs and relative errors obtained from the MCMC light-curve analysis.}
\label{tab:koi1599ttvB}
\begin{tabular}{crrrrr}
\hline\hline
Planet  & Transit  & Transit time & TTV & $+1\sigma$ & $-1\sigma$  \\
\smallskip
[KOI] & number &  [BJD-2454900] & [min] & [min] & [min] \\
\hline\hline 
 1599.02 &  57 &   850.07208     &  -19.886     &   26.32     &   27.67     \\
 &          58 &   863.68848     &  -25.747     &   11.09     &   11.74     \\
 &          59 &   877.30487     &  -80.165     &   54.96     &   52.08     \\
 &          60 &   890.92133     &  -92.448     &   55.38     &   27.55     \\
 &          61 &   904.53772     &   16.157     &   32.64     &   63.01     \\
 &          62 &   918.15411     &  -35.309     &   28.63     &   18.12     \\
 &          63 &   931.77051     &   31.694     &   30.90     &   30.87     \\
 &          65 &   959.00330     &  -10.267     &   19.63     &   23.08     \\
 &          66 &   972.61970     &   2.851      &   36.83     &   33.88     \\
 &          67 &   986.23608     &  -0.346      &   43.42     &   38.72     \\
 &          68 &   999.85248     &   41.242     &   51.36     &   25.76     \\
 &          69 &   1013.46887     &   40.003     &   8.12      &   15.26     \\
 &          73 &   1067.93445     &  -30.370     &   29.15     &   27.43     \\
 &          74 &   1081.55090     &   55.022     &   25.70     &   17.77     \\
 &          76 &   1108.78369     &   19.613     &   42.65     &   67.22     \\
 &          77 &   1122.40015     &   90.288     &   82.45     &   61.21     \\
 &          78 &   1136.01648     &   19.886     &   23.07     &   26.80     \\
 &          79 &   1149.63293     &   93.802     &   47.79     &   34.76     \\
 &          80 &   1163.24927     &   35.021     &   16.96     &   15.75     \\
 &          81 &   1176.86572     &   47.333     &   18.26     &   16.81     \\
 &          82 &   1190.48206     &   62.467     &   20.33     &   91.28     \\
 &          83 &   1204.09851     &   4.6800     &   19.02     &   12.90 \\
 &          84 &   1217.71484     &   35.381     &   25.11     &   31.38     \\
 &          85 &   1231.33130     &   36.446     &   32.13    &   24.77     \\ 
 &          86 &   1244.94775     &  -9.461     &   52.92     &   34.07     \\
 &          87 &   1258.56409     &   92.909     &   19.60     &   41.59     \\
 &          88 &   1272.18054     &   57.370     &   47.48     &   58.45     \\
 &          89 &   1285.79688     &   100.901     &   61.20     &   43.57     \\
 &          90 &   1299.41333     &   26.770     &   19.12     &   60.91     \\
 &          91 &   1313.02966     &   61.272     &   73.15     &   50.44     \\
 &          92 &   1326.64612     &   47.160     &   17.01     &   12.73     \\
 &          93 &   1340.26245     &   29.491     &   63.12     &   99.36     \\
 &          94 &   1353.87891     &  -80.093     &   23.10     &   31.26     \\
 &          95 &   1367.49524     &   62.712     &   23.46     &   36.91     \\
 &          96 &   1381.11169     &   80.611     &   32.72     &   37.64     \\
 &          97 &   1394.72815     &   30.442     &   15.51     &   21.10     \\
 &          99 &   1421.96094     &  -18.360     &   31.67     &   494.97     \\
 &         100 &   1435.57727     &   4.018     &   23.95     &   18.48     \\
 &         101 &   1449.19373     &  -5.155     &   19.63     &   26.14     \\
 &         102 &   1462.81006     &   2.203     &   23.88     &   64.70     \\
 &         103 &   1476.42651     &   12.658     &   14.01     &   13.78     \\
 &         104 &   1490.04285     &  -21.082     &   31.70     &   35.78     \\
 &         105 &   1503.65930     &  -11.635     &   12.33     &   19.02     \\
   \hline
1599.01     &  0 &   73.14220     &   -263.002     &   29.48     &   18.81     \\
 &           1 &   93.55050     &   -242.870     &   21.77     &   33.16     \\
 &           2 &   113.95880     &  -205.661     &   37.12     &   20.82     \\
 &           3 &   134.36710     &  -181.958     &   17.97     &   18.54     \\
 &           4 &   154.77541     &  -165.600     &   27.56     &   42.97     \\
 &          5 &   175.18370     &   -77.818     &   20.39     &   29.22     \\
 &           6 &   195.59200     &   -91.858     &   20.13     &   22.23     \\
 &           7 &   216.00030     &   -40.910     &   23.54     &   28.34     \\
 &           8 &   236.40860     &   -48.312     &   33.67     &   17.61     \\
 &           9 &   256.81690     &    19.584     &   80.48     &   50.67     \\
 &          10 &   277.22520     &   9.677     &   24.72     &   21.86     \\
 &          11 &   297.63351     &   9.620     &   33.22     &   46.18     \\
 &          12 &   318.04181     &   59.414     &   14.00     &   41.34     \\ 
 &          13 &   338.45010     &   10.757     &   46.18     &   17.38     \\ 
 & 	        14 &   358.85840     &   69.509     &   21.72     &   17.70     \\
 &          15 &   379.26670     &   108.950     &   29.38     &   28.18     \\
  \hline
&  & & & & Continue \\ 
\hline\hline
\end{tabular}
\end{table}
\begin{table}
\centering
\caption{Transit midpoint times, TTVs and relative errors obtained from the MCMC light-curve analysis.}
\label{tab:koi1599ttvC}
\begin{tabular}{clllll}
\hline\hline
Planet  & Transit  & Transit time & TTV & $+1\sigma$ & $-1\sigma$  \\
\smallskip
[KOI] & number &  [BJD-2454900] & [min] & [min] & [min] \\
\hline\hline   
1599.01   &   16 &   399.67500     &   100.051     &   15.51     &   39.77     \\
 &          17 &   420.08331     &   89.410     &   21.90     &   24.70     \\
 &          18 &   440.49161     &   122.342     &   43.59     &   24.51     \\
 &          19 &   460.89990     &   129.082     &   16.20     &   23.24     \\
 &          20 &   481.30820     &   115.474     &   19.70     &   27.14     \\
 &          21 &   501.71649     &   94.291     &   24.71     &   47.33     \\ 
 &          22 &   522.12482     &   134.525     &   22.61     &   42.60     \\
 &          23 &   542.53308     &   142.891     &   20.02     &   16.80     \\
 &          25 &   583.34967     &   153.274     &   30.27     &   43.06     \\
 &          26 &   603.75800     &   98.986     &   15.48     &   20.53     \\
 &          27 &   624.16632     &   152.554     &   26.31     &   8.09     \\
 &          28 &   644.57459     &   126.850     &   25.76     &   37.27     \\
 &          30 &   685.39117     &   120.715     &   22.84     &   15.25     \\
 &          31 &   705.79950     &   116.525     &   77.17     &   13.69     \\
 &          32 &   726.20783     &   105.106     &   24.68     &   30.43     \\
 &          33 &   746.61609     &   65.866     &   37.24     &   28.93     \\
 &          34 &   767.02441     &   71.755     &   24.83     &   51.75     \\
 &          35 &   787.43268     &   75.802     &   62.48     &   22.02     \\
 &          36 &   807.84100     &   94.061     &   72.29     &   73.74    \\
 &          37 &   828.24933     &   63.245     &   19.31     &   11.03     \\
 &          38 &   848.65759     &   47.808     &   29.88     &   41.18     \\
 &          39 &   869.06592     &   65.088     &   36.92     &   41.64     \\
 &          40 &   889.47418     &     18.677     &   22.92     &   31.45     \\
 &          41 &   909.88250     &     -0.562     &   32.80     &   23.72     \\
 &          42 &   930.29077     &      5.112     &   15.11     &   16.57     \\
 &          43 &   950.69910     &      5.904     &   18.07     &   26.68     \\
 &          44 &   971.10742     &    -42.509     &   31.90     &   33.24     \\
 &          45 &   991.51569     &    -42.206     &   28.71     &   33.93     \\
 &          46 &   1011.92401     &   -51.624     &   39.14     &   41.36     \\ 
 &          48 &   1052.74060     &   -19.598     &   77.99    &   21.70     \\
 &          49 &   1073.14893     &   -15.062     &   21.54     &   18.81     \\
 &          50 &   1093.55725     &   -53.136     &   81.56     &   64.84     \\
 &          51 &   1113.96545     &   -46.843     &   21.87     &   14.08     \\
 &          52 &   1134.37378     &   -73.930     &   21.89     &   21.18     \\
 &          53 &   1154.78210     &   -82.642     &   19.21     &   18.82     \\
 &          54 &   1175.19043     &   -56.938     &   15.13     &   22.42     \\
 &          55 &   1195.59875     &  -128.750     &   27.35     &   30.90     \\
 &          56 &   1216.00696     &   -68.242     &   17.19    &   30.82     \\
 &          57 &   1236.41528     &   -98.654     &   14.96     &   30.11     \\
 &          58 &   1256.82361     &   -77.242     &   11.17     &   27.06     \\
 &          59 &   1277.23193     &  -107.136     &   26.60     &   47.51     \\
 &          60 &   1297.64026     &   -71.280     &   22.09     &   16.46     \\
 &          61 &   1318.04846     &  -113.717     &   33.83     &   45.85     \\
 &          62 &   1338.45679     &  -28.843     &   17.27     &   43.55     \\
 &          63 &   1358.86511     &  -64.094     &   25.73     &   13.69     \\
 &          64 &   1379.27344     &   1.454     &   26.96     &   8.04    \\
 &          65 &   1399.68164     &  -7.978     &   23.59     &   21.33     \\
 &          67 &   1440.49829     &   21.542     &   10.83     &   12.99     \\
 &          68 &   1460.90662     &   87.394     &   35.29     &   47.29     \\
 &          69 &   1481.31494     &   42.768     &   64.24     &   19.24     \\
 &          70 &   1501.72314     &   48.989     &   21.43     &   13.38     \\
 &          71 &   1522.13147     &   51.250     &   38.33    &   28.32     \\
  \hline
&  & & & & Continue \\ 
\hline\hline
\end{tabular}
\end{table}
%
\end{document}

%% file: macros.tex
\def\idm#1{{\mbox{\scriptsize #1}}}
\def\vec#1{{\pmb #1}}
\def\kepler{{\em Kepler}}
\def\code#1{{\sc #1}}
\def\Y{\langle Y \rangle}

\def\vec#1{{\pmb #1}}  
\DeclareMathAlphabet{\mathsfit}{\encodingdefault}{\sfdefault}{m}{sl}
\SetMathAlphabet{\mathsfit}{bold}{\encodingdefault}{\sfdefault}{bx}{sl}

\newcommand{\au}{\mbox{au}} 
\newcommand{\msun}{\mbox{M}_{\odot}}
\newcommand{\Rsun}{\mbox{R}_{\odot}}
\newcommand{\mE}{\mbox{M}_{\oplus}}
\newcommand{\mJ}{\mbox{M}_{\idm{Jup}}}
\newcommand{\RE}{\mbox{R}_{\oplus}}

\newcommand{\Mmean}{\mathcal{M}}

\newcommand{\Tsp}{T_{\idm{s-p}}}
\newcommand{\ratio}{\langle P_2/P_1\rangle_{\idm{obs}}}
\newcommand{\Pinobs}{\langle P_1\rangle_{\idm{obs}}}
\newcommand{\Poutobs}{\langle P_2\rangle_{\idm{obs}}}
\newcommand{\Pobss}{\langle P\rangle_{\idm{obs}}}
\newcommand{\emean}{\langle e\rangle}
\newcommand{\Asp}{A_{\idm{s-p}}}
\newcommand{\Ares}{A_{\idm{res}}}
\newcommand{\Ainsp}{A_{\idm{s-p,1}}}
\newcommand{\Ainres}{A_{\idm{res,1}}}
\newcommand{\Aoutsp}{A_{\idm{s-p,2}}}
\newcommand{\Aoutres}{A_{\idm{res,2}}}
\newcommand{\phisp}{\Phi_{\idm{s-p}}}
\newcommand{\phiinsp}{\Phi_{\idm{s-p,1}}}
\newcommand{\phioutsp}{\Phi_{\idm{s-p,2}}}
\newcommand{\phires}{\Phi_{\idm{res}}}
\newcommand{\phiinres}{\Phi_{\idm{res,1}}}
\newcommand{\phioutres}{\Phi_{\idm{res,2}}}
\newcommand{\phieq}{\phi_1^{\idm{(eq)}}}
\newcommand{\Nsp}{N_{\idm{s-p}}}
\newcommand{\Nres}{N_{\idm{res}}}
\newcommand{\Tres}{T_{\idm{res}}}

\newcount\exno

\definecolor{myred}{rgb}{0.89, 0.0, 0.13}
\definecolor{myblue}{rgb}{0.1,0.0,0.6} 
\definecolor{mybrown}{rgb}{0.9,0.4,0.3}
\newcommand\ednote[1]{{\global\advance\exno by 1\
    \color{myred}$\spadesuit$({\bfseries\the\exno}).
\color{myblue}\bfseries\em #1}}

\newcommand\hide[1]{}

\def\kepler{{\sc Kepler}}
\def\K2{{\sc K2}}

\def\idm#1{{\mbox{\scriptsize #1}}}
\newcommand\Chi{{\chi^2_\nu}}

\newcommand\gccm{{\,g$\cdot${}cm$^{-3}$}}